\documentclass[twocolumn,showpacs,preprintnumbers,nofootinbib,prd,
superscriptaddress,10pt]{revtex4-1}

\usepackage{amsmath,amssymb}
\usepackage[normalem]{ulem}
\usepackage{textcomp}
\usepackage{hyperref}
\usepackage{bm}
\usepackage{graphicx}
\usepackage{psfrag}
\usepackage[usenames,dvipsnames]{xcolor}
\usepackage{aas_macros}
\usepackage{multirow}
\usepackage[utf8]{inputenc}

\allowdisplaybreaks[4]

\newcommand{\pd}{\partial}

\newcommand{\uvec}[1]{\bm{\hat{#1}}}
\newcommand{\dvec}[1]{\dot{\bm{#1}}}

\newcommand{\duvec}[1]{\dot{\bm{\hat{#1}}}}

\newcommand{\be}{\begin{equation}}
\newcommand{\ee}{\end{equation}}
\newcommand{\beq}{\begin{eqnarray}}
\newcommand{\eeq}{\end{eqnarray}}
\newcommand{\sub}[1]{_{\text{#1}}}
\newcommand{\super}[1]{^{\text{#1}}}

\def\ph{\phi_{lmn}}
\def\Qlm{Q_{lmn}}

\begin{document}

\title{Science with the space-based interferometer eLISA.\\I: Supermassive black hole binaries}

\date{\today}

\author{Antoine Klein}
\affiliation{Department of Physics and Astronomy, The University of 
Mississippi, University, MS 38677,
USA}
\author{Enrico Barausse}
\affiliation{Sorbonne Universit\'es, UPMC Univesit\'e Paris 6, UMR 7095, 
Institut d'Astrophysique de Paris, 98 bis Bd Arago, 75014 Paris, France}
\affiliation{CNRS, UMR 7095, Institut d'Astrophysique de Paris, 98 bis Bd Arago, 
75014 Paris, France}
\author{Alberto Sesana}
\affiliation{School of Physics and Astronomy, The University of Birmingham, 
Edgbaston, Birmingham B15 2TT, UK}
\author{Antoine Petiteau}
\affiliation{APC, Universit\'e Paris Diderot, Observatoire de Paris, Sorbonne 
Paris Cit\'e,
10 rue Alice Domon et L\'eonie Duquet, 75205 Paris Cedex 13, France}
\author{Emanuele Berti}
\affiliation{Department of Physics and Astronomy, The University of 
Mississippi, University, MS 38677, USA}
\affiliation{CENTRA, Departamento de F\'isica, Instituto Superior
T\'ecnico, Universidade de Lisboa, Avenida Rovisco Pais 1,
1049 Lisboa, Portugal}
\author{Stanislav Babak}
\affiliation{Max Planck Institute for Gravitational Physics, Albert Einstein 
Institute, Am M\"uhlenberg 1, 14476 Golm, Germany}
\author{Jonathan Gair}
\affiliation{Institute of Astronomy, University of Cambridge, Cambridge, CB3 0HA, UK}
\affiliation{School of Mathematics, University of Edinburgh, The King's Buildings, Peter Guthrie Tait Road, Edinburgh, EH9 3FD, UK}
\author{Sofiane Aoudia}
\affiliation{Laboratoire de Physique Th\'{e}orique, Facult\'{e} des Sciences Exactes Universit\'{e} de Bejaia, 06000 Bejaia, Algeria}
\author{Ian Hinder}
\affiliation{Max Planck Institute for Gravitational Physics, Albert Einstein 
Institute, Am M\"uhlenberg 1, 14476 Golm, Germany}
\author{Frank Ohme}
\affiliation{School of Physics and Astronomy, Cardiff University, Queens Building, CF24 3AA, Cardiff, United Kingdom}
\author{Barry Wardell}
\affiliation{Department of Astronomy, Cornell University, Ithaca, NY
  14853, USA}
\affiliation{School of Mathematical Sciences and Complex \& Adaptive Systems Laboratory,
University College Dublin, Belfield, Dublin 4, Ireland}

\begin{abstract}
  We compare the science capabilities of different eLISA mission
  designs, including four-link (two-arm) and six-link (three-arm)
  configurations with different arm lengths, low-frequency
  noise sensitivities and mission durations. For each of these
  configurations we consider a few representative massive black hole
  formation scenarios. These scenarios are chosen to explore two
  physical mechanisms that greatly affect eLISA rates, namely (i)
  black hole seeding, and (ii) the delays between the merger of two
  galaxies and the merger of the black holes hosted by those galaxies.
  We assess the eLISA parameter estimation accuracy using a Fisher
  matrix analysis with spin-precessing, inspiral-only waveforms. We
  quantify the information present in the merger and ringdown by
  rescaling the inspiral-only Fisher matrix estimates using the
  signal-to-noise ratio from non-precessing inspiral-merger-ringdown
  phenomenological waveforms, and from a reduced set of precessing
  numerical relativity/post-Newtonian hybrid waveforms. We find that
  all of the eLISA configurations considered in our study should
  detect some massive black hole binaries. However, configurations
  with six links and better low-frequency noise will provide much more
  information on the origin of black holes at high redshifts and on
  their accretion history, and they may allow the identification of
  electromagnetic counterparts to massive black hole mergers.
\end{abstract}


\maketitle

\section{Introduction}
Gravitational waves (GWs) are a generic prediction of general relativity (GR)~\cite{1918SPAW.......154E} and of other relativistic gravitational theories~\cite{Gair:2012nm,Yunes:2013dva,Berti:2015itd}.
Indirect evidence for the existence of GWs comes from observations of binary systems involving at least one pulsar~\cite{Hulse:1974eb}, which allow us to track the orbital period very accurately over long timescales and to observe small secular changes
due to the emission of GWs. The observed damping is in sub percent-level agreement with the predictions of GR's quadrupole formula for GW emission~\cite{Damour:1991rd,Damour:1990wz}, and overall the combined measurements of secular changes are consistent with the predictions of GR within 0.05\% \cite{2006Sci...314...97K}.

A world-wide experimental effort towards a direct detection of GWs is also 
underway. Ground-based, km-scale laser interferometers target the GW emission 
from a variety of sources, including the late inspiral of neutron star binaries; 
the inspiral, merger and ringdown of systems comprised of two stellar-mass black 
holes, or a neutron star and a stellar-mass black hole; supernova explosions and 
isolated pulsars.  These ground-based interferometers work as a network of 
``second-generation'' detectors, as opposed to the first generation of 
ground-based interferometers (i.e., the initial LIGO and Virgo experiments), 
which were active from 2002 to 2010.  They include the two Advanced 
LIGO~\cite{ligo} interferometers in the US (which are currently taking data in 
science mode) and the French-Italian detector Advanced Virgo~\cite{virgo} (which 
will undergo commissioning in 2016). Within the next few years the Japanese 
interferometer KAGRA~\cite{kagra} and a LIGO-type detector in 
India~\cite{indigo} will join this network.

At the same time, pulsar timing arrays~\cite{1990ApJ...361..300F} are targeting 
the GW signal from binaries of massive black holes (MBHs) with masses $\sim 
10^8$--$10^{10} M_\odot$ at separations of hundreds to thousands of 
gravitational radii. Low-frequency GWs emitted by these systems passing between 
a pulsar and the Earth leave a characteristic imprint in the time of arrival of 
the radio pulses. This signal can be disentangled from other sources of noise 
and unambiguously identified by timing an ensemble of ultra-stable millisecond 
pulsars (i.e.,~a pulsar timing array)~\cite{1983ApJ...265L..39H}.  The European 
Pulsar Timing Array (EPTA)~\cite{2013CQGra..30v4009K}, the Parkes Pulsar Timing 
Array (PPTA)~\cite{2013CQGra..30v4007H} and the North American Nanohertz 
Observatory for Gravitational Waves (NANOGrav)~\cite{2013CQGra..30v4008M}, 
joining together in the International Pulsar Timing Array 
(IPTA)~\cite{2010CQGra..27h4013H,2013CQGra..30v4010M}, are constantly improving 
their sensitivity in the frequency range $\sim10^{-9}-10^{-6}$ Hz, and have 
already significantly constrained the stochastic GW signal from the massive 
black hole binary (MBHB) 
population~\cite{2015MNRAS.453.2576L,2015arXiv150803024A,Shannon:2015ect}.

A common feature of current ground-based interferometers and pulsar-timing arrays is that they primarily probe the low-redshift Universe. For instance, pulsar-timing arrays are mostly sensitive to MBHBs at $0.2\lesssim z\lesssim1.5$, while the range of ground-based detectors depends on the particular family of sources being observed, but never exceeds $z\sim 2$~\cite{Dominik:2014yma,Belczynski:2014iua,Belczynski:2015tba}. The two classes of experiments explore hugely different frequency ranges for the GW signal ($f\sim 10^{-9}$--$10^{-6}$ Hz for pulsar-timing array, and $f\sim 10-10^3$ Hz for ground-based interferometers), and the GW frequency band in between these widely separated bands is difficult to probe from the ground. Pulsar-timing arrays are intrinsically limited at the high-frequency end by the Nyquist frequency, set by the typical interval between subsequent pulsar observation sessions (typically a few weeks, corresponding to roughly $10^{-6}$ Hz).
Second-generation ground-based detectors are intrinsically limited at 
frequencies lower than $f\sim 10 $ Hz by ``seismic noise'' (the main noise 
source is actually given not by geological activity, but by vibrations due to 
surface events: human activity, waves, wind, etc.). While third-generation 
ground-based interferometers (such as the Einstein Telescope~\cite{ET}), which 
have been proposed for construction in the next decade, might abate seismic 
noise by going underground, they will not be able to probe frequencies lower 
than $f\sim 0.1$ Hz~\cite{Harms:2013raa}.

However the frequency window between $10^{-7}$ Hz and $\sim 1$ Hz is expected to be populated by a rich variety of astrophysical sources. For instance, our current understanding of galaxy formation and
 an increasing body of observational evidence support the idea that MBHs (with masses from $10^{5} M_\odot$ -- or even lower -- up to $10^{9}-10^{10} M_\odot$) are hosted in the centers
of almost all galaxies~\cite{Kormendy:1995er}. When galaxies merge to form 
bigger systems [as predicted by the $\Lambda$ cold dark-matter ($\Lambda$CDM) 
model and supported by observations], these MBHs are expected to form tight 
binaries~\cite{1980Natur.287..307B}, inspiral and finally merge into a perturbed 
black hole, which sheds away these perturbations in the ringdown 
phase~\cite{Berti:2005ys,Berti:2009kk}. The inspiral, merger and ringdown phases 
are expected to be the main GW source
in the $10^{-8}-1$ Hz spectrum. Their detection will provide precious information on the co-evolution between MBHs and their host galaxies (and henceforth on the hierarchical formation of
structures in the $\Lambda$CDM model), on the very origin of MBHs at high redshift, on the dynamics of gas and accretion onto MBHs, and on the strong-field, highly relativistic dynamics of GR (see e.g.~\cite{Plowman:2009rp,Plowman:2010fc,Gair:2010bx,Sesana:2010wy}, and~\cite{Barausse:2014oca} for a review of the science achievable with GW observations of MBHBs).
Moreover, if these sources could also be identified in the electromagnetic band, they may be used as ``standard sirens'' to probe the expansion history of the Universe~\cite{Schutz:1986gp,2005ApJ...629...15H}.

Other sources that may populate this unprobed frequency band include binary 
systems formed of a compact object (i.e., a stellar-mass black hole, a neutron 
star or a white dwarf) orbiting a MBH~\cite{AmaroSeoane:2007aw}. The observation 
of the GW signal  from these
systems, known as extreme mass ratio inspirals (EMRIs), would provide a way to test GR and the geometry of MBHs with unprecedented accuracy, as well as a way to understand
the dynamics of stellar objects in galactic nuclei. Also present in this band will be countless almost monochromatic sources from our own Galaxy, 
e.g. neutron-star or white-dwarf binary systems at wide separations. White-dwarf
binary systems will be particularly numerous, especially at $f\sim 10^{-4}-10^{-5}$ Hz~\cite{0264-9381-18-19-306}, and may give rise to a background GW signal only partially resolvable as individual sources.
Finally, more exotic sources may also be present, for instance stochastic background signals of cosmological origin, arising from new physics at the TeV energy scale or beyond, such as cosmic strings or a first-order electroweak phase transition~\cite{2009PhRvD..79h3519C}.

Given its tremendous potential for fundamental physics and astrophysics, the European Space Agency (ESA)
has selected the observation of the Universe at GW frequencies around one mHz 
as one of the three main science themes of the ``Cosmic Vision 
Program''~\cite{cosmic_vision}. Indeed, a call for mission
proposals for the ``Gravitational Universe'' science theme is expected for late 2016, and the L3 launch slot in 2034 has been reserved for the selected mission. The main candidate mission
for this call (for which a decision will be made by 2018-19, so as to allow sufficient time for industrial production before the nominal 2034 launch date) is 
the evolving Laser Interferometer Space Antenna (eLISA)~\cite{Seoane:2013qna}, named after the ``classic LISA'' concept of the late 90's and early 2000s~\cite{Bender:1996cv}. 
The eLISA mission concept consists of a constellation of three spacecraft, trailing the Earth around the Sun at a distance of 
about fifteen degrees. Each spacecraft will contain one or two
test masses in almost perfect free fall, and laser transponders which will allow measurements of the relative proper distances of the test masses in different spacecraft
via laser interferometry. This will allow the detection of the effect of possible GW signals (which would change the distance between the test masses).
The most technically challenging aspect of the mission will be to maintain the 
test masses in almost perfect free fall. For this reason, a scaled-down version 
of one of eLISA's laser links will be tested by the ``LISA Pathfinder'' mission. 
Pathfinder was launched by ESA in December 2015, and it will
provide crucial tests of how well eLISA's low-frequency acceleration noise can be suppressed.

There are, however, other aspects to the eLISA mission that are yet to be evaluated and decided upon by ESA, within the constraints imposed by the allocated budget for the ``Gravitational Universe'' science theme. A ``Gravitational Observatory Advisory Team'' (GOAT)~\cite{goat} 
has been established by ESA to advise on the scientific and technological issues pertaining to an eLISA-like mission.
Variables that affect the cost of the mission include: (i) the already
mentioned low-frequency acceleration noise; (ii) the mission lifetime,
which is expected to range between one and several years, with longer
durations involving higher costs because each component has to be
thoroughly tested for the minimum duration of the mission, and may also require higher fuel consumption, since the orbital stability of the triangular constellation sets an upper
limit on the mission duration and therefore achieving a longer mission may 
require the constellation to be further from the Earth; (iii) the length $L$ of 
the
constellation arms, which may range from one to several million km,
with longer arms involving higher costs to put the constellation into
place and to maintain a stable orbit and slowly varying distances
between the spacecraft; (iv) the number of laser links between the
spacecraft, i.e., the number of ``arms'' of the interferometer (with
four links corresponding to two arms, i.e., only one interferometer,
and six links to three arms, i.e., two independent interferometers at
low frequencies~\cite{cutler-98}): giving up the third arm would cut
costs (mainly laser power, industrial production costs), while
possibly hurting science capabilities (especially source localization)
and allowing for no redundancy in case of technical faults in one of
the laser links.

This paper is the first in a series that will evaluate the impact of
these four key design choices on the scientific performance of eLISA.
Here we focus on the main scientific target of eLISA, namely the
inspiral, merger and ringdown of MBHBs. We assess how the number of
observed sources, their distance, and the accuracy with which their
parameters can be extracted from the data change under different
design choices. In subsequent papers in this series we will repeat
this exercise for other sources/science capabilities of eLISA, namely
(a) EMRIs and their science impact; (b) the measurement of the
expansion history of the Universe (``cosmography''); (c) Galactic
white-dwarf binaries and their science impact; (d) 
the stochastic background from a first order phase transition in the early universe.

The plan of the paper is as follows. In Sec.~\ref{design} we review
the different mission designs used in our analysis, and how the
corresponding noise curves are produced. In Sec.~\ref{BH-evolution} we
describe our MBH evolution models. In Sec.~\ref{WF-PE} we review the
gravitational waveforms, instrument response model and data analysis
tools used in the present work. In Sec.~\ref{results} we present the
results of the parameter estimation study in detail. Some conclusions
and a discussion of possible directions for future work are given in
Sec.~\ref{conclusions}. Appendix~\ref{app:Hybrid} describes the
construction of the precessing-binary hybrid waveforms that we used to
extrapolate our inspiral-only error estimates on sky location and
distance determination. In Appendix~\ref{app:ringdown} we give a
simple analytical prescription to estimate the error on the remnant
spin from the ringdown radiation. Unless otherwise specified,
throughout this paper we adopt geometrical units ($G=c=1$).

\section{eLISA mission designs}
\label{design}

We investigated six detector noise
curves 
by varying two key parameters characterizing the noise, namely: (1)
the arm length $L$, chosen to be either 1, 2 or $5\times 10^6$~km (A1, A2 and 
A5, respectively); (2) the low-frequency
acceleration, that is either projected from the expected performance
of LISA Pathfinder (N2) or 10 times worse, assuming a very pessimistic
outcome of LISA Pathfinder (N1).
The laser power and diameter of the telescope have been adjusted based
on the interferometer arm length to get similar sensitivities at high
frequencies. 
We considered a
laser power of 0.7~W for configuration A1 and 2~W for configurations A2
and A5;
the telescope mirror size has been chosen to be 25~cm for A1, 28~cm for
A2, 40~cm for A5. Note that fixing a 2~W laser and 40~cm telescope
improves the high-frequency performance of some configurations, but
this only affects the high-frequency noise, and it has a very mild
impact on the study presented here.

\begin{figure*} \includegraphics[scale=0.6,clip=true,angle=0]{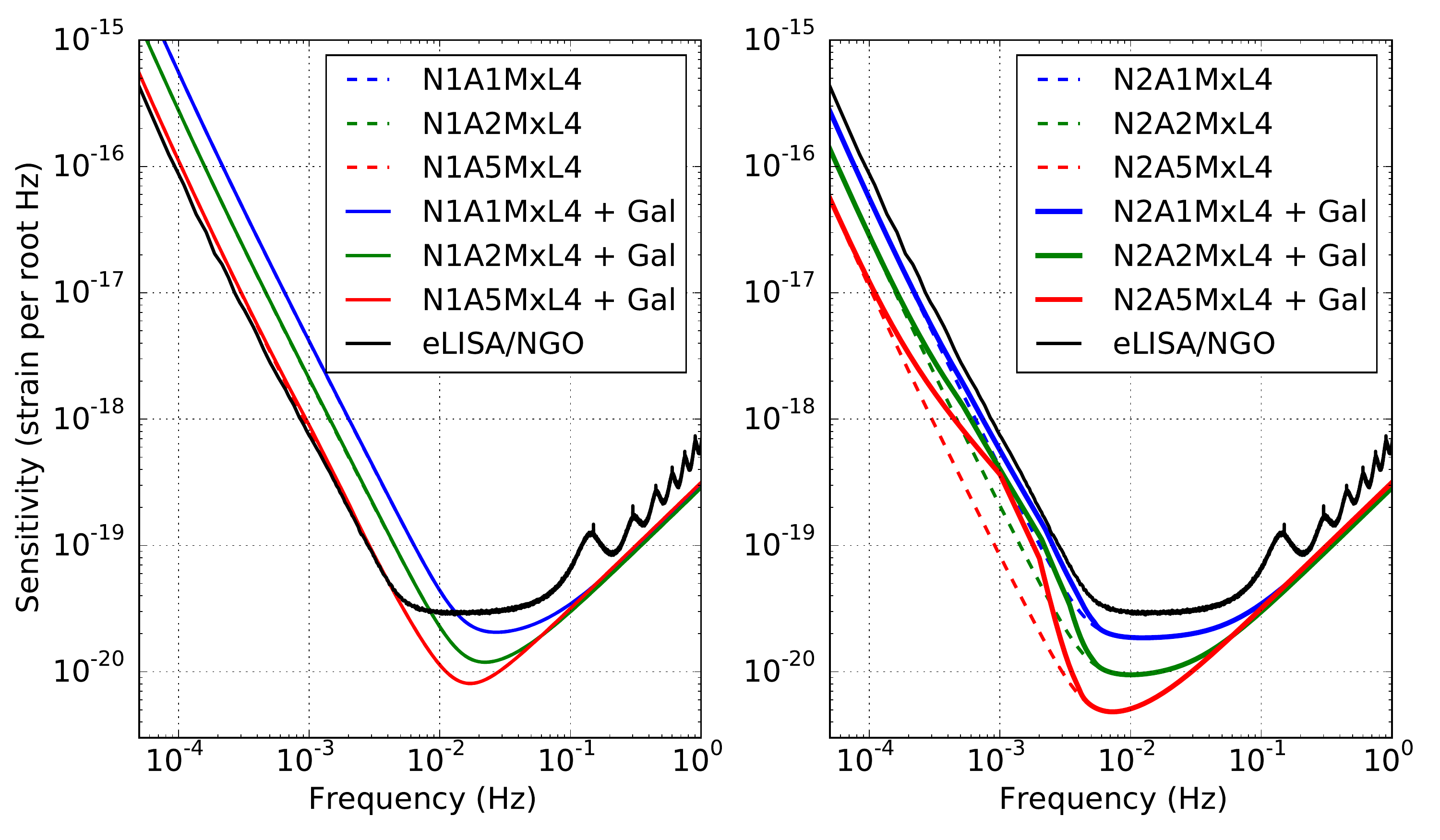} 
\caption{Analytic fits to the sensitivity curves for different configurations 
investigated in this paper.  In both panels the curves running from the top down 
are for A1 (blue), A2 (green) and A5 (red) configurations, and all curves have 
the same high-frequency noise by design. The (black) curve with wiggles at high 
frequency is the numerical sensitivity for eLISA/NGO, which is shown in both 
panels for reference. Dashed curves include only instrumental noise, while solid 
lines represent the total noise (including the contribution from CWD confusion 
noise). The left panel shows all configurations with pessimistic (N1) 
acceleration noise levels; the right panel assumes optimistic (N2) acceleration 
noise levels.}
\label{fig1}
\end{figure*}

Analytic fits to the six sky-averaged sensitivity curves obtained in
this way are of the form
\begin{eqnarray}
  S_n(f)&=&\frac{20}{3}\frac{4S_{n,\rm acc}(f)+S_{n,\rm sn}(f)+S_{n,\rm
      omn}(f)}{L^2} \nonumber \\
&\times& \left[1+\left(\frac{f}{0.41\frac{c}{2L}}\right)^2\right]. 
\label{eq:sens}
\end{eqnarray}
Here $L$ is the arm length in meters, and $S_{n,\rm acc}(f)$,
$S_{n,\rm sn}(f)$ and $S_{n,\rm omn}$ denote the noise components due
to low-frequency acceleration, shot noise and other measurement noise,
respectively. We use the following values:
\begin{multline}
  S_{n,\rm acc}(f)=\\
\begin{cases}
  9\times10^{-28} \frac{1}{(2\pi f)^4}\left(1+\frac{10^{-4}{\rm
        Hz}}{f}\right)\,\rm{m}^2\,{\rm Hz}^{-1} & \mbox{for N1},\\  
  9\times10^{-30} \frac{1}{(2\pi f)^4}\left(1+\frac{10^{-4}{\rm
        Hz}}{f}\right)\,\rm{m}^2\,{\rm Hz}^{-1} & \mbox{for N2},  
\nonumber
\end{cases}
\end{multline}
\begin{equation}
  S_{n,\rm sn}(f)=
\begin{cases}
  1.98\times10^{-23} \rm{m}^2\,{\rm Hz}^{-1} & \mbox{for A1},\\  
  2.22\times10^{-23} \rm{m}^2\,{\rm Hz}^{-1} & \mbox{for A2},\\  
  2.96\times10^{-23} \rm{m}^2\,{\rm Hz}^{-1} & \mbox{for A5},
\nonumber
 \end{cases}
\end{equation}
\begin{equation}
S_{n,\rm omn}(f)=2.65\times10^{-23} \rm{m}^2\,{\rm Hz}^{-1} \,\,\, \mbox{for all configurations}.
\nonumber
\label{eq:fit}
\end{equation}
%
Here $S_{n,\rm omn}$ -- for ``other measurement noise,'' as estimated
in the eLISA study of~\cite{Seoane:2013qna} -- might vary across
configurations, however we keep it fixed according to the most
conservative choice. The noise curves obtained in this way are shown
in Fig.~\ref{fig1}. 
%

Each analytic curve was compared to its numerical counterpart
generated by the {\sc LISACode} simulator. We found very good
agreement in all cases, the only difference being that the
analytic fit does not reproduce the high-frequency oscillatory
behavior. This is because the analytic fit assumes the long-wavelength
approximation, which breaks down at
$f = c/(2\pi L) \approx 0.05 (1\rm{Gm}/L)$~Hz. Still, the analytic
approximation is sufficient for our purposes, since most of the
relevant sources emit radiation at $f\lesssim 0.05$~Hz.

In addition to the instrumental noise, we expect an astrophysical
foreground coming from the compact white dwarf (CWD) binaries in our
Galaxy. Millions of galactic binaries emit almost monochromatic GW
signals which superpose with random phase, creating an unresolved
stochastic foreground above a few mHz. Sufficiently loud signals
(standing above the background) and all individual signals at high
frequencies can be identified and removed from the
data~\cite{Blaut:2009si,Littenberg:2011zg,Crowder:2007ft}. Based on
the population synthesis model of~\cite{Nelemans:2003ha}, we can
estimate the unresolvable (stochastic) part of the GW signal generated
by the population of galactic white dwarf binaries and produce a
piecewise analytic fit of this signal, which is given below for each
configuration{\footnote{We provide here fits for the N2
    configurations only, since the CWD background was found to give negligible
contribution to the noise budget in all N1 baselines.}}:
\begin{widetext}
\begin{eqnarray}
  S_{\rm gal, N2A1}(f)&=&
\begin{cases}
     f^{-2.1} \times 1.55206 \times10^{-43} & 10^{-5} \le f < 5.3\times 
10^{-4}, 
\\
     f^{-3.235} \times 2.9714 \times 10^{-47} &  5.3\times 10^{-4} \le f < 
2.2\times 10^{-3}, \\
     f^{-4.85} \times   1.517\times10^{-51} &  2.2\times 10^{-3} \le f < 4 
\times 10^{-3}, \\
     f^{-7.5} \times  6.706 \times10^{-58} & 4 \times 10^{-3} \le f < 
5.3\times10^{-3}, \\
     f^{-20.0} \times  2.39835 \times 10^{-86} & 5.3\times10^{-3} \le f \le 
10^{-2}
\nonumber
 \end{cases}\\
  S_{\rm gal, N2A2}(f)&=&
\begin{cases}
f^{-2.1} \times 1.3516 \times10^{-43} & 10^{-5} \le f < 5.01\times 10^{-4}, \\
     f^{-3.3} \times 1.4813 \times 10^{-47} &  5.01 \times 10^{-4} \le f < 
2.07\times 10^{-3}, \\
     f^{-5.2} \times   1.17757\times10^{-52} &  2.07\times 10^{-3} \le f < 3.4 
\times 10^{-3}, \\   
     f^{-9.1} \times  2.7781 \times10^{-62} & 3.4 \times 10^{-3} \le f < 
5.2\times10^{-3}, \\    
     f^{-20.0} \times  3.5333 \times 10^{-87} & 5.2\times10^{-3} \le f \le 
10^{-2}
\nonumber
 \end{cases}\\
   S_{\rm gal, N2A5}(f)&=& \frac{20}{3}
\begin{cases}
f^{-2.3} \times 10^{-44.62} & 10^{-5} \le f <  10^{-3}, \\
     f^{-4.4} \times  10^{-50.92} &  10^{-3} \le f <  10^{-2.7}, \\
     f^{-8.8} \times  10^{-62.8} &  10^{-2.7} \le f < 10^{-2.4}, \\   
     f^{-20.0} \times 10^{-89.68} & 10^{-2.4} \le f \le 10^{-2}
 \end{cases}
 \label{E:GWB}
\end{eqnarray}
\end{widetext}
The fit for the LISA-like configuration $S_{gal,\rm A5}(f)$ is taken
from~\cite{Timpano:2005gm}.
This astrophysical stochastic GW foreground is added to the
instrumental noise in quadrature\footnote{The
    CWD unresolved background depends on the mission duration: as the
    observation time increases, more individual CWDs
    can be identified and subtracted. The fits of Eq.~(\ref{eq:fit})
    were derived
    for a two-year mission lifetime (M2). We expect the background to be
    slightly lower for a five-year mission (M5), but for simplicity we
    omit this effect and use Eq.~(\ref{eq:fit}) for both M2 and M5
    configurations.},
  and the resulting curves are also shown in Figure~\ref{fig1}.  Note that 
  the GW foreground is below the instrumental nose for N1 configurations.
  For  each acceleration noise (N1/N2) we consider either four or six laser
  links (L4/L6), and we assume a mission lifetime of either two or
  five years (M2/M5). This amounts to a total of 24 mission
  configurations, labeled as N$i$A$j$M$k$L$l$ (where $i=1,2$,
  $j=1,2,5$, $k=2,5$, $l=4,6$).

The configuration N2A5M5L6 corresponds to Classic LISA. Configuration N2A1M2L4 
corresponds approximately to the New Gravitational Observatory 
(NGO)~\cite{elisa1-12} concept, which was proposed to ESA during the selection 
process for the L1 large satellite mission. and which was also used to 
illustrate the science case in The Gravitational 
Universe~\cite{Seoane:2013qna}. Figure~\ref{fig1} shows that configuration 
N2A1M2L4 differs from NGO only in a multiplicative factor $1/0.65$, which was 
included
in the NGO design as a safety margin. In Sec.~\ref{results} we will
therefore use N2A1MxL4 as a proxy for NGO, and ``normalize'' our
science performance results to this configuration.

\section{Supermassive black hole evolution}
\label{BH-evolution}

Sound observational evidence~\cite{msigma} as well as theoretical
considerations (see e.g.~\cite{GalFormreview1,GalFormreview2}) suggest
that the evolution of MBHs on cosmological timescales is inextricably
coupled to the evolution of their host galaxies. Methods to follow
this MBH-galaxy coevolution include Eulerian and smoothed-particle
hydrodynamics simulations, as well as semianalytical galaxy formation
models (see~\cite{GalFormreview1,GalFormreview2} for recent reviews of
the theory of galaxy formation with an overview of these techniques).
While hydrodynamical simulations have a better handle of the
small-scale physics, subgrid dissipative phenomena such as star
formation and feedback are not yet treatable self-consistently by
simulations, and it is unlikely that they will be in the near future.
The same is true for the scale of the horizon of MBHs, on which key
phenomena for the cosmological evolution of these objects, such as
mergers and accretion, take place.

Clearly, the same drawback applies to semianalytical models, in which
not only the subgrid physics, but also the scales that would be within
reach of hydrodynamical simulations are treated with simplified
prescriptions depending on a (limited) number of free parameters that
are calibrated against observations. For our purposes, semianalytical
models have the advantage of being computationally more convenient,
which allows us to explore the parameter space of galaxy formation and
MBH evolution with large statistics.

\begin{figure*}[th]
\centering 
\includegraphics[width=14cm,angle=0.]{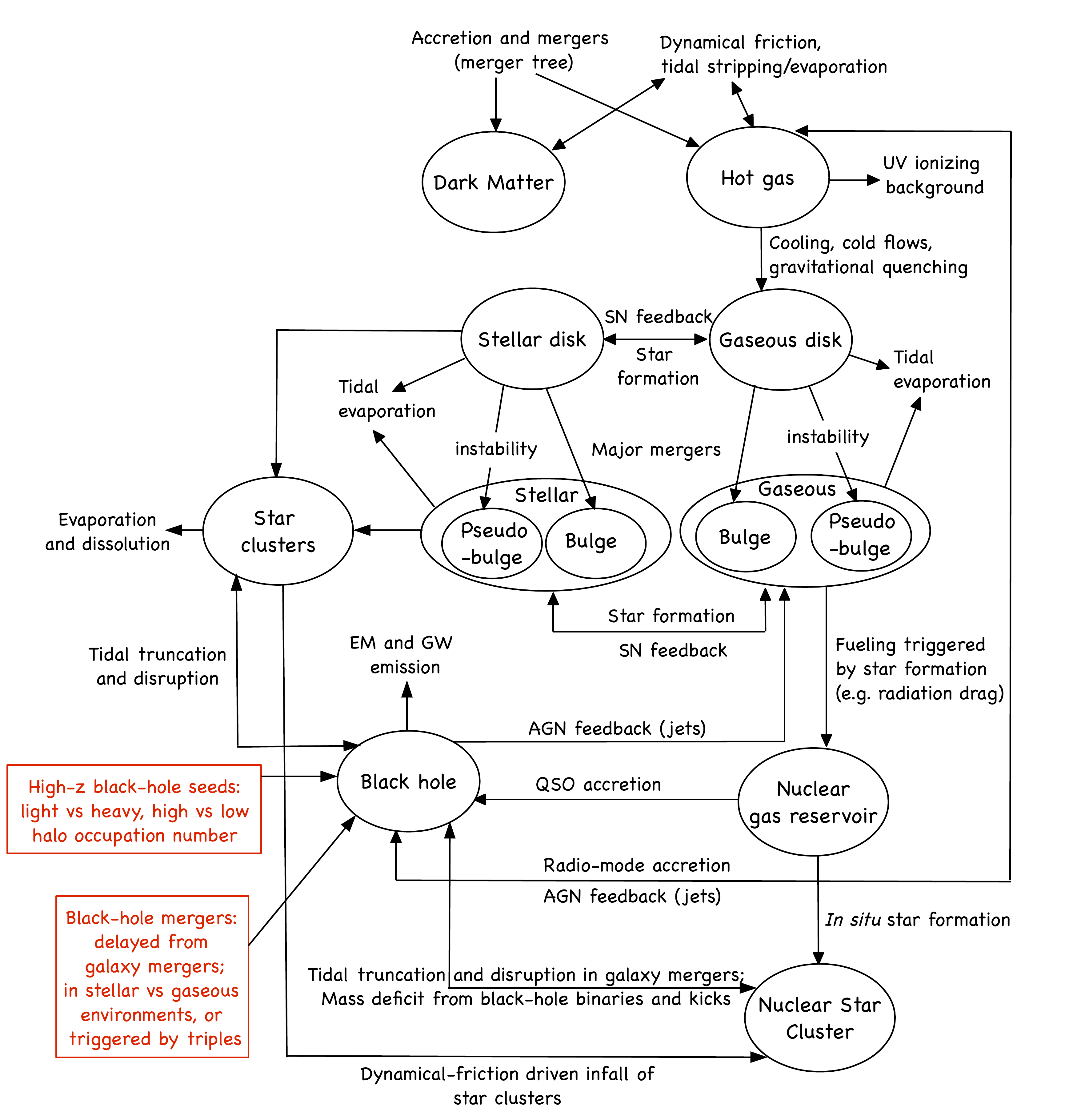}
\caption{Schematic summary of the model of~\cite{mymodel}, with the improvements 
of~\cite{spin_model,letter,newpaper}. Red boxes on the left highlight the 
elements that heavily affect rates (black hole seeding and delays), for which we 
consider multiple options in this paper. 
\label{fig:model}}
\end{figure*}

Here we adopt the semianalytical model of~\cite{mymodel}, which was
later improved in~\cite{spin_model} (where the prescriptions for the
MBH spin evolution and the star formation were improved)
and~\cite{letter,newpaper} (which implemented the cosmological
evolution of nuclear star clusters, and accounted for the delay
between a galaxy merger and that of the MBHs hosted by the two
galaxies). Our model follows the evolution of baryonic structures
along a dark-matter merger tree produced by an extended
Press-Schechter formalism, suitably modified to reproduce the results
of N-body simulations following~\cite{mergertree}. The model evolves
the hot unprocessed inter-galactic medium; the cold, metal-enriched
inter-stellar medium (in both its galactic disk and bulge components);
the stellar galactic disk and the stellar spheroid; the nuclear gas
and the nuclear star cluster; and, of course, the MBHs.
These components are linked by a number of gravitational and
non-gravitational interactions, which are summarized graphically in
Fig.~\ref{fig:model}. We refer to~\cite{mymodel,spin_model,newpaper}
for detailed descriptions of our implementation of the various
processes represented in this diagram. Highlighted in red are the key
assumptions that we will vary in this paper, and that we discuss
below: black hole seeding and delays.

\subsection{Black hole seeding}
\label{sec:seeds}
One of the crucial uncertainties in rate predictions is the birth
mechanism of MBHs, that are thought to grow from high-redshift
``seeds'' whose exact nature is still debated.  

The ``light-seed'' scenario assumes that these seeds may be the
remnants of population III (popIII) stars forming in the
low-metallicity environments characterizing the Universe at
$z\approx 15-20$~\cite{MadauRees2001}. While the mass of these first
stars (and therefore that of their remnants) is uncertain, it can be
of the order of a few hundred $M_\odot$~\cite{2002ApJ...567..532H}
(although recent simulations favor more fragmentation and lower
masses, see e.g.~\cite{2011ApJ...737...75G}). In this ``light-seed''
scenario, we draw the popIII star mass from a log-normal distribution
centered at $300 M_\odot$ with a rms of 0.2 dex and an exclusion
region between 140 and 260 $M_\odot$: in this mass range popIII stars
explode as supernovae due to the electron-positron pair instability,
without forming a black hole~\cite{2002ApJ...567..532H}. Outside this
mass range, a black hole will generally form at the end of the popIII
star's evolution. We assume that the mass of this black hole is about
$2/3$ the mass of the initial star~\cite{2002ApJ...567..532H}. Also,
because active star formation at $z\approx 15-20$ is only expected in
the deepest potential wells, we place a black hole seed only in the
rare massive halos collapsing from the 3.5$\sigma$ peaks of the
primordial density field~\citep{MadauRees2001,volonteri_haardt_madau},
between $z=15$ and $z=20$ (this latter being the initial redshift of
our merger trees). Because in light-seed scenarios it is difficult to
reproduce the active galactic nuclei (AGN) luminosity function at high 
redshifts unless
super-Eddington accretion is allowed~\cite{supereddington}, we assume
the maximum MBH accretion rate to be
$\dot{M}=A_{\rm Edd} \dot{M}_{\rm Edd}$, with $A_{\rm Edd}$ a free
parameter that we set to $\approx 2.2$ as
in~\cite{spin_model,newpaper,letter}.

In the ``heavy-seed'' scenario, MBHs already have masses
$\sim 10^5 M_\odot$ at high redshifts $z\sim 15-20$. These seeds may
arise from the collapse (due e.g. to bar instabilities) of
protogalactic disks. This would funnel large quantities of cold gas to
the nuclear region, where a black hole seed might then form.  Several
flavors of this scenario have been proposed (see
e.g.~\cite{2004MNRAS.354..292K,2006MNRAS.370..289B,
  2006MNRAS.371.1813L,2008MNRAS.383.1079V}). In this paper we adopt a
particular model, namely that of~\cite{2008MNRAS.383.1079V}. Like in
the light-seed model, we start our evolutions at $z=20$ and stop seed
formation at $z=15$, where we assume that the Universe has been metal
enriched by the first generation of stars, which results in quenching
of the seed formation due to the enhanced radiative cooling.
The model of~\cite{2008MNRAS.383.1079V} has a free parameter, the
critical Toomre parameter $Q_c$ at which the protogalactic disks are
assumed to become unstable. By changing $Q_c$ one varies the
probability that a halo hosts a black hole seed (i.e., the halo
occupation number). Plausible values of $Q_c$ range from 1.5 to 3,
with larger values corresponding to larger halo occupation numbers,
but one must have $Q_c\gtrsim 2$ to ensure that a significant fraction
of massive galaxies host a MBH at $z=0$~\cite{2008MNRAS.383.1079V}.

\subsection{Delays}
 \label{sec:delays}
Another ingredient that is particularly important for calculating MBH merger 
rates is the delay between
 MBH mergers and galaxy mergers. When two dark-matter halos coalesce, the 
galaxies that they host initially maintain their
identity, because they are smaller and more compact than the halos. The galaxies 
are then brought together by dynamical friction
on typical timescales of a few Gyr. During this time, environmental effects such 
as tidal stripping and tidal evaporation
remove mass from the smaller galaxy, which in turn affects the dynamical 
evolution of the system: see~\cite{mymodel} 
for more details about our treatment of dynamical friction, tidal stripping and 
evaporation. 

After the two galaxies have merged, the MBHs they host
are slowly brought to the center of the newly formed galaxy by dynamical 
friction against the stellar background. As a result, the MBHs eventually form
a bound system (a ``hard'' binary), i.e., one such that their relative velocity 
exceeds the velocity dispersion of the stellar background. From this
moment on, the MBHB will further harden by three-body interactions with stars. 
It is unclear if this mechanism alone 
can bring the binary to the small separations ($\lesssim 10^{-3}$ pc) where GW 
emission can drive 
the system to merger within a Hubble time. This is known as the ``last
parsec problem''~\cite{begelman}. Recently, however, it has been
suggested that triaxiality of the galaxy potential (resulting
e.g. from a recent galaxy merger) would allow three-body stellar
interactions to harden the binary to the GW-dominated regime on a typical 
timescale of a few 
Gyr~\cite{yu,lastPc1,2014CQGra..31x4002V,lastPc2,2015arXiv150505480V}. Galaxy 
rotation has also been suggested as a possible mechanism helping the binary 
reach GW-dominated separations~\cite{2015arXiv150506203H}. 
Moreover, if the nuclear region contains a significant amount of gas in a disk 
geometry, planet-like migration might
drive the binary to merger on significantly shorter timescales, typically of 
$\sim 10^7-10^8$ yr~\cite{bence,2014SSRv..183..189C} (but see e.g. 
\cite{2009MNRAS.398.1392L} for possible complications arising in this scenario). 
Finally, if an MBHB stalls and in the meantime another galaxy merger happens, a 
third MBH may be added to the system. Triple interactions are expected to 
trigger the merger of the two most massive MBHs and ejection of the lightest 
one on timescales $\sim 10^8$ yr~\cite{hoffman}. Nevertheless, this process is 
likely to be effective at inducing coalescence of the inner binary only for 
systems with masses $\gtrsim 10^6-10^7 M_\odot$. Below that threshold, the 
lightest MBH may be ejected before the triple interactions trigger the merger of 
the inner binary, especially if its mass is much lower than that of the inner 
binary~\cite{hoffman,newpaper}.

We refer to~\cite{newpaper} for a detailed description of our implementation of 
these delays. To highlight their impact on our results for MBH merger rates, we 
consider 
both models where the delays are included, and models where no delays are 
present (i.e., the MBHs merge at the same time as their host galaxies).

\subsection{Population models used in our study}
\label{subsec:models}

\begin{figure*}
\centering
\begin{tabular}{ccc}
\includegraphics[width=\columnwidth,clip=true,angle=0]{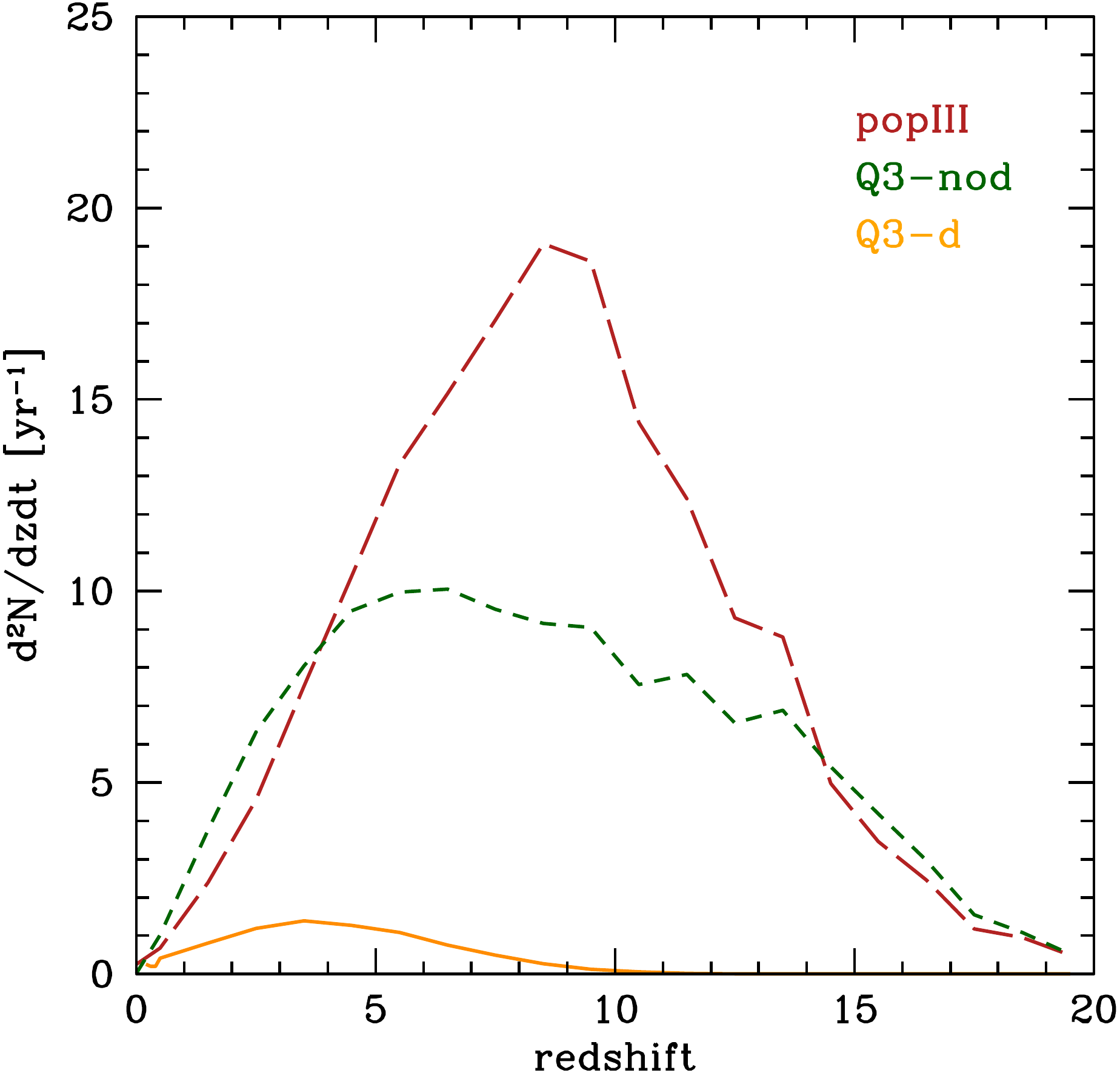}&
\includegraphics[width=\columnwidth,clip=true,angle=0]{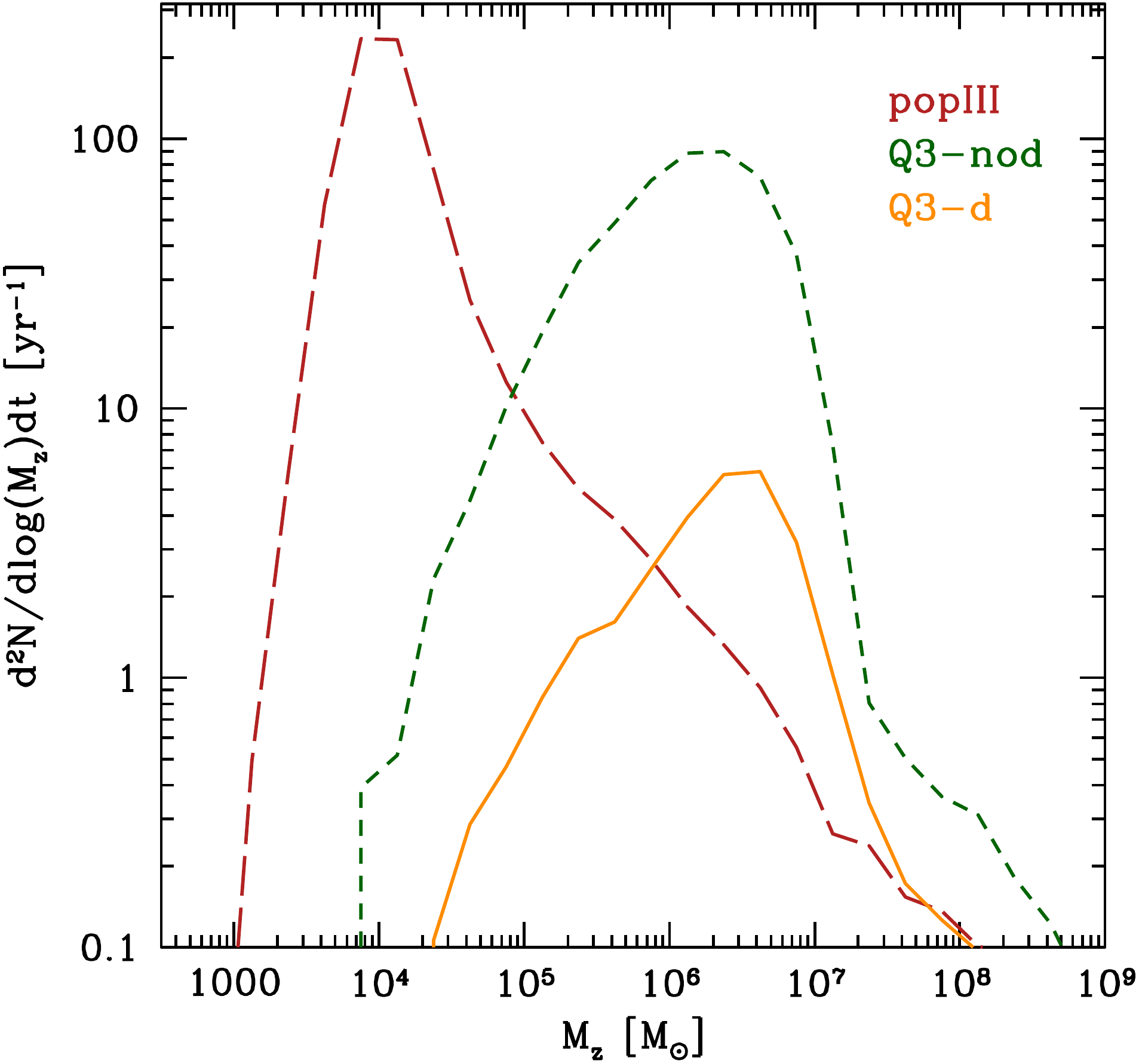}\\
\end{tabular}
\caption{Predicted merger rates per unit redshift (left panel) and per unit 
total redshifted mass $M_z=(m_1+m_2)(1+z)$ (right panel) for the three models 
described in the text.}
\label{fig:pop_rate}
\end{figure*}

We have generated possible realizations of the MBH population based on
this semianalytical galaxy-formation model. We present results for
variants of the model which are representative of the possible
combinations that can be obtained by varying the prescriptions for
seeding and delays presented above.  More specifically, we focus on
three models:
\begin{itemize}
\item[(1)] Model popIII: This model assumes light MBH seeds from
popIII stars, while accounting for the delays between MBH and
galaxy mergers (cf. Sec.~\ref{sec:delays}). The inclusion of these
delays makes the model more realistic, and certainly
more conservative. We verified that typical eLISA event rates 
change by less than a factor of two when setting the delays to zero,
hence we decided to omit the variant of this light-seed model in which delays 
are not present.
\item[(2)] Model Q3-d: This assumes heavy MBH seeds from the
collapse of protogalactic disks, while accounting for the delays
between MBH and galaxy mergers. The halo occupation fraction of the
seeds at high redshifts is determined by the critical Toomre parameter for
disk instability, which we set to $Q_c=3$
(cf. Sec.~\ref{sec:seeds}). Note however that setting $Q_c=2$ only
decreases the merger rates by a factor $\sim 2$.
As in the case of the
popIII model, the inclusion of delays makes this
model more ``realistic'' (and conservative).
\item[(3)] Model Q3-nod: This is the same as model {Q3-d}, but
without accounting for the delays between galaxy and MBH mergers. For
this reason, this model should be considered an ``optimistic''
(upper bound) scenario for eLISA event rates.
\end{itemize}
For each of the three models above we simulate about 1300
galaxies/galaxy clusters, with dark-matter masses ranging from
$10^{10} M_\odot$ to $10^{16} M_\odot$.
By tracking self-consistently the mass and spin evolution of MBHs and
their interaction (e.g. via feedback and accretion) with the galactic
host, our model allows us to predict the masses, spin magnitudes and
spin orientations of the MBHs when they form a GW-driven binary
system. As such, while MBHBs often present partially aligned,
high spins in our simulations, systems with low and/or misaligned spins are
also possible in the three models listed above

The merger rates for popIII models are rather insensitive to the
inclusion of delays, but this is not true for the heavy-seed models.
This is illustrated in Fig.~\ref{fig:pop_rate}, which shows the
predicted MBH merger rates as a function of mass and redshift in the
three models considered above. Note that while the popIII and Q3-nod
models predict a high merger rate up to $z>15$, very few or no events
at $z>10$ are expected in the Q3-d model.

The difference in merger rates among the various seed models is
due to two factors. First, different models have different mass
functions and occupation numbers at high $z$. Second, whether a MBHB stalls or
merges depends on the details of its interactions with the stars,
nuclear gas, and other MBHs, which depend critically on the MBH masses 
(cf.~\cite{newpaper} for more details on our assumptions regarding these 
interactions). 
The different MBH mass functions at high redshifts in the light- and heavy-seed
models imply that many more binaries can ``stall'' at high redshifts
in the heavy-seed scenarios. Nevertheless, since our treatment of the delays is
quite simplified (in particular when it comes to modeling triple MBH
systems and the interaction with nuclear gas), models Q3-d and Q3-nod can be 
thought of as bracketing
the possible range of merger rates.

\subsection{Population completeness}

Because our models follow the coevolution of MBHs with galaxies
including both their dark-matter and baryonic constituents
(cf. Fig.~\ref{fig:model}), at fixed resolution for the dark-matter
merger trees our simulations become computationally expensive for
high redshifts and very massive galaxies. For galaxies with
dark-matter halo masses $M_H> 10^{13} M_\odot$ at $z=0$, we find
that it becomes difficult to resolve the halos where MBHs form at high
redshifts within acceptable computational times. Therefore it is
possible that we may ``miss'' merging binaries at high redshift, when simulating 
the
most massive halos at $z=0$. 

\begin{figure}
\centering
\includegraphics[width=\columnwidth,clip=true,angle=0]{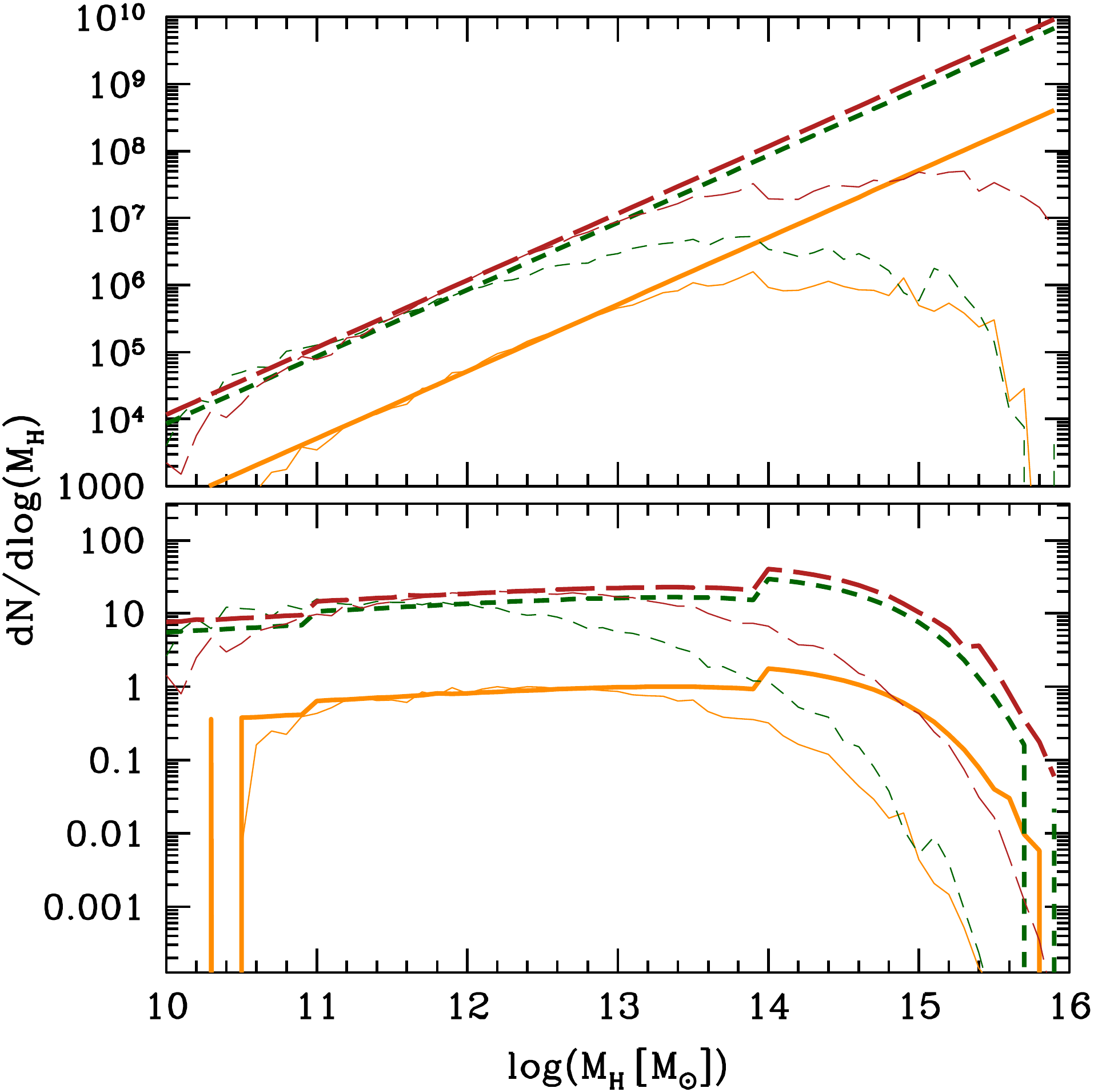}
\caption{\label{fig:correct} Contribution of each halo mass to the
  total merger rate for models popIII (long-dashed brown lines), Q3-nod
  (short-dashed green lines) and Q3-d (solid orange lines). Lines in the
  top panel are proportional to the number of mergers {\rm per halo}
  (i.e., removing the Press \& Schechter weights), whereas lines in the
  bottom panel represent the halo contribution to the cosmic merger rate
  $dN/d{\rm log}M_H$. In both panels, thin lines are the results of our
  MBH population models; thick lines are extrapolations assuming a linear
  relation between the number of mergers and the halo mass.}
\end{figure}

We quantify this effect in Figure~\ref{fig:correct}, which shows the
number of MBH mergers as a function of halo mass. The linear trend
seen at $M_H<10^{13} M_\odot$ is easily explained. Suppose that seeds
form in halos of mass $M_S$. A halo of mass $M_0$ has formed from
roughly $N=M_0/M_S$ halos of mass $M_S$. This implies a number of
seeds proportional to $M_0$. Suppose for simplicity that we start off
with $2^n$ seeds. If we consider a perfect hierarchy in which two
remnants of a previous round of mergers keep merging with each other
until there is only one MBH left, the number of mergers is
$\sum_{i=0}^{n-1}2^i=2^n-1$, which approximately matches the number of
seeds (i.e., $2^n$), and which is therefore proportional to $M_0$.
Although simplistic, this argument highlights the reason why the trend
shown in Fig.~\ref{fig:correct} may hint at a lack of resolution in
our simulations for $M_H>10^{13} M_\odot$.

To assess the impact of this issue on our results, we computed the MBH
merger rate per unit (dark matter) mass in the low-mass halos, and
used it to correct the merger rates at larger halo masses (cf. the
thick lines in Fig.~\ref{fig:correct}). The results of this exercise
(reported in Table \ref{tabres}) suggest that this lack of resolution
may lead to our simulations missing up to a factor of two in terms of
merger events. In this sense, the event rates in our study are
therefore conservative. To further confirm this finding, we also ran a
few test simulations with increased halo resolution. These
higher-resolution runs show that the number of mergers is essentially
resolution-independent at $M_H<10^{13} M_\odot$, but that MBH mergers
are much more numerous for $M_H>10^{13} M_\odot$, in line with the
expected
linear trend with $M_H$. We also confirmed the expectation that the missing
events are mostly seed-mass, high-$z$ MBHBs. This does not make a large
difference for eLISA rates in the popIII model, but it may
increase more significantly the number of detections in the Q3-nod and Q3-d
scenarios, where the MBH seeds are massive enough to be within the
instrument's sensitivity range.

\begin{table}
\begin{center}
\begin{tabular}{c|ccc}
\hline 
\hline
Model & Total rate & Extrapolated rate & Ratio\\
\hline
popIII   &  175.36  &  332.65  &  1.89\\
Q3-d     &  8.18    &   14.06 &   1.72\\
Q3-nod   & 121.80   &  240.96 &   1.98\\
\hline
\end{tabular}
\end{center}
\caption{MBHB merger rates (total number of mergers per year) along the cosmic 
history predicted by our population models. The second column reports the rates 
found in the simulations, the third column reports the rate corrected as 
described in the text and in Fig.~\ref{fig:correct}, and the fourth column is 
the ratio of the two (i.e., the expected level of incompleteness of the 
populations adopted in this study).}
\label{tabres}
\end{table}

\section{Gravitational waveforms and data analysis}
\label{WF-PE}

An accurate waveform model, encapsulating the complexity produced by a
potentially precessing spinning MBHB, is required in order to make a
realistic assessment of eLISA's capabilities. Since many MBHBs
merge within the detector band, including merger and ringdown in the
computation is also crucial. Unfortunately, 
inspiral-merger-ringdown (IMR) waveform models for precessing-spinning
binaries suitable for efficient parameter estimation are still under active
development~\cite{Pan:2013rra,Hannam:2013oca,Ossokine:2015vda}. 
We therefore employ the following strategy:

\begin{itemize}
\item[(a)] our core analysis is based on a generic precessing
  inspiral-only gravitational waveform [the shifted uniform asymptotics (SUA) 
waveform
  described in ~\cite{kcy-14}, see next section], with the detector
response modeled as in~\cite{cutler-98}  (see also 
\cite{Lang:1900bz,Lang:2007ge,Lang:2011je} for similar studies in the context 
of 
LISA);
\item[(b)] since the signal-to-noise ratio (SNR) $\rho$ of the inspiral
  phase depends only mildly on spins, we compared our precessing
  waveforms to the restricted second order post-Newtonian (2PN)
  waveform described in~\cite{Berti:2004bd,Berti:2005qd}, ensuring
  that they yield comparable SNR distributions and detection rates;
\item[(c)] results are then rescaled with the aid of the spin-aligned IMR
  waveform family of ~\cite{santamaria-et-al-10} (commonly referred to as
``PhenomC'') in combination
with a
  restricted set of dedicated precessing IMR hybrid waveforms,
  which are constructed from numerical relativity (NR) simulations stitched
  to a post-Newtonian (PN) model of the early inspiral.
\end{itemize}

In the following, we first describe our core inspiral precessing
waveform model (Section~\ref{sec:SUA}). Then we briefly summarize the
basics of the adopted Fisher matrix analysis
(Section~\ref{sec:fisher}) and our IMR rescaling
(Section~\ref{sec:rescaling}).

\subsection{Inspiral-precessing waveform model}
\label{sec:SUA}

The spacecraft in all eLISA configurations considered in this study
share the same orbits, modulo a rescaling proportional to the detector
arm length. If we choose an orthogonal reference system
$(\uvec{x}, \uvec{y})$ in the orbital plane tied to the detector arms,
in a fixed Solar System frame
$(\uvec{x}', \uvec{y}', \uvec{z}')$ tied to the
ecliptic (with $\uvec{z}'$ perpendicular to the ecliptic) we can
write~\cite{elisa1-12,cutler-98}
\begin{align}
\uvec{x} &= \left( \frac{3}{4} - \frac{1}{4} \cos 2 \Phi \right) \uvec{x}'
 - \frac{1}{4} \sin 2 \Phi \uvec{y}' + \frac{\sqrt{3}}{2} \cos \Phi 
\uvec{z}', \\
\uvec{y} &= - \frac{1}{4} \sin 2 \Phi \uvec{x}' + \left( \frac{3}{4} + 
\frac{1}{4} \cos 2 \Phi \right) \uvec{y}'
  + \frac{\sqrt{3}}{2} \sin \Phi 
\uvec{z}'.
\end{align}
A third vector $\uvec{z} = \uvec{x} \times \uvec{y}$ completes the
three-dimensional orthogonal reference system. The constellation
drifts away from the Earth at the rate of 7.5 degrees per year, so that
$\dot{\Phi} = 2\pi (352.5/360) / \text{yr}$.

Consider a binary with orbital angular momentum direction $\uvec{L}$,
located in a direction specified by the unit vector $\uvec{N}$ in the
Solar System frame.
The response of a single (four-link) detector to the GWs emitted by such a 
binary can be described as~\cite{cutler-98}
\begin{align}
 h &= \frac{\sqrt{3}}{2} \left( F_+ h_+ + F_\times h_\times \right),
\end{align}
where
\begin{align}
 F_+(\theta, \phi, \psi) &= \frac{1}{2} \left( 1 + \cos^2 \theta \right) \cos 2 
\phi \cos 2 \psi \nonumber\\
&- \cos \theta \sin 2 \phi \sin 2 \psi, \\
 F_\times (\theta, \phi, \psi) &= F_+(\theta, \phi, \psi - \pi/4), \\
 \cos \theta &= \uvec{N} \cdot \uvec{z}, \\
 \tan \phi &= \frac{\uvec{N}\cdot \uvec{y}}{\uvec{N}\cdot \uvec{x}}, \\
 \tan \psi &= \frac{\left[ \uvec{L} - \left( \uvec{L} \cdot \uvec{N} \right) 
\uvec{N} \right] \cdot \uvec{z}}{\left(\uvec{N} \times \uvec{L}\right) \cdot 
\uvec{z}}.
\end{align}

In the six-link case, we model the response of a second independent
detector as
\begin{align}
 h\super{(II)} &= \frac{\sqrt{3}}{2} \left( F_+\super{(II)} h_+ + 
F_\times\super{(II)} h_\times \right), \\ 
F_{+,\times}\super{(II)} (\theta, \phi, \psi) &= F_{+,\times} (\theta, \phi - 
\pi/4, \psi) .
\end{align}

In addition, we have to take into account the fact that the barycenter
of eLISA is traveling around the Sun. Instead of modeling this using
a Doppler phase as in~\cite{cutler-98}, we prefer to time shift the
waveform accordingly:
\begin{align}
 h(t) &= F_+(t) h_+(t - t_D) + F_\times(t) h_\times(t - t_D), \\
 t_D &= \frac{R}{c} \sin \theta' \cos( \Phi - \phi' ),
\end{align}
where $R = 1$ AU, and $(\theta', \phi')$ are the spherical
angles of $\uvec{N}$ in the Solar System frame. The two descriptions
are equivalent in the limit where the orbital frequency varies slowly
with respect to the light-travel time across the orbit of the
constellation: $\dot{\omega}/ \omega \ll c/R$. While this condition is
satisfied in the early inspiral, it breaks down near merger.

We decompose the time-domain waveform $h(t)$ into a sum of orbital harmonics
as
\begin{align}
 h_{+,\times} &= \sum_n A_{+,\times}^{(n)}(\iota) e^{i n \varphi}, \\
 \cos \iota &= - \uvec{L} \cdot \uvec{N}, \\
 \varphi &= \phi_C + \phi_T, \\
 \phi_C &= \phi\sub{orb} - 3 v^3 \left( 2 - v^2 \right) \log v,
\end{align}
where $\iota$ is the inclination angle (defined with a minus sign to
agree with the common convention in the literature),
$A_{+,\times}^{(n)}(\iota)$ can be found at 2.5PN in~\cite{abiq-04}
and at 3PN in~\cite{bfis-08,blanchet-lrr}, $\phi\sub{orb}$ is the
orbital phase of the binary, $v = (M\omega)^{1/3}$ (with $M$ the total binary 
mass) is a post-Newtonian
parameter, $\phi_C$ is the carrier phase and $\phi_T$ is the Thomas
phase, taking into account the fact that the orbital plane is
precessing~\cite{acst-94} and satisfying
\begin{align}
 \dot{\phi}_T &= - \frac{\cos \iota}{1 - \cos^2 \iota} \left(\uvec{L} \times 
\uvec{N} \right) \cdot \duvec{L}.
\end{align}

We use a signal $h(t)$ in the so-called TaylorT4-form at 3.5PN order,
i.e., we integrate the following equations of motion: 
\begin{align}
 M \dot{\phi}\sub{orb} &= v^3, \label{eq:phidot-T4} \\
 M \dot{v} &= v^{9} \sum_{n = 0}^7 a_n v^n, \label{eq:vdot-T4}
\end{align}
together with the equations of precession at 3.5PN spin-orbit~\cite{bmfb-13} 
and 2PN spin-spin orders~\cite{racine-08}
\begin{align}
 M\duvec{L} &= - v^6 \left( \bm{\Omega}_1 + 
\bm{\Omega}_2 \right),  \\
 M\dvec{s}_A &= \mu_B v^5 \bm{\Omega}_A,\\
 \bm{\Omega}_A &= \left( C_{A,0} +  C_{A,2} v^2 + C_{A,4} v^4 + D v 
\right) 
\uvec{L} \times \bm{s}_A \nonumber\\
&+ \frac{1}{2} v \, \bm{s}_B \times \bm{s}_A. \label{eq:sAdot}
\end{align}
Here $\bm{s}_A = \bm{S}_A/m_AM$ are the dimensionless reduced spins, $\mu_A = 
m_A/M$ are the dimensionless individual masses, and the couplings are
\begin{align}
C_{A,0} &=  2 \mu_A + 
\frac{3}{2} 
\mu_B, 
\\
C_{A,2} &=  3 \mu_A^3 + 
\frac{35}{6} \mu_A^2 \mu_B + 4 \mu_A \mu_B^2 + \frac{9}{8} \mu_B^3, 
\\
C_{A,4} &=  \frac{27}{4} 
\mu_A^5 + \frac{31}{2} \mu_A^4 \mu_B + \frac{137}{12} \mu_A^3 \mu_B^2 
\nonumber\\
&+ 
\frac{19}{4} 
\mu_A^2 \mu_B^3 + \frac{15}{4} \mu_A \mu_B^4 + \frac{27}{16} \mu_B^5 , \\
D &= - \frac{3}{2} \uvec{L} \cdot \left( \bm{s}_1 + \bm{s}_2 \right).
\end{align}

To compute the Fourier transform of the waveform, we use a SUA 
transformation~\cite{kcy-14}. We first
separate each harmonic of the waveform into parts varying on the
orbital timescale and parts varying on the precession timescale:
\begin{align}
 h^{(n)}(t) &= h^{(n)}\sub{prec}(t) h^{(n)}\sub{orb}(t), \\
 h^{(n)}\sub{prec}(t) &= \Big\{ F_+(t) A_+^{(n)} (t - t_D) \nonumber\\
 &+  
F_\times(t) A_\times^{(n)}(t - t_D) \Big\} e^{i n \phi_T (t - t_D)}, \\
h^{(n)}\sub{orb} (t) &= e^{i n \phi_C(t- t_D)}.
\end{align}

The Fourier transform of the signal is then given by
\begin{align}
 \tilde{h}(f) &= \sum_n \tilde{h}^{(n)}(f), \\
 \tilde{h}^{(n)}(f) &= \sqrt{2\pi} T e^{i [ 2 \pi f t_0 - 
n\phi_C(t_0 - t_D) - \pi/4]} \nonumber\\
&\times \sum_{k = 0}^{k\sub{max}} \frac{a_k}{2} 
\left[ h^{(n)}\sub{prec} (t_0 + k T) + h^{(n)}\sub{prec} (t_0 - k T) \right],
\end{align}
where $a_k$ are constants satisfying the system
\begin{align}
 \frac{(-i)^p}{2^p p!} &= \sum_{k=0}^{k\sub{max}} a_k \frac{k^{2p}}{(2p)!}
\end{align}
for $p\in \{0, \ldots, k\sub{max} \}$, and $t_0$ and $T$ are defined through 
\begin{align}
 2 \pi f &= n \dot{\phi}\sub{orb} ( t_0 - t_D), \\
 T &= \sqrt{\frac{1}{n \ddot{\phi}\sub{orb} ( t_0 - t_D) }}.
\end{align}

In our simulations we used the value $k\sub{max} = 3$ as a good
compromise between computational efficiency and waveform
accuracy~\cite{kcy-14}.

\subsection{Fisher matrix analysis}
\label{sec:fisher}

A careful estimate of the likely errors in eLISA measurements of MBHB 
parameters will
ultimately require numerical simulations and full evaluations of the 
multi-dimensional posterior probability distributions, (see e.g. 
\cite{2015PhRvD..92f4001P}), but these
techniques are computationally expensive. The simple Fisher matrix
analysis described in this section allows us to efficiently estimate
errors on ensembles of thousands of systems in different MBHB
population scenarios, and it is expected to be sufficiently accurate
in the high SNR regime.

We first define the detector-dependent inner product
\begin{align}
 (a|b) &= 4 \text{Re} \int_0^\infty \frac{\tilde{a}(f) \tilde{b}^*(f)}{S_n(f)} 
df,
\end{align}
where a tilde denotes Fourier transform, a star denotes complex
conjugation, and $S_n(f)$ is the one-sided noise power spectral
density of the detector, equal to $3/20$ times the sky-averaged sensitivity for 
each configuration given in Eq.~\eqref{eq:sens}~\cite{Berti:2005ys}.
The SNR of the signal $h$ is given by
\begin{align}
 \rho^2 &= (h|h).
\end{align}

The Fisher information matrix $\Gamma$ for the signal $h$ has elements
\begin{align}
 \Gamma_{ij} = \bigg( \frac{\pd h}{\pd \theta^i} \bigg| \frac{\pd h}{\pd 
\theta^j} \bigg), 
\end{align}
where $\bm{\theta}$ is the vector of source parameters. The combined Fisher 
matrix for several independent detectors is the
sum of the single-detector Fisher matrices,
$\Gamma = \sum_i \Gamma^{(i)}$, and the combined squared SNR is the
sum of the individual squared SNR $\rho_i$:
$\rho^2 = \sum_i \rho_i^2$.

The correlation matrix $\Sigma$ is the inverse of the Fisher matrix,
$\Sigma = \Gamma^{-1}$. The estimated expectation of the statistical
error on a parameter $\Delta \theta^i$ is given by the corresponding diagonal
element of the correlation matrix,
$(\Delta \theta^i)^2 = \Sigma^{ii}$. The estimated error in some function of
the parameters is obtained by linear propagation of errors:
\begin{align}
 (\Delta \alpha)^2 &= \sum_{i,j} \frac{\pd \alpha}{\pd \theta^i} \frac{\pd 
\alpha}{\pd \theta^j} \Sigma^{ij}.
\end{align}

The signal from a MBHB in a quasi-circular orbit is described by fifteen
parameters: the sky location of the source in ecliptic coordinates
(co-latitude, $\theta$, and longitude, $\phi$), the luminosity distance,
$D_{l}$, the time at coalescence, $t_{c}$, the total redshifted mass,
$M_z=m_{1z}+m_{2z}$, the symmetric mass ratio,
$\eta\equiv m_{1z}m_{2z}/M_z^2$, the initial phase, $\phi_0$, the
dimensionless spin parameters, $\chi_{1}$ and $\chi_{2}$, the direction
of the spins (two polar angles, $\theta_{\chi_{1}}$ and $\theta_{\chi_{2}}$
and two azimuthal angles, $\phi_{\chi_{1}}$ and $\phi_{\chi_{2}}$), the
inclination, $\iota$, of the orbital angular momentum with respect to
the line of sight, and the polarization angle, $\Psi$. Because the
system is precessing, the latter six parameters must be specified at
some reference time $t_0$.

In this study we focus in particular on 
\begin{itemize}
\item[(a)] the errors in the two redshifted masses ($\Delta{m_{1z}}, 
\Delta{m_{2z}}$);
\item[(b)] the error in the sky location, related to the errors on the
  $\theta$ and $\phi$ angles via
  $\Delta \Omega=2\pi \sin\theta \sqrt{\Delta \theta \Delta \phi -
    (\Sigma^{\theta\phi})^2}$;
\item[(c)] the error in the luminosity distance, $\Delta{D_l}$;
\item[(d)] the errors in the magnitudes of the two individual spins
  ($\Delta{\chi_{1}}$, $\Delta{\chi_{2}}$) and the errors on their
  misalignment angles relative to the orbital angular momentum at the
  innermost stable circular orbit ($\Delta \theta_{\chi_{1}}$,
  $\Delta\theta_{\chi_{2}}$).
\end{itemize}
Additionally, we use simple analytical expressions (described in
Appendix~\ref{app:ringdown}) to estimate the accuracy in measuring the
spin of the MBH remnant $\Delta{\chi_{r}}$ from the radiation emitted
in the ringdown phase.

We compute the Fisher matrices using the SUA waveform model, but we
additionally model the effect of merger and ringdown by rescaling the
errors as described in the next section. The rescaling cannot take
into account the fact that the eigenvectors of the Fisher matrix are
different in the merger-ringdown part and in the inspiral part, so we
expect the results of this calculation to be generally conservative.

\subsection{Inspiral-merger-ringdown rescaling}
\label{sec:rescaling}

\begin{figure}
\centering
\includegraphics[width=6.5cm,clip=true,angle=0]{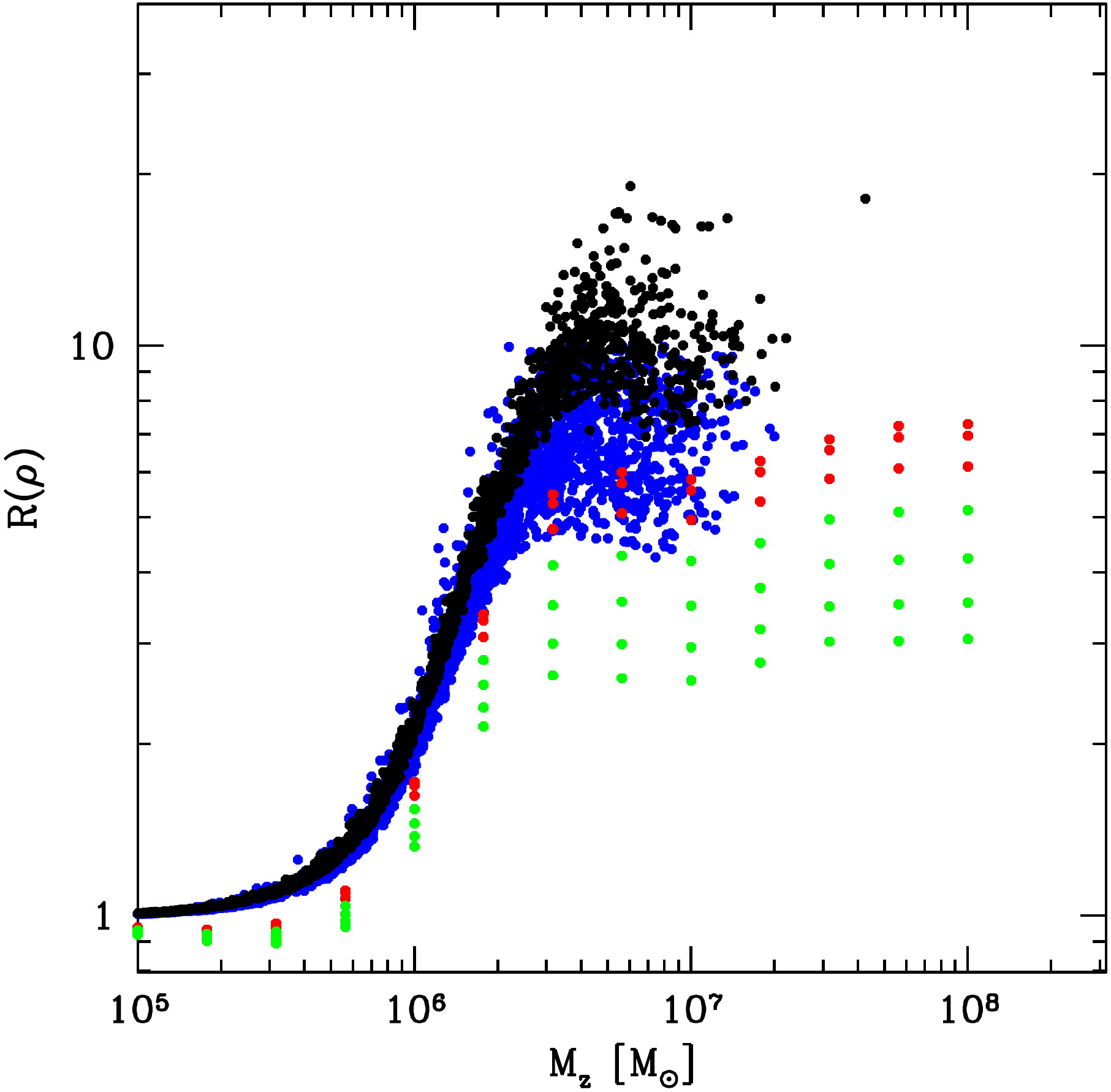}
\caption{SNR gain ${\cal R}(\rho)$ as a function of redshifted total
  mass $M_z$. PhenomC waveforms applied to one realization of the
  Q3-nod population model are represented by black
  ($0.5<m_{2z}/m_{1z}<1$) and blue ($0<m_{2z}/m_{1z}<0.5$) dots. The red
  and green dots are computed using non-spinning PhenomC waveforms at
  a fixed $M_z$ for decreasing values of $m_{2z}/m_{1z}$ (from top to
  bottom); red dots are for $m_{2z}/m_{1z}>0.5$ and green dots are for
  $m_{2z}/m_{1z}<0.5$. This calculation refers to the detector
  configuration that we labeled N2A1M2L6.}
\label{fig:SNgain}
\end{figure}
\begin{figure}
\centering
\includegraphics[width=6.5cm,clip=true,angle=0]{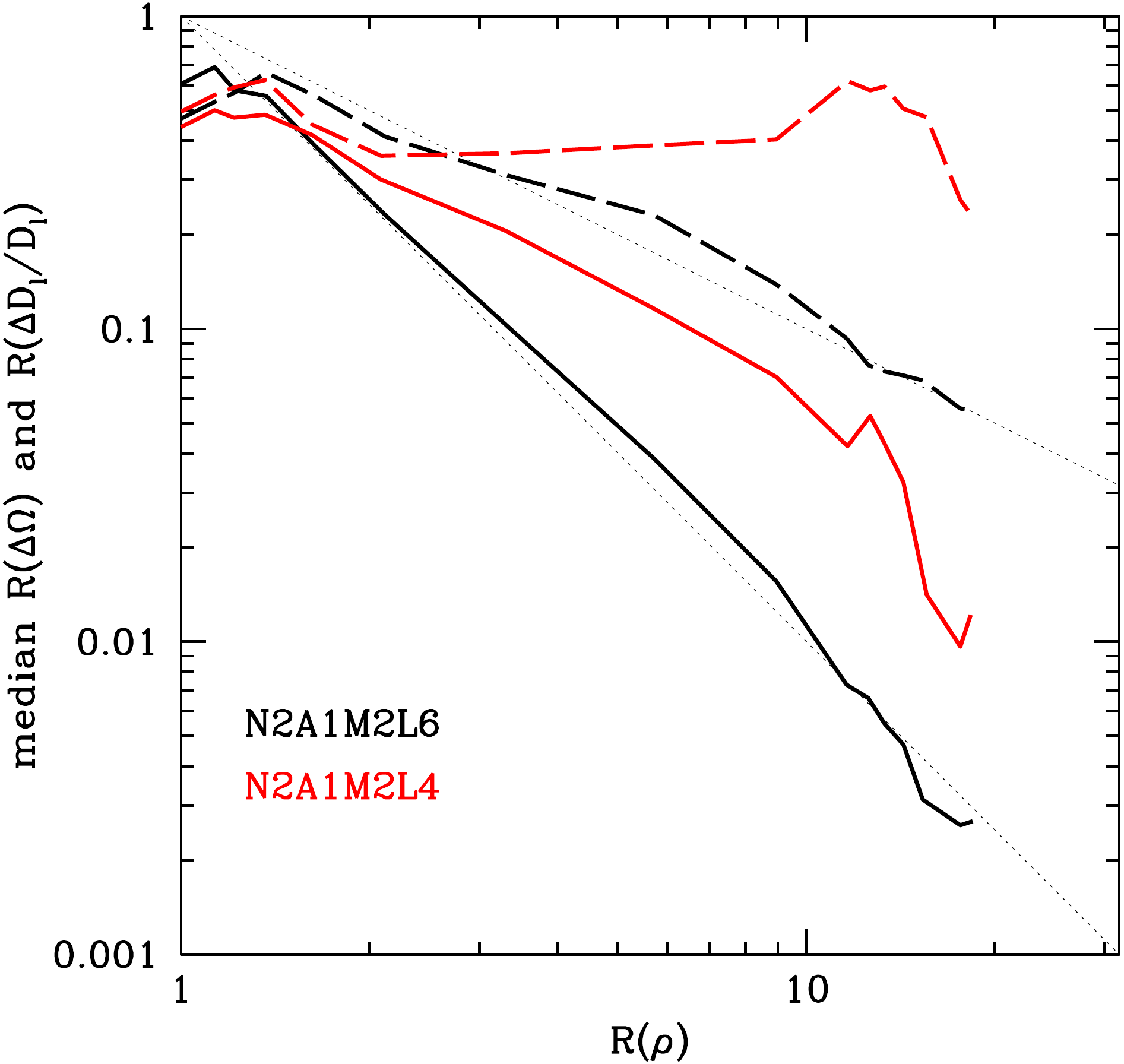}
\caption{Median improvement in sky localization ${\cal R}(\Delta{\Omega})$ 
(solid lines) and luminosity distance errors ${\cal R}(\Delta{D_l}/D_l)$ 
(dashed 
lines) as a function of the SNR gain ${\cal R}(\rho)$. Black and red lines are 
for a six-link and four-link detector configuration, respectively, as indicated 
in figure. Dotted lines represent linear and quadratic scalings to guide the 
eye.
}
\label{fig:IMRgain}
\end{figure}

There is a very limited literature trying to estimate the effect of
merger and ringdown on parameter estimation for space-based
detectors~\cite{Babak:2008bu,McWilliams:2011zs,Littenberg:2012uj}.
These works focused on specific choices for the intrinsic parameters
of the binary, which makes it hard to use their conclusions in
population studies.
Babak et al.~\cite{Babak:2008bu} first claimed that adding merger and
ringdown can provide roughly an order-of-magnitude improvement in
angular resolution. McWilliams et al.~\cite{McWilliams:2011zs} studied
the improvement in the estimation of various parameters as a function
of time as one transitions from the innermost stable circular orbit to
the post-merger phase. They found negligible improvement in the
determination of the system's mass, but their Figure 1 shows that
angular accuracy improves by a factor $\sim 4$ for a four-link
configuration, and by a factor $\sim 5-6$ for a six-link
configuration, hinting that the inclusion of merger and ringdown may
have a greater impact on angular resolution for six-link configurations.
Their Figure~5 claims ``two to three orders of magnitude improvement
in a mix of the five angular parameters and in $\ln(D_l)$'' when the SNR 
improves by a factor
$\sim 10$ due to the inclusion of merger and ringdown.

Here we estimate the impact of merger and ringdown by extrapolating
our results using spin-aligned PhenomC IMR
waveforms~\cite{santamaria-et-al-10} in combination with a restricted
set of dedicated precessing IMR hybrid waveforms, constructed from NR
simulations stitched to a PN model of the early inspiral.

We first use the IMR PhenomC waveform model to rescale the SUA
waveform SNR as follows. For each MBHB in our catalog, we construct
the PhenomC waveform corresponding to the given MBHB parameters. Since
the PhenomC model is valid for systems with spins (anti-)aligned with
the orbital angular momentum, this requires computing an ``effective
spin'' obtained by projecting the two individual spins along the
orbital angular momentum. Note that the SNR produced by PhenomC is a
very good proxy (i) when spins are partially (anti)aligned with the
orbital momentum and (ii) for systems with low spin magnitude. The
population of MBHBs considered here usually have nearly aligned spins,
especially in the small-seed scenario.
From the PhenomC waveform, we compute (for each detector
configuration) the SNR produced by the inspiral portion of the
waveform alone ($\rho_{\rm PhenC,I}$) and by considering also merger
and ringdown ($\rho_{\rm PhenC,IMR}$). The ratio of the two,
${\cal R}(\rho)=\rho_{\rm PhenC,IMR}/\rho_{\rm PhenC,I}$, defines the
gain due to the inclusion of merger and ringdown. This quantity is
shown in Fig.~\ref{fig:SNgain} for the Q3-nod model and the N2A1M2L6
configuration. Note that ${\cal R}(\rho)\rightarrow 1$ for total
redshifted masses $M_z<3\times 10^5 M_\odot$, but the gain becomes
much larger (ranging between $\sim4$ and $\sim 20$) for
$M_z>3\times 10^6 M_\odot$: at these redshifted masses the inspiral
is out of band, and the merger-ringdown contribution to the signal
is dominant. The plot also shows ${\cal R}(\rho)$ for a family of
nonspinning PhenomC waveforms. Although the trend is the same, the
average gain is smaller in this case, because highly spinning MBHBs
are louder GW sources in the merger-ringdown phase.

Next, we need to check how the parameter estimation accuracy scales with
${\cal R}(\rho)$. To this end we constructed a set of six 
analytic IMR precessing waveforms by ``stitching''
NR simulations to PN approximations of the early
inspiral phase (see Appendix~\ref{app:Hybrid}
for details).  Each
waveform is constructed for fixed values of the symmetric mass ratio
$\eta$, of the initial phase $\phi_0$ and of the six parameters
defining the spin magnitudes and orientations.  We are therefore left
with seven free parameters determining the response of the detector to a
given waveform:
\begin{equation}
{\cal C}= \{ \theta, \phi, M_z, t_c, D_l, \iota, \Psi\}.
\label{paramC}
\end{equation}
We perform $10^4$ Monte-carlo drawings of ${\cal C}$ by assuming
isotropic distributions in all angles, a flat distribution in $t_c$
between one week and two years, a flat-in-log distribution in $M_z$ between
$10^5M_\odot$ and $7 \times 10^7M_\odot$, and a flat distribution in $D_l$
between 1~Gpc and~250 Gpc.
We then perform an error analysis using the sub-matrix of the
``complete'' Fisher matrix that corresponds to these parameters.
For each event, we first compute parameter errors for the inspiral
portion of the waveform ($\Delta{{\cal C}_{\rm I}}$), and then for the
full hybrid precessing waveform ($\Delta{{\cal C}_{\rm IMR}}$).
The ratio
${\cal R}(\Delta{\cal C})=\Delta{{\cal C}_{\rm IMR}}/\Delta{{\cal C}_{\rm I}}$
is then compared to the ratio
${\cal R}(\rho)=\rho_{\rm IMR}/\rho_{\rm I}$.
For a fixed ${\cal R}(\rho)$ we find a fairly wide range
${\cal R}(\Delta{\cal C})$, depending on the parameters of the system
(i.e., sky location, inclination, etc.), and we consider the median
of the distribution of ${\cal R}(\Delta{\cal C})$ as a function of
${\cal R}(\rho)$. The results for the median $\Delta{\Omega}$ and
$\Delta{D_l}/D_l$ are shown in Fig.~\ref{fig:IMRgain} for the detector
configurations N2A1M2L4 and N2A1M2L6, and for the waveform $Q_2$ in
Table~\ref{tab:configurations}.
The figure indicates that: (i) for a
six-link detector, ${\cal R}(\Delta{\Omega})\propto[{\cal R}(\rho)]^{-2}$
and ${\cal R}(\Delta{D_l}/D_l)\propto[{\cal R}(\rho)]^{-1}$, as one would
expect from analytical scalings; (ii) for a four-link detector,
${\cal R}(\Delta{\Omega})\propto[{\cal R}(\rho)]^{-1}$ and
${\cal R}(\Delta{D_l}/D_l)\approx 0.5$. 
This latter result is indicative of parameter degeneracy preventing an
optimal scaling. Since mass ratios and spins have been fixed, we
cannot use this model to scale errors on these parameters. However, we
notice that in the merger-ringdown phase, the waveform is
characterized by the mass and spin of the MBH remnant (and not of the
individual progenitors). It is therefore unlikely that a full error
analysis on IMR waveforms would lead to significant improvements in
the errors $\Delta{m_{1z}}, \Delta{m_{2z}}$, $\Delta{\chi_{1}}$,
$\Delta{\chi_{2}}$.

Based on these scaling estimates, we tentatively extrapolate the results 
obtained by
from the precessing inspiral-only waveforms as follows:
\begin{enumerate}
\item For each MBHB we compute the ratio
  ${\cal R}(\rho)=\rho_{\rm PhenC,IMR}/\rho_{\rm PhenC,I}$ using a
  PhenomC waveform;
\item We rescale the SNR computed using SUA waveforms by the factor
  ${\cal R}(\rho)$;
\item We finally rescale the errors $\Delta{\Omega}$ and
  $\Delta{D_l}/D_l$ as described above (but we do not apply any
  correction to the mass and spin determination errors), to get
  what we refer to as an ``SUA IMR'' estimate.
\end{enumerate}

We caution that the scaling is based on the analysis of a
seven-parameter Fisher sub-matrix using a restricted number of
selected waveforms, and we could not check whether it holds when the
full set of 15 parameters is considered. As such, the IMR results
presented below should only be taken as roughly indicative of the
effect of adding merger and ringdown. A more rigorous and
comprehensive study taking into account the impact of systematic
errors is a topic for future work, and it will be crucial to assess
the accuracy of our rough estimates and to improve upon them.

\section{Results}
\label{results}

\begin{table*}
  \begin{tabular}{|c|cc|cc|cc|cc|cc|cc|}
\hline
\multirow{3}{*}{Config ID}&\multicolumn{6}{|c|}{SUA 
(IMR)}&\multicolumn{6}{|c|}{restricted 2PN}\\
\cline{2-13}
&\multicolumn{2}{|c|}{popIII}&\multicolumn{2}{|c|}{Q3-nod}&\multicolumn{2}{|c|}{
Q3-d}&\multicolumn{2}{|c|}{popIII}&\multicolumn{2}{|c|}{Q3-nod}&\multicolumn{2}{
|c|}{Q3-d}\\
\cline{2-13}
&all&$z>7$&all&$z>7$&all&$z>7$&all&$z>7$&all&$z>7$&all&$z>7$\\
\hline
N2A5M5L6 &659.7(660.4) &401.1(401.1) &595.6(611.8) &342.6(358.0) &40.4(40.8) 
&3.6(3.6)    &665.8 &402.7 &610.2 &357.0 &40.4 &3.6\\
\hline
N2A5M5L4 &510.7(511.8) &277.5(277.5) &555.6(608.7) &306.4(355.0) &40.2(40.8) 
&3.4(3.6)    &507.6 &278.5 &602.4 &349.8 &40.4 &3.6\\
\hline
N2A2M5L6 &356.8(357.9) &160.1(160.1) &558.8(609.4) &307.6(355.9) &40.2(40.8) 
&3.6(3.6)    &359.3 &162.6  &593.8 &341.8 &40.4 &3.6\\
\hline
N2A2M5L4 &233.1(235.0) &78.8(78.8)   &495.9(598.1) &253.2(346.1) &39.8(40.8) 
&3.4(3.6)    &223.4 &76.8  &557.5 &309.6 &39.9 &3.6\\
\hline
N2A1M5L6 &157.6(159.5) &34.9(34.9)   &498.1(602.9) &251.6(350.0) &39.1(40.8) 
&3.1(3.6)    &152.4 &34.6   &570.5 &320.0 &40.4 &3.6\\
\hline
N2A1M5L4 &97.2(99.9)   &16.4(16.4)   &417.9(574.1) &186.8(327.5) &37.9(40.6) 
&2.8(3.4)    &96.3  &14.9   &519.1 &278.2 &39.1 &3.3\\
\hline
N1A5M5L6 &246.6(249.3) &86.8(86.8)   &416.2(598.3) &177.5(345.5) &37.5(40.8) 
&2.5(3.6)    &245.9 &87.0  &533.0 &283.9 &39.9 &3.6\\
\hline
N1A5M5L4 &153.9(158.7) &36.1(36.1)   &342.9(565.4) &125.6(317.7) &33.7(40.7) 
&2.0(3.5)    &149.1 &35.6   &470.8 &231.6  &38.7 &3.4\\
\hline
N1A2M5L6 &118.7(122.1) &22.5(22.5)   &255.7(554.2) &66.5(305.0)  &27.8(40.8) 
&1.1(3.6)    &120.3 &21.9   &398.2 &167.5  &36.8 &2.4\\
\hline
N1A2M5L4 &70.6(78.0)  &8.0(8.1)      &189.7(484.1) &37.3(249.0)  &22.4(40.6) 
&0.7(3.4)    &69.5  &7.8   &316.7 &113.4  &31.1 &1.8\\
\hline
N1A1M5L6 &48.8(58.6)  &3.9(4.1)      &142.1(456.4) &17.0(223.0)  &16.8(40.1) 
&0.5(3.4)    &56.1  &4.1   &262.0 &69.6  &29.2 &1.1\\
\hline
N1A1M5L4 &28.4(38.2)  &1.3(1.5)      &95.3(371.4)  &6.1(161.5)   &11.7(38.5) 
&0.3(2.9)    &35.4  &1.4   &193.5 &39.3  &24.0 &0.7\\
\hline
\end{tabular}
\caption{\label{tabmbhrates} Number of detected MBH mergers for the three MBH 
population models discussed in the text, using the two different waveform 
models 
(SUA and restricted 2PN) discussed in the text. Numbers in parentheses (IMR) 
are 
for SUA waveforms, with the SNR rescaled to account for the contribution of 
merger and ringdown as described in Section~\ref{WF-PE}. For each model we 
report both the overall number of detections and only those at $z>7$, assuming 
a 
five-year mission lifetime and a detection threshold $\rho=8$ on the SNR. 
Approximate rates for a two-year mission can be obtained by multiplying 
entry 
by 0.4.}
\end{table*}

\begin{table*}
  \begin{tabular}{|c|ccc|ccc|ccc|ccc|ccc|}
\hline
\multirow{3}{*}{Config ID}
&\multicolumn{3}{|c|}{$\Delta{m}_{1z,2z}/m_{1z,2z}<0.01$}
&\multicolumn{3}{|c|}
{$\Delta\chi_1<0.01$}&\multicolumn{3}{|c|}{$\Delta\chi_2<0.1$}
&\multicolumn{3}{|c|}{$\Delta\theta_{\chi_{1,2}} < 10$ deg}
&\multicolumn{3}{|c|}{
$\Delta\chi_r<0.1$}\\
\cline{2-16}
&popIII&Q3-nod &Q3-d &popIII&Q3-nod &Q3-d &popIII&Q3-nod &Q3-d &popIII&Q3-nod 
&Q3-d &popIII&Q3-nod &Q3-d\\
\hline
N2A5M2L6 &146.6&141.8&13.3&45.3&76.8&2.6 &41.8 &44.7 &3.9 &21.0 &40.9 &9.4 &3.5 
&31.4 &10.9\\ 
\cline{1-16}
N2A5M2L4 &94.6 &108.5&11.3&32.4&60.5&2.1 &21.2 &27.2 &2.5 &11.5 &19.1 &4.8 
&3.0&18.5&10.7\\
\cline{1-16}
N2A2M2L6 &71.4 &99.6 &10.9&28.3&54.4&2.0 &17.1 &22.2 &2.1 &11.7 &18.9 &5.1 &3.3 
&27.0 &10.5\\ 
\cline{1-16}
N2A2M2L4 &40.7 &69.1 &8.4 &19.6&40.8&1.5 &8.2 &11.1 &1.1 &6.0 &7.7 &2.3 
&2.9&17.0&10.2\\
\cline{1-16}
N2A1M2L6 &30.4 &66.4 &8.5 &18.7&39.3&1.5 &7.4 &10.8 &1.0 &6.1 &9.2 &2.9 &3.1 
&21.3 &9.5\\
\cline{1-16}
N2A1M2L4 &15.3 &41.2 &6.3 &13.4&27.6&1.0 &3.8 &4.9 &0.6 &3.1 &3.0 &1.0 
&2.9&12.3&9.3\\
\hline
N1A5M2L6 &40.7 &49.3 &7.0 &20.5&29.8&0.9 &7.3 &8.0 &0.6 &5.7 &6.8 &1.9 &3.0 
&22.1 &10.5\\ 
\cline{1-16}
N1A5M2L4 &18.7 &29.8 &4.7 &14.6&20.3&0.6 &3.6 &3.7 &0.4 &2.5 &2.2 &0.6 
&2.7&16.5&10.3\\ 
\cline{1-16}
N1A2M2L6 &11.6 &20.4 &3.2 &12.6&12.6&0.2 &2.2 &2.4 &0.2 &1.8 &2.2 &0.6 &2.7 
&15.0 &9.2\\
\cline{1-16}
N1A2M2L4 &4.4  &10.1 &2.3 &7.5 &8.2 &0.1 &1.1 &1.0 &0.1 &0.8 &0.6 &0.2 
&2.6&12.1&9.2\\
\cline{1-16}
N1A1M2L6 &3.3  &8.7  &2.4 &4.8 &5.7 &0.1 &0.6 &0.6 &0.1 &0.7 &0.6 &0.2 &2.2 
&9.1 &6.6\\ 
\cline{1-16}
N1A1M2L4 &1.6  &3.8  &1.0 &2.4 &3.3 &0.0 &0.3 &0.4 &0.1 &0.3 &0.2 &0.1 
&2.1&7.8&6.4\\
\hline
\end{tabular}
\caption{\label{tabmbhspin2} Number of MBHBs detected with specific values of 
the mass and spin errors. $\Delta{m}_{1z,2z}/m_{1z,2z}$ is the relative error 
on each of the MBH masses, $\Delta\chi_1$ and $\Delta\chi_2$ are the absolute 
errors on the individual spin magnitudes, $\Delta\theta_{\chi_{1,2}}$ is the 
absolute error on each of the spin misalignment angles with respect to the 
orbital angular momentum at the innermost stable circular orbit, and 
$\Delta\chi_r$ is the error on the magnitude on the remnant MBH spin. Numbers 
are for a two-year mission lifetime.}
\end{table*}
\begin{table*}
  \begin{tabular}{|c|ccc|ccc|ccc|ccc|ccc|}
\hline
\multirow{3}{*}{Config ID}
&\multicolumn{3}{|c|}{$\Delta{m}_{1z,2z}/m_{1z,2z}<0.01$}
&\multicolumn{3}{|c|}
{$\Delta\chi_1<0.01$}&\multicolumn{3}{|c|}{$\Delta\chi_2<0.1$}
&\multicolumn{3}{|c|}{$\Delta\theta_{\chi_{1,2}} < 10$ deg}
&\multicolumn{3}{|c|}{
$\Delta\chi_r<0.1$}\\
\cline{2-16}
&popIII&Q3-nod &Q3-d &popIII&Q3-nod &Q3-d &popIII&Q3-nod &Q3-d &popIII&Q3-nod 
&Q3-d &popIII&Q3-nod &Q3-d\\
\hline
N2A5M5L6 &510.5 &406.6 &33.5 &114.4&199.5&6.9  &153.1 &130.2 &10.4 &63.7 &111.6 
&24.2 &8.8 &78.3 &27.2\\ 
\cline{1-16}
N2A5M5L4 &366.8 &328.5 &28.7 &89.1 &160.0&5.4  &81.2 &82.7 &7.4 &35.6 &56.8 
&12.1 &7.4&46.8&26.1\\ 
\cline{1-16}
N2A2M5L6 &255.6 &300.0 &27.4 &73.6 &140.5&4.7  &61.4 &66.2 &6.0 &34.7 &53.6 
&13.2 &8.2 &67.4 &26.2\\ 
\cline{1-16}
N2A2M5L4 &157.0 &219.6 &21.0 &52.0 &106.3&3.7  &30.3 &34.8 &3.6 &16.3 &22.1 
&5.8 &7.4&40.4&25.4\\
\cline{1-16}
N2A1M5L6 &101.4 &214.0 &20.7 &46.1 &101.3&3.3  &24.5 &32.8 &3.4 &16.9 &24.7 
&7.1 &7.8 &52.9 &24.0\\
\cline{1-16}
N2A1M5L4 &53.3  &142.4 &16.0 &32.8 &69.9 &2.3  &11.1 &16.5 &1.9 &7.4 &8.1 &2.6 
&7.2&30.8&23.1\\
\hline
N1A5M5L6 &148.7 &164.6 &15.5 &52.1 &73.8 &2.2  &25.3 &23.3 &1.9 &15.3 &17.7 
&4.7 &7.5 &55.0 &26.3\\
\cline{1-16}
N1A5M5L4 &79.0  &104.9 &10.7 &36.0 &53.2 &1.5  &10.1 &11.6 &1.0 &6.3 &5.8 &1.4 
&6.9&37.3&24.7\\
\cline{1-16}
N1A2M5L6 &52.9  &75.8  &8.4  &31.1 &33.4 &0.6  &6.0 &5.7 &0.6 &4.8 &4.5 &1.7 
&6.9 &38.2 &23.3\\
\cline{1-16}
N1A2M5L4 &25.5  &43.9  &4.9  &22.5 &20.9 &0.5  &2.4 &3.2 &0.3 &1.8 &1.7 &0.3 
&6.5&26.3&19.1\\
\cline{1-16}
N1A1M5L6 &14.3  &34.4  &4.0  &15.1 &13.0 &0.4  &1.6 &2.3 &0.2 &1.8 &1.1 &0.4 
&5.5 &23.1 &16.6\\
\cline{1-16}
N1A1M5L4 &7.7   &16.7  &1.9  &6.5  &8.0  &0.2  &0.4 &1.3 &0.1 &0.5 &0.4 &0.0 
&5.5&14.2&13.9\\
\hline
\end{tabular}
\caption{\label{tabmbhspin5}  Same as Table~\ref{tabmbhspin2}, but for a 
five-year mission lifetime.}
\end{table*}

\begin{table*}
  \begin{tabular}{|c|ccc|ccc|ccc|ccc|}
\hline
\multirow{3}{*}{Config ID} &\multicolumn{6}{|c|}{$\Delta\Omega<10$ deg$^2$ \& 
$\Delta{D_l}/D_l<0.1$ \& $z<5$} &\multicolumn{6}{|c|}{$z>7$ \& 
$\Delta{D_l}/D_l<0.3$}\\
\cline{2-13}
&\multicolumn{3}{|c|}{SUA}&\multicolumn{3}{|c|}{SUA 
IMR}&\multicolumn{3}{|c|}{SUA}&\multicolumn{3}{|c|}{SUA IMR}\\
\cline{2-13}
&popIII&Q3-nod &Q3-d &popIII&Q3-nod &Q3-d &popIII&Q3-nod &Q3-d &popIII&Q3-nod 
&Q3-d\\
\hline
N2A5M2L6 &14.5 &34.8 &6.0 &16.1 &47.4 &10.1   &71.6 &117.2 &1.2 &71.6 &141.1 
&1.4 \\  
\cline{1-13}
N2A5M2L4 &3.2 &8.7 &1.1 &4.8 &16.0 &4.9   &10.2 &54.4 &0.6 &30.4 &96.8  &1.0 \\ 
\cline{1-13}
N2A2M2L6 &6.8 &23.2 &3.8 &9.2 &35.2 &9.5   &20.8 &82.6&0.9 &20.8  &134.4 &1.4 
\\  
\cline{1-13}
N2A2M2L4 &1.6 &4.2 &0.4 &2.6 &5.8 &1.6   &2.8 &18.0 &0.2 &10.1 &54.0 &0.7 \\
\cline{1-13}
N2A1M2L6 &3.4 &14.9 &2.5 &5.7 &26.4 &7.8   &3.9 &50.9&0.6 &3.9 &120.1 &1.3 \\
\cline{1-13}
N2A1M2L4 &0.6 &1.7 &0.1 &1.0 &2.6 &0.5   &0.5 &0.8 &0.0 &2.6 &41.8 &0.2 \\
\hline
N1A5M2L6 &4.0 &13.7 &1.9 &7.0 &27.3 &7.5   &9.8  &30.5 &0.4 &9.9  &111.9 &1.2 
\\ 
\cline{1-13}
N1A5M2L4 &0.7 &1.6 &0.0 &1.2 &2.6 &0.2   &1.3  &2.2 &0.0  &5.2  &9.0 &0.2 \\ 
\cline{1-13}
N1A2M2L6 &1.9 &5.1 &0.8 &4.4 &18.0 &5.5   &2.3 &6.6 &0.2 &2.4 &77.7 &1.0 \\
\cline{1-13}
N1A2M2L4 &0.4 &0.5 &0.0 &0.6 &1.0 &0.1   &0.2 &0.4 &0.0 &1.0 &2.0 &0.0 \\
\cline{1-13}
N1A1M2L6 &0.7 &1.5 &0.2 &2.7 &9.8 &3.9   &0.2 &0.1 &0.0 &0.5 &0.4 &0.6 \\
\cline{1-13}
N1A1M2L4 &0.2 &0.2 &0.0 &0.2 &0.3 &0.0   &0.0 &0.0 &0.0 &0.0 &0.0 &0.0 \\
\hline
\end{tabular}
\caption{\label{tabmbhsky2} Number of MBHBs detected within specific values of 
the sky location and luminosity distance errors, as reported in the table 
headers. A mission lifetime of two years is assumed.}
\end{table*}
\begin{table*}
  \begin{tabular}{|c|ccc|ccc|ccc|ccc|ccc|}
    \hline
    \multirow{3}{*}{Config ID} &\multicolumn{6}{|c|}{$\Delta\Omega<10$ deg$^2$ 
\& $\Delta{D_l}/D_l<0.1$ \& $z<5$} &\multicolumn{6}{|c|}{$z>7$ \& 
$\Delta{D_l}/D_l<0.3$}\\
\cline{2-13}
&\multicolumn{3}{|c|}{SUA}&\multicolumn{3}{|c|}{SUA 
IMR}&\multicolumn{3}{|c|}{SUA}&\multicolumn{3}{|c|}{SUA IMR}\\
\cline{2-13}
&popIII&Q3-nod &Q3-d &popIII&Q3-nod &Q3-d &popIII&Q3-nod &Q3-d &popIII&Q3-nod 
&Q3-d\\
\hline
N2A5M5L6 &41.0 &90.6 &14.8 &45.0 &119.6 &26.1    &207.1 &299.4 &3.4 &207.1 
&352.4 &3.6 \\ 
\cline{1-13}
N2A5M5L4 &10.5 &23.9 &3.5 &15.7 &43.9 &13.4    &35.3  &147.6 &1.6 &100.6 &258.8 
&2.7 \\ 
\cline{1-13}
N2A2M5L6 &21.0 &62.9 &9.3 &26.4 &94.2 &23.1    &60.6 &210.0 &2.3 &60.6 &338.4 
&3.6 \\
\cline{1-13}
N2A2M5L4 &3.9 &11.0 &1.4 &6.4 &16.4 &3.7    &9.7 &53.1 &0.9 &31.4 &147.4 &1.7 \\
\cline{1-13}
N2A1M5L6 &10.7 &37.5 &6.0 &15.2 &68.4 &19.2    &12.1 &134.1 &1.6 &12.1 &306.0 
&3.4 \\ 
\cline{1-13}
N2A1M5L4 &1.9 &4.6 &0.4 &3.0 &7.8 &1.4    &1.9 &13.4 &0.1 &6.3 &64.6 &0.9 \\
\hline
N1A5M5L6 &12.3 &34.3 &4.4 &18.9 &72.2 &18.0    &26.9 &79.1 &1.3 &26.9 &286.7 
&3.4 \\ 
\cline{1-13}
N1A5M5L4 &1.9 &4.5 &0.3 &3.4 &6.4 &1.0    &4.2 &5.8 &0.1 &14.4 &26.8 &0.3 \\ 
\cline{1-13}
N1A2M5L6 &5.5 &14.3 &2.4 &12.0 &45.8 &13.5    &6.1 &17.2 &0.5 &6.3 &197.7 &2.4 
\\
\cline{1-13}
N1A2M5L4 &0.8 &1.2 &0.0 &1.2 &2.0 &0.1    &1.3 &0.4 &0.0 &2.7 &4.9 &0.1 \\
\cline{1-13}
N1A1M5L6 &2.0 &4.6 &0.9 &7.9 &24.9 &9.0    &1.0 &2.1 &0.1 &1.3 &110.8 &1.7 \\
\cline{1-13}
N1A1M5L4 &0.2 &0.5 &0.0 &0.7 &0.8 &0.0    &0.1 &0.0 &0.0 &0.6 &0.6 &0.0 \\ 
\hline
\end{tabular}
\caption{\label{tabmbhsky5} Same as Table~\ref{tabmbhsky2}, but for a five-year 
mission lifetime.}
\end{table*}

The models presented in Section~\ref{BH-evolution} were used to
generate Monte Carlo catalogs of the population of coalescing MBHBs,
for a total observation time of 50 years (i.e., in terms of plausible
eLISA lifetimes, 10 realizations of five-year catalogs, or 25
realizations of two-year catalogs). The MBH masses, redshifts, spin
orientations and magnitudes were chosen according to the output of our
semianalytical galaxy formation model, by using appropriate smoothing
kernels. The other ``extrinsic'' parameters (sky location,
inclination, polarization angles, time and GW phase at merger), which
are not provided by our model, are randomized by assuming either
uniform distributions or isotropic angular distributions.

The gravitational waveforms used to model the signal of each merger
event in the catalogs were described in Section \ref{WF-PE}. Our
parameter estimation analysis is based on the SUA model introduced
in Section~\ref{sec:SUA}, which includes precession and higher
harmonics, taking advantage of the information on the spin magnitudes
and orientations provided by our galaxy formation model.
The implementation of precession makes it necessary to taper the
waveform toward merger by introducing a window function. This causes a
partial damping of the SUA waveform amplitude close to merger. To
quantify how this affects the detection rates, we compared with a 2PN
model \cite{Berti:2004bd,Berti:2005qd} that has no spins and no higher
harmonics. The 2PN waveform does not carry any information related to
spin precession, and it was not used in our parameter estimation
calculations.
The impact of merger and ringdown was quantified by the extended SUA
IMR model, constructed as detailed in
Section~\ref{sec:rescaling}. Recall that the SUA IMR model is only
used to rescale errors in the sky location and luminosity distance.

In the following we compare the performance of all 12 eLISA baselines
described in Section~\ref{design} for mission durations of two and
five years. As already mentioned, a longer integration time allows the
resolution and subtraction of more individual CWDs.
This effect is expected to have an impact on the recovery of MBHB
signals, but we neglect it here and use Eq.~(\ref{eq:fit}) for the CWD
noise for both two- and five-year mission lifetimes. Note that even
with this approximation, the performance of the instrument does not
always scale trivially with the mission lifetime. This is because
signals are long-lasting in the detector band (especially for massive
nearby sources), and longer observations lead to a slightly
better-than-linear improvement in the detector performance. This is
especially true for the most sensitive baselines, for which several
sources generate detectable GWs in the detector band for more than two years.
As already mentioned, in the discussion of the results we average
over 10 independent realizations
of the eLISA MBHB data stream in the case of a five-year mission,
and over 25 realizations in the case of a two-year mission. Therefore
all of our results should be understood as accurate within some
Poissonian noise, which we omit in all figures and tables to improve
readability.

\subsection{Detection rates}
\label{sec:rates}
 
\begin{figure*}
\centering
\begin{tabular}{ccc}
\includegraphics[width=8.0cm,clip=true,angle=0]{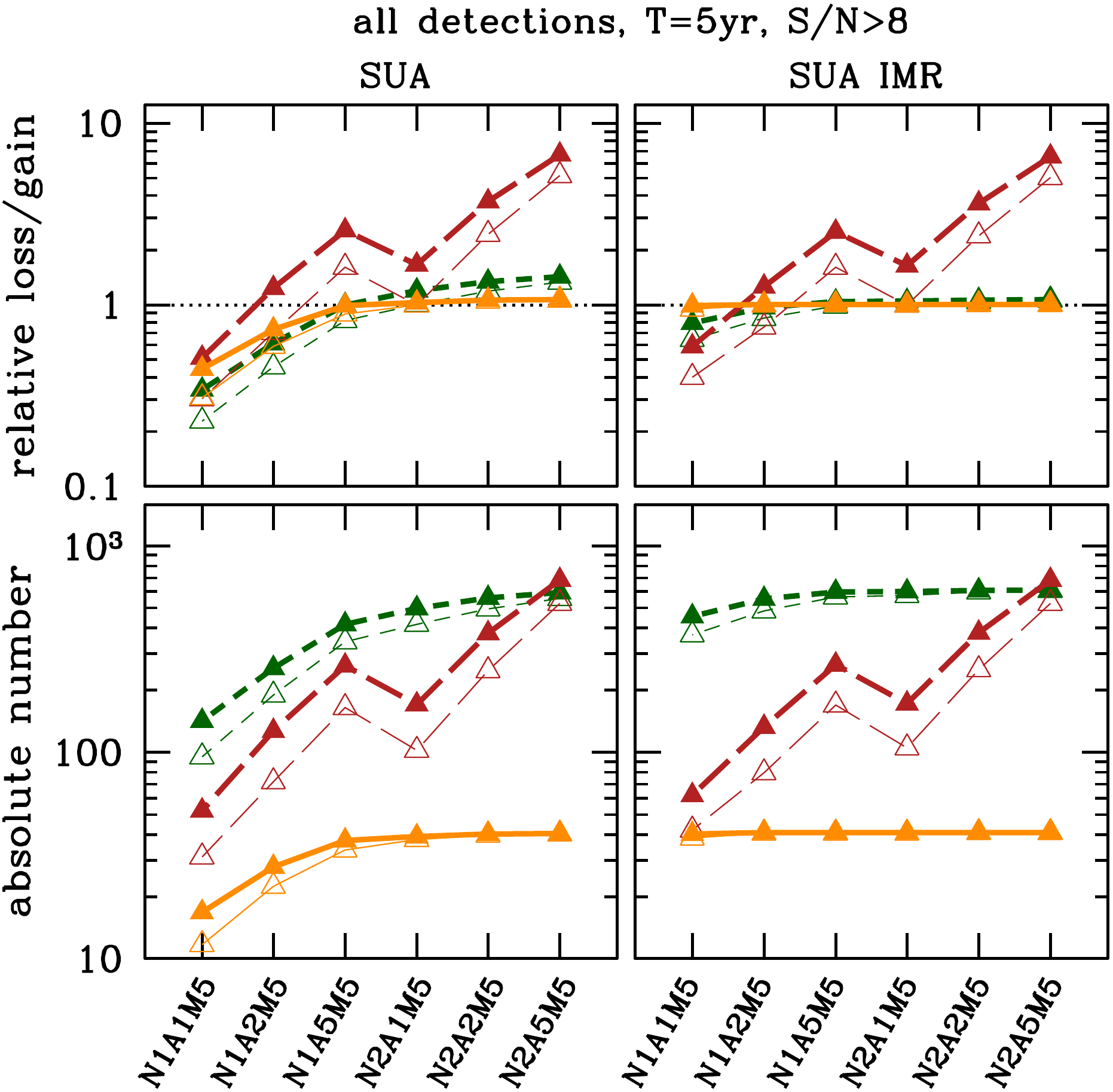}&
\includegraphics[width=0.5cm,clip=true,angle=0]{gap}&
\includegraphics[width=8.0cm,clip=true,angle=0]{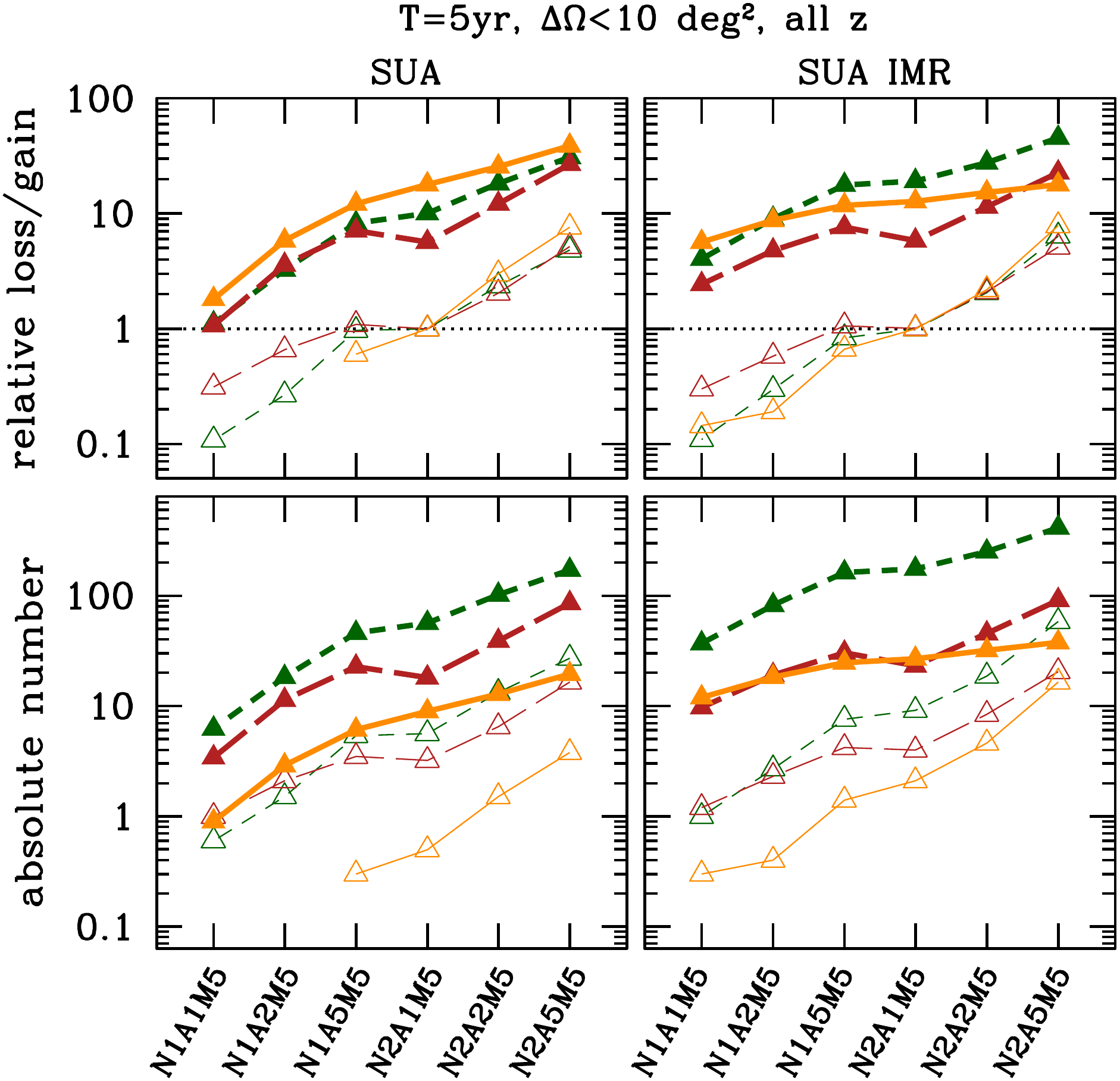}\\
\end{tabular}
\caption{Total number of detections (i.e., sources with $\rho>8$, left
  plot) and total number of detections with $\Delta{\Omega}<10$ deg$^2$
  (right plot) assuming a five year mission (M5). In each plot, the left
  and right panels are for inspiral and IMR-rescaled waveforms,
  respectively. The bottom panels represent the absolute number of
  detections for different eLISA configurations, while the top panels
  represent the gain/loss of a given configuration with respect to the
  standard NGO design, i.e., the ratio [number of sources for
  (N$i$A$j$M$k$L$\ell$)]$/$[number of sources with
  (N2A1M$k$L4)]. Long-dashed brown lines are for model popIII, solid
  orange lines for model Q3-d, and short-dashed green lines for model
  Q3-nod. Thick lines with filled triangles are for six links (L6),
  while thin lines with open triangles are for four links (L4).}
\label{fig:omega_10degsq}
\end{figure*}

The number of events that would be observed (with a threshold $\rho=8$ on the 
SNR) by different eLISA
configurations in a five-year mission (averaged over the 10 catalog
realizations), as well as the
number of events with $z>7$ (roughly corresponding to the furthest
observed quasar at the moment of writing \cite{2011Natur.474..616M}),
are presented in Table
\ref{tabmbhrates}. These numbers are calculated by using three models:
the inspiral-only restricted 2PN waveform model, the (inspiral-only)
SUA model, and the SUA model with the merger-ringdown correction
(SUA IMR). Detection rates scale linearly with the mission
duration, well within the Poissonian error due to the stochastic
nature of our cosmological models. Therefore, to a very good
approximation, a two-year mission would observe a number of events that
can be obtained by multiplying the values in Table~\ref{tabmbhrates} by a
factor of 0.4.

In the popIII scenario, Table~\ref{tabmbhrates} shows that both the
overall detection rates and the detection rates at $z>7$ are
remarkably similar for all waveform models. This is because detectable
mergers in the popIII scenario are dominated by low-mass systems
(cf. Fig.~\ref{fig:pop_rate}), and therefore neither the tapering of
the SUA waveform caused by the window function nor the addition of
the merger and ringdown makes a significant difference in the SNR,
compared to a simple 2PN waveform.

In the heavy-seed scenarios (Q3-d and Q3-nod) the number of detections
is instead waveform-dependent, especially as the detector becomes less
sensitive. This is also to be expected, because most detectable events
have total MBHB redshifted mass $10^5<M_z<10^6$. These binaries merge
well inside the eLISA band, so the SNR is very sensitive to the final
portion of the inspiral (and to whether we include merger or not). For
SUA waveforms the tapering at the end of the inspiral tends to
suppress the SNR, resulting in fewer detections than with restricted
2PN waveforms, which have a hard cutoff at the innermost stable
circular orbit.
On the contrary, adding merger to the SUA waveforms significantly
boosts the SNR, resulting in more detections than with restricted 2PN
waveforms. The inclusion of the merger is especially important for
events with $z>7$ and less sensitive detector configurations --
particularly those with high low-frequency noise (N1) and/or short arm
length (A1, A2), for which the SNR is dominated by the high-frequency
part of the waveform. This consideration highlights the importance of
having accurate IMR waveform models {\em even for
  detection}, and not just for parameter estimation.

A comparative view of the performance of the different designs is
given in the left panel of Fig.~\ref{fig:omega_10degsq}. In this
figure (and in the following ones) thick lines with filled triangles
refer to six-link configurations (L6), while thin lines with open
triangles refer to four-link configurations (L4).
Long-dashed brown lines refer to model popIII, solid orange lines to
model Q3-d, and short-dashed green lines to model Q3-nod.
The bottom panels represent the absolute number of detections as a
function of the eLISA configuration, while the top panels represent
the gain/loss of a given configuration with respect to the standard
NGO design \cite{elisa1-12}, i.e., the ratio [number of sources for
(N$i$A$j$M$k$L$m$)]$/$[number of sources for (N2A1M$k$L4)].
The figure shows that in terms of event rates alone, four- or six-link
configurations yield relatively similar results: roughly speaking, the
SNR of an event only increases by a factor of $\sqrt{2}$ as we move from
a four-link (single-detector) configuration to a six-link
(two-detector) configuration. However, the arm length (A1, A2 or A5)
and the level of the low-frequency noise (either N1 or N2) are of key
importance.  Either of these factors can modify the event rates by
more than a factor of ten, depending on the MBH population scenario.  For
instance, the N1A1 configurations are likely to see just a few tens of
MBHBs in a five-year mission in the ``conservative/realistic'' popIII and
Q3-d models. Even more dramatically, as can be seen from Table
\ref{tabmbhrates}, these same configurations are likely to see at most
a handful of binaries at $z>7$ in the popIII and Q3-d models. This
could severely jeopardize the mission's potential to investigate the
origin of MBH seeds at high redshifts.

The right panel of Fig.~\ref{fig:omega_10degsq} shows an example of
the potential advantages of a six-link configuration in terms of
science return. We compare the number of sources that can be localized
in the sky within 10 square degrees, a figure of merit indicative of
how many detections can be used for electromagnetic follow-up
observations (a 10 deg$^2$ error box is comparable to the SKA and LSST
fields of view). On average, six-link configurations perform about ten
times better than their four-link counterparts. The difference is even
larger when the SUA IMR scaling is adopted, because the improvement in
parameter estimation is more prominent for six links (cf.
Section~\ref{sec:rescaling}). Note that any six-link configuration
performs better than NGO for all the considered MBHB population
models, highlighting the importance of adopting this feature in the
mission design. Including merger somewhat mitigates the difference
across designs for six-link configurations, but a factor $\sim 10$
difference still persists between the best and the worst configuration (see
e.g. the top-right plot in the right panel of
Fig.~\ref{fig:omega_10degsq}).

\subsection{Parameter estimation}


\begin{figure}
\includegraphics[width=8.0cm,clip=true,angle=0]{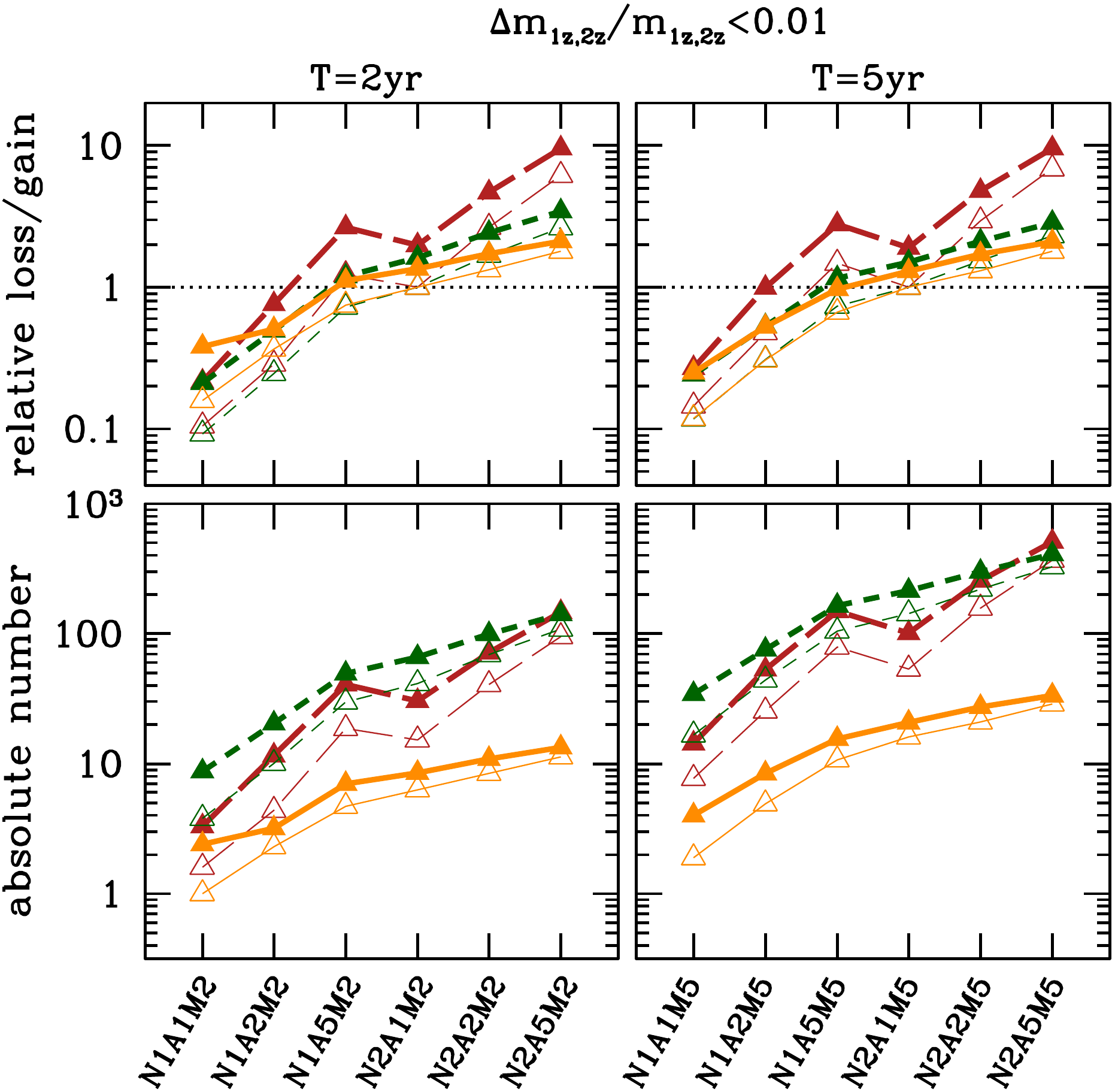}
\caption{Number of detections with fractional errors of less than 0.01 for both 
redshifted masses. Left and right panels are for a mission lifetime of two 
years 
and five years, respectively. Inspiral-only waveforms have been used in all 
cases. The bottom panels represent the number of sources as a function of the 
eLISA configuration, while the top panels represent the gain/loss of a given 
configuration with respect to NGO, i.e., the ratio  [number of sources for
  (N$i$A$j$M$k$L$\ell$)]$/$[number of sources with
  (N2A1M$k$L4)]. Long-dashed brown lines are for model popIII, solid orange 
lines  for model Q3-d, and short-dashed green lines for model Q3-nod. Thick 
lines with filled triangles are for six links (L6), while thin lines with open 
triangles are for four links (L4).}
\label{fig:m12}
\end{figure}
\begin{figure*}
\centering
\begin{tabular}{ccc}
\includegraphics[width=8.0cm,clip=true,angle=0]{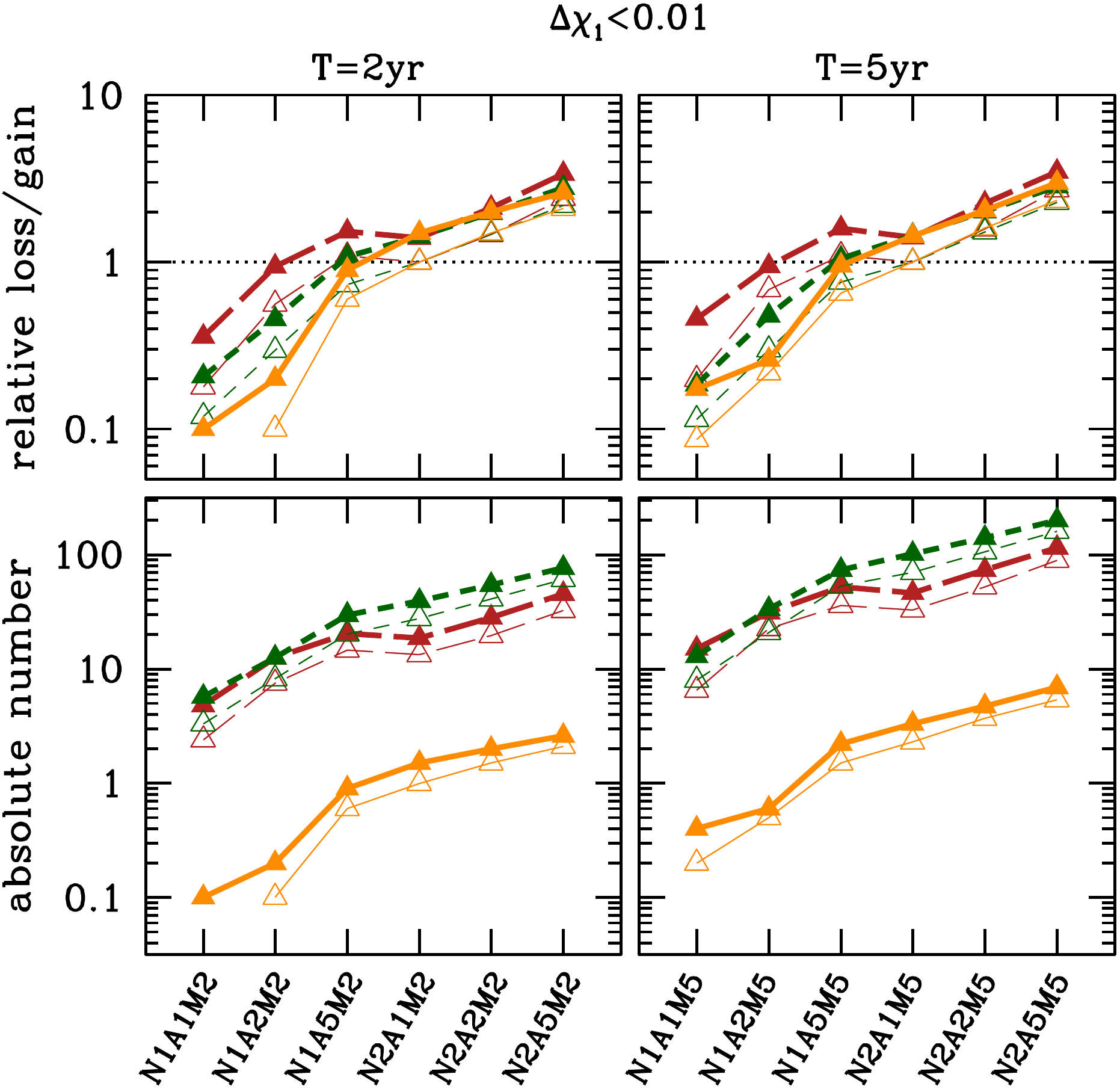}&
\includegraphics[width=0.5cm,clip=true,angle=0]{gap}&
\includegraphics[width=8.0cm,clip=true,angle=0]{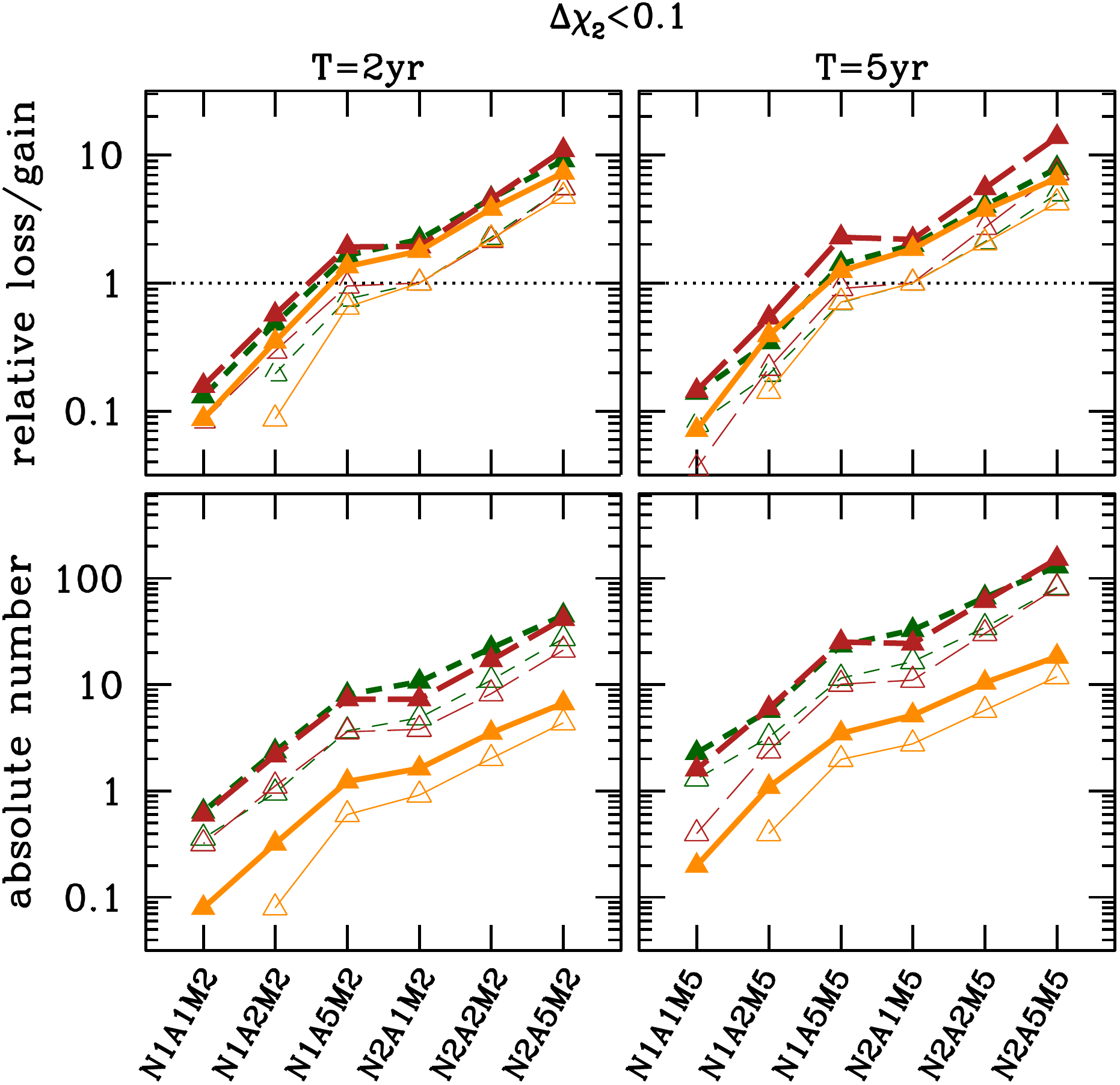}\\
\end{tabular}
\caption{Number of detections with absolute error on the primary MBH spin 
smaller than 0.01 (left plot), and with absolute error on the secondary MBH 
spin 
smaller than 0.1. In each plot, left and right panels are for a mission 
lifetime 
of two years and five years, respectively; inspiral-only waveforms have been 
used in all cases. The bottom panels represent the number of sources as a 
function of the eLISA configuration, while the top panels represent the 
gain/loss of a given configuration with respect to NGO, i.e. the ratio  [number 
of sources for
  (N$i$A$j$M$k$L$\ell$)]$/$[number of sources with
  (N2A1M$k$L4)]. Long-dashed brown lines are for model popIII, solid orange 
lines  for model Q3-d, and short-dashed green lines for model Q3-nod. Thick 
lines with filled triangles are for six links (L6), while thin lines with open 
triangles are for four links (L4).}
\label{fig:s12}
\end{figure*}
\begin{figure} \centering
  \includegraphics[width=7.0cm,clip=true,angle=0]{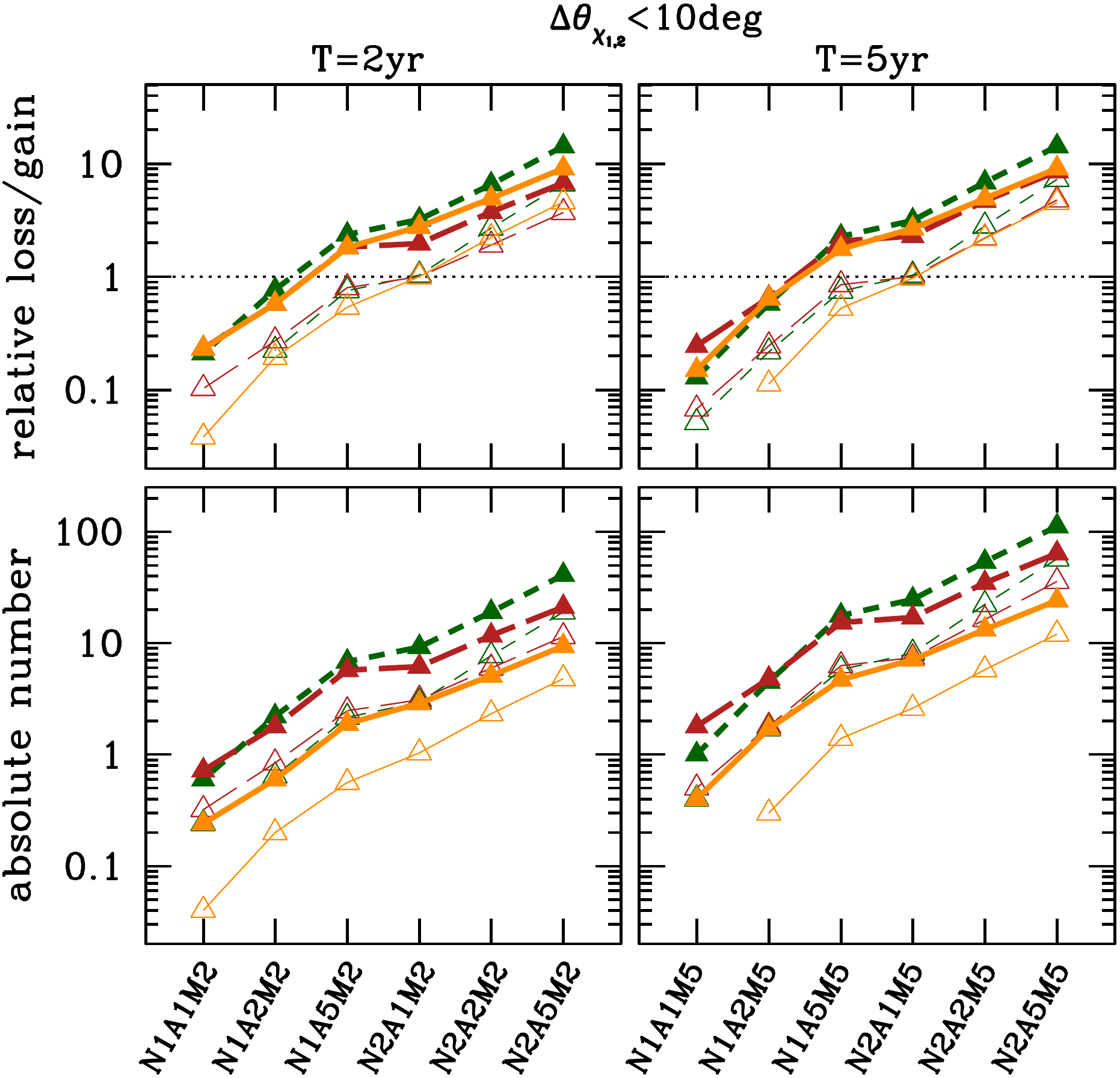}
  \caption{\label{fig:angle} Number of detections such that the absolute error 
in the measurement of both misalignment angles $\theta_{\chi_{1}}$ and 
$\theta_{\chi_{2}}$ at the innermost stable circular orbit is less than 10 
degrees. Left and right panels are for a mission lifetime of two years and five 
years, respectively. Inspiral-only waveforms have been used in all cases. The 
bottom panels represent the number of sources as a function of the eLISA 
configuration, while the top panels represent the gain/loss of a given 
configuration with respect to NGO, i.e., the ratio  [number of sources for 
(N$i$A$j$M$k$L$\ell$)]$/$[number of sources with (N2A1M$k$L4)]. Long-dashed 
brown lines are for model popIII, solid orange lines  for model Q3-d, and 
short-dashed green lines for model Q3-nod. Thick lines with filled triangles 
are 
for six links (L6), while thin lines with open triangles are for four links 
(L4).}
\end{figure}
\begin{figure} \centering
  \includegraphics[width=7.0cm,clip=true,angle=0]{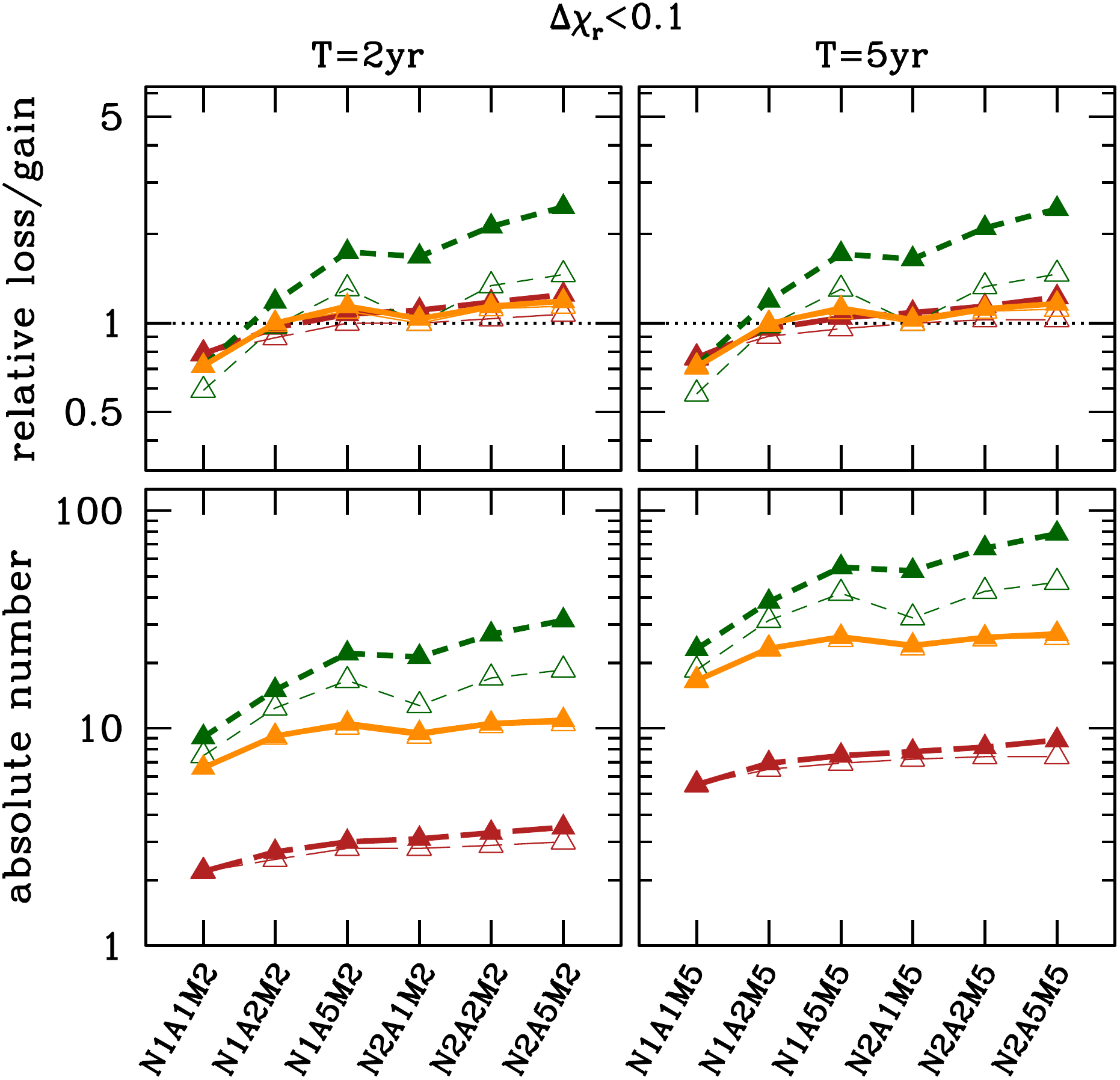}
  \caption{\label{fig:chir} Same as Fig.~\ref{fig:s12}, but for the
    remnant spin $\chi_r$.}
\end{figure}

We assess the accuracy with which various eLISA configurations can
estimate MBHB parameters using the Fisher matrix approach described in
Sec.~\ref{sec:fisher}, either with inspiral-only SUA waveforms or
including a merger-ringdown correction as described in
Sec.~\ref{sec:rescaling}. As a sanity check, we verified that
qualitatively similar trends for the parameter estimation errors are
found with an independent Fisher matrix code employing restricted 2PN,
non-spinning waveforms~\cite{Berti:2004bd} (although the absolute
errors are typically larger for the 2PN models, which omit spin
precession information).

Our main goal is to assess the scientific return of the mission, so we
report mostly the number of systems for which selected parameters
  can be measured within a certain error, rather than the average (or
median) absolute errors on those parameters. This representation is
more directly linked to the mission's science goal of testing the
formation and evolution of the MBH population, which requires
parameters to be measured with reasonable precision for a large sample
of the astrophysical MBH population. For other mission goals, it might
be more appropriate to quote the absolute errors: for instance, in
order to test the black hole no-hair theorem of GR using MBH mergers,
a single MBHB with very well determined parameters (remnant spin
and dominant quasinormal mode
frequencies~\cite{Berti:2005ys,Berti:2007zu}) might be enough. We will
keep this in mind below (e.g. when we report the absolute error with
which the final remnant spin can be measured), but we defer a more
complete analysis of absolute errors and tests of GR to future work.

\begin{figure*}
\centering
\begin{tabular}{ccc}
\includegraphics[width=8.0cm,clip=true,angle=0]{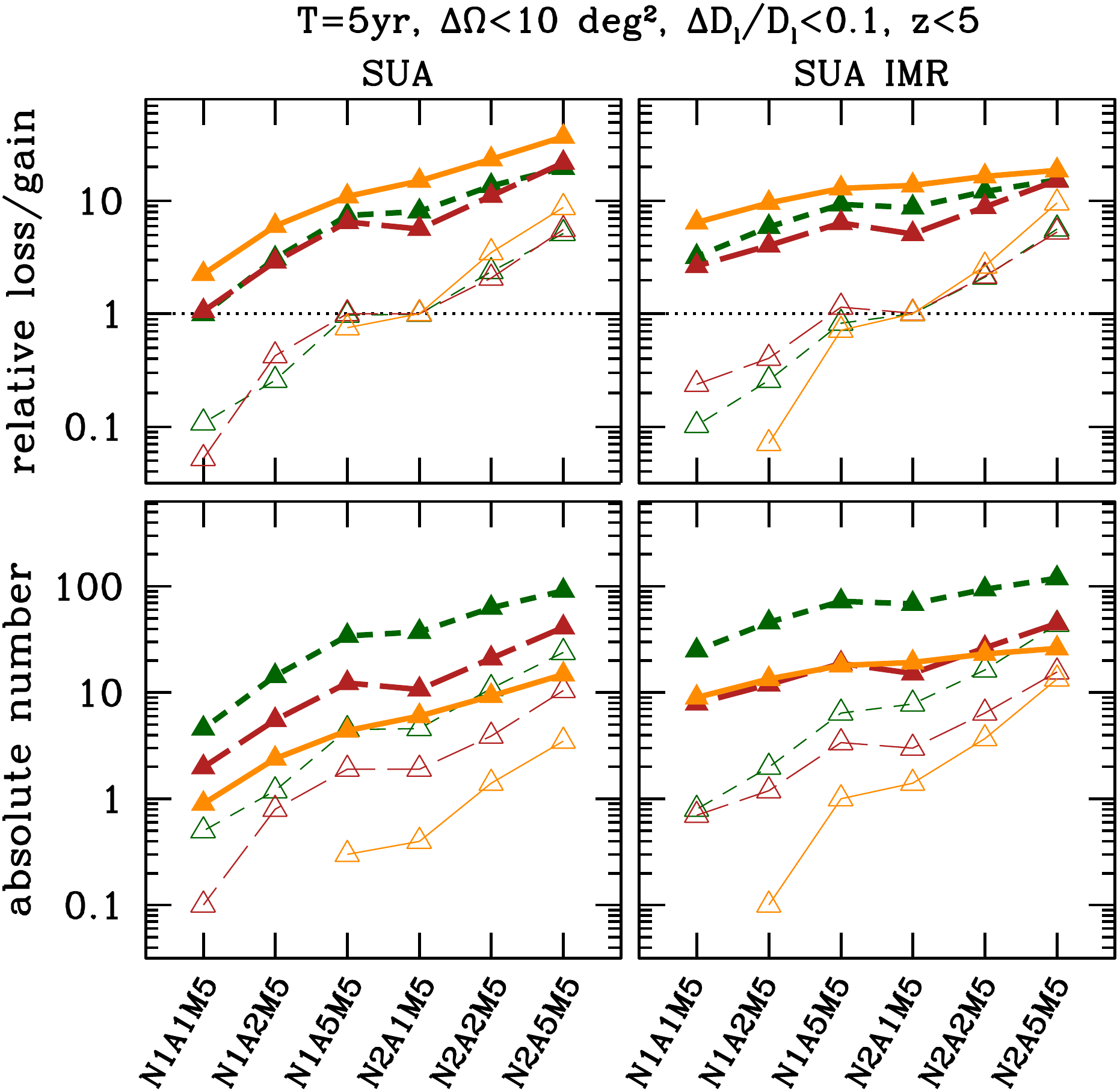}&
\includegraphics[width=0.5cm,clip=true,angle=0]{gap}&
\includegraphics[width=8.0cm,clip=true,angle=0]{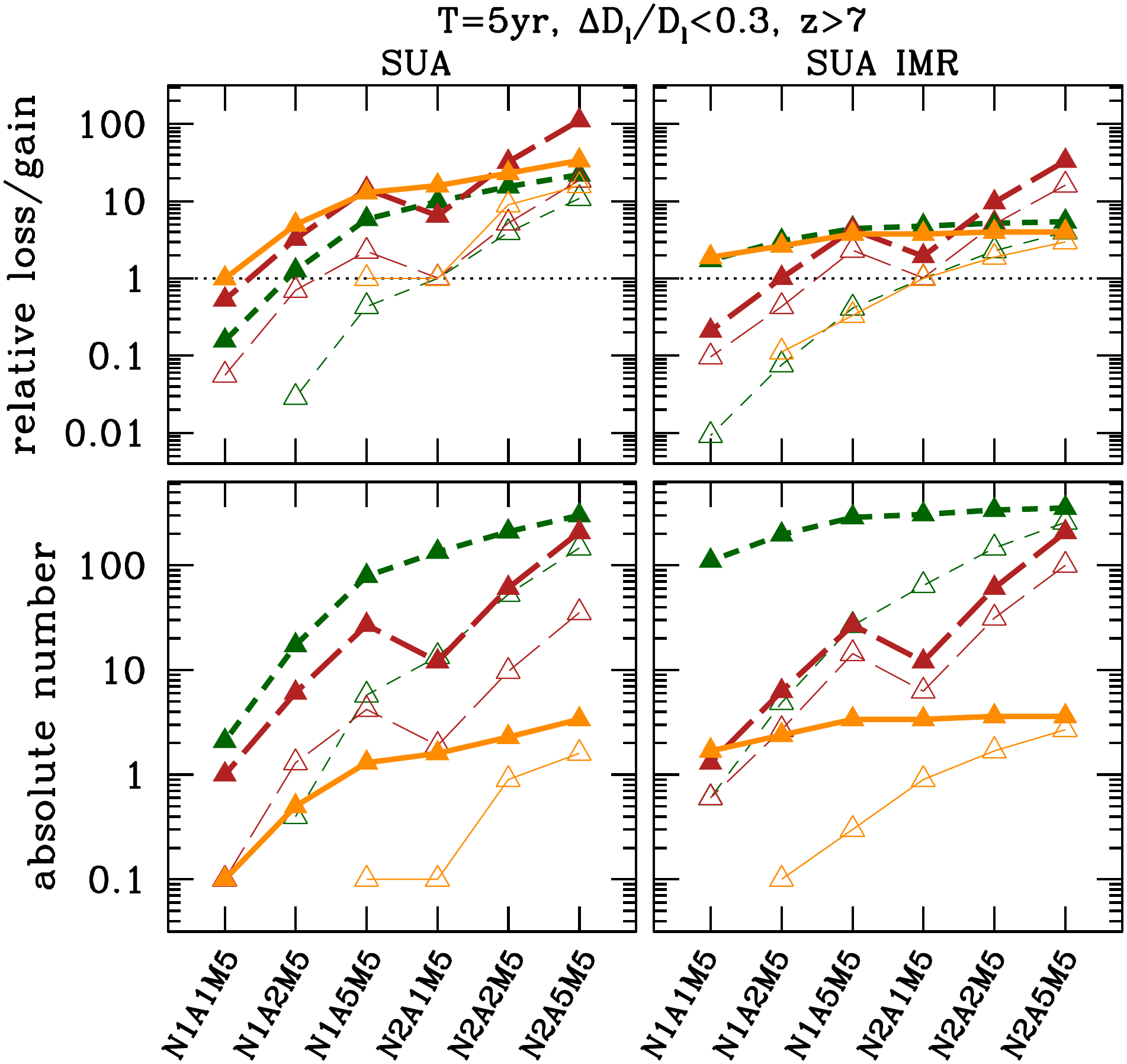}\\
\end{tabular}
\caption{\label{fig:omega_bis} Left plot: total number of detections at $z<5$ 
with 
$\Delta{\Omega}<10$~deg$^2$ and $\Delta{D_l}/{D_l}<0.1$; right plot: total 
number 
of detections at $z>7$ with $\Delta{D_l}/{D_l}<0.3$. In each plot, left and 
right 
panels are for inspiral and IMR rescaled waveform, respectively; five years of 
observations are assumed. The bottom panels represent the number of sources as 
a 
function of the eLISA configuration, while the top panels represent the 
gain/loss of a 
given configuration with respect to NGO, i.e., the ratio  [number of sources for
  (N$i$A$j$M$k$L$\ell$)]$/$[number of sources with
  (N2A1M$k$L4)]. Long-dashed brown lines are for model 
popIII, solid orange lines for model Q3-d, and short-dashed green lines  
for model Q3-nod. Thick lines with filled triangles are for six links (L6), 
while 
thin lines with open triangles are for four links (L4).}
\end{figure*}
\begin{figure*}
\centering
\begin{tabular}{ccc}
\includegraphics[width=7.5cm,clip=true,angle=0]{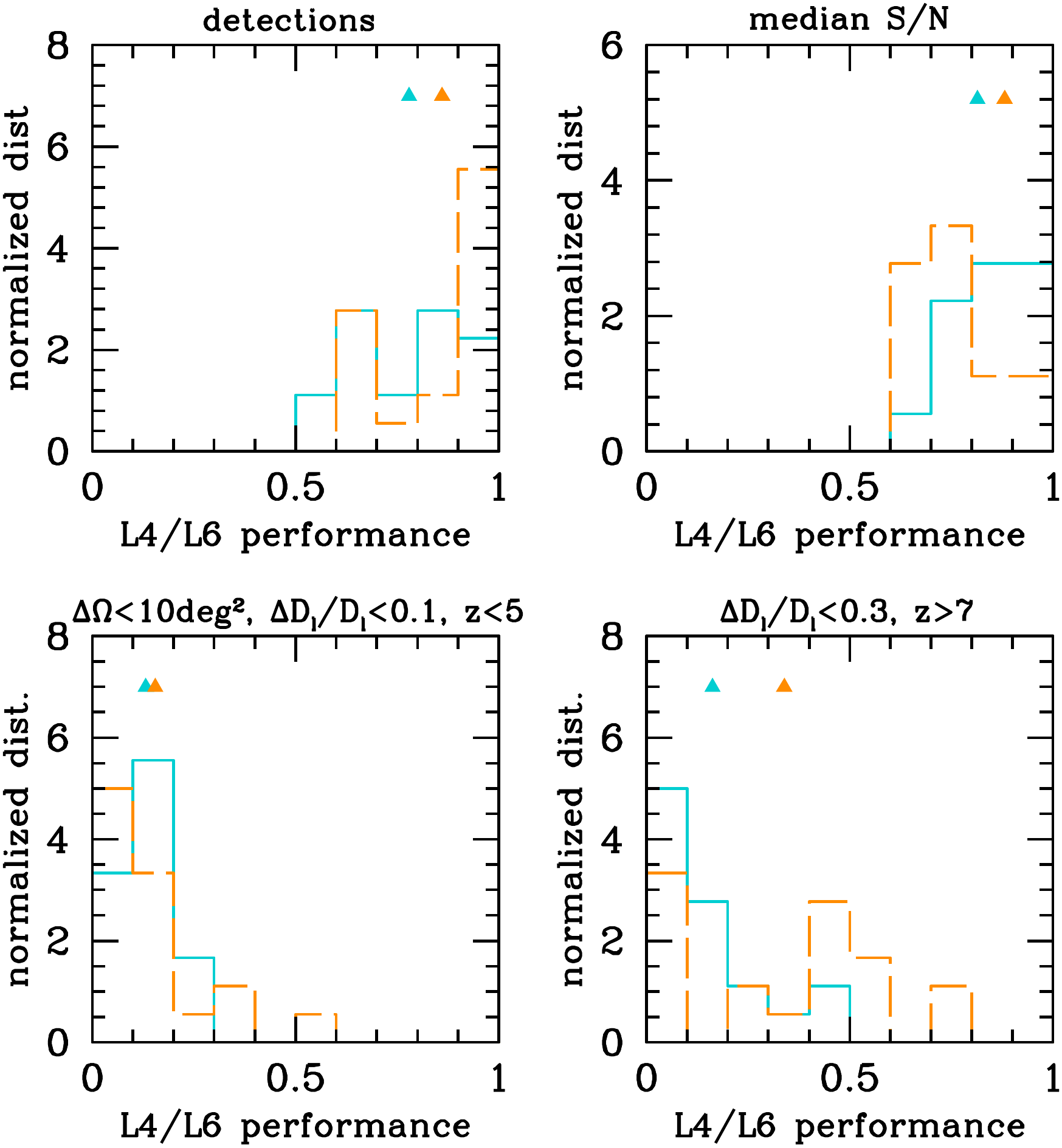}&
\includegraphics[width=0.8cm,clip=true,angle=0]{gap}&
\includegraphics[width=7.5cm,clip=true,angle=0]{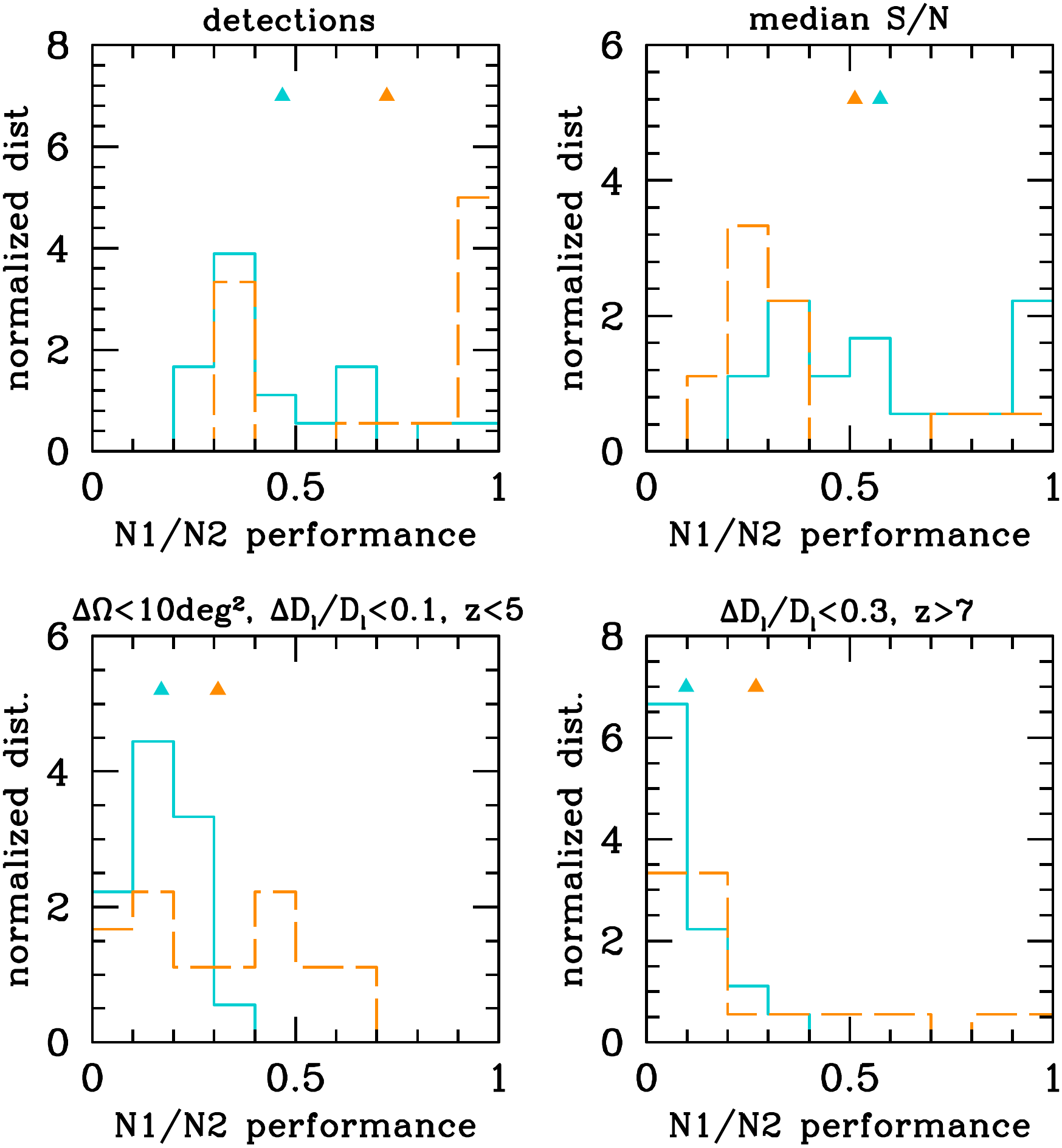}\\
\includegraphics[width=0.8cm,clip=true,angle=0]{gap}&
\includegraphics[width=0.8cm,clip=true,angle=0]{gap}&
\includegraphics[width=0.8cm,clip=true,angle=0]{gap}\\
\includegraphics[width=7.5cm,clip=true,angle=0]{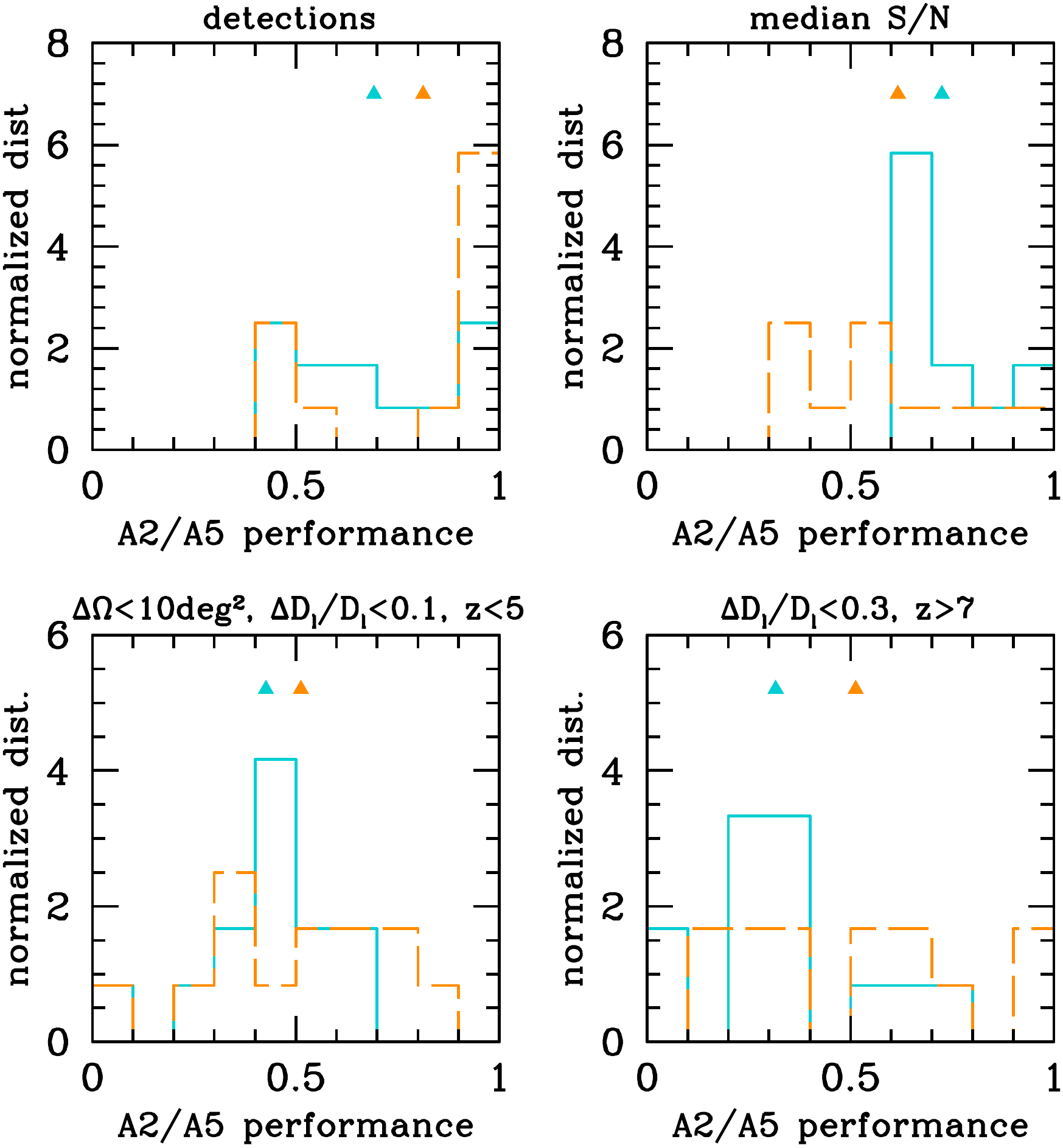}&
\includegraphics[width=0.8cm,clip=true,angle=0]{gap}&
\includegraphics[width=7.5cm,clip=true,angle=0]{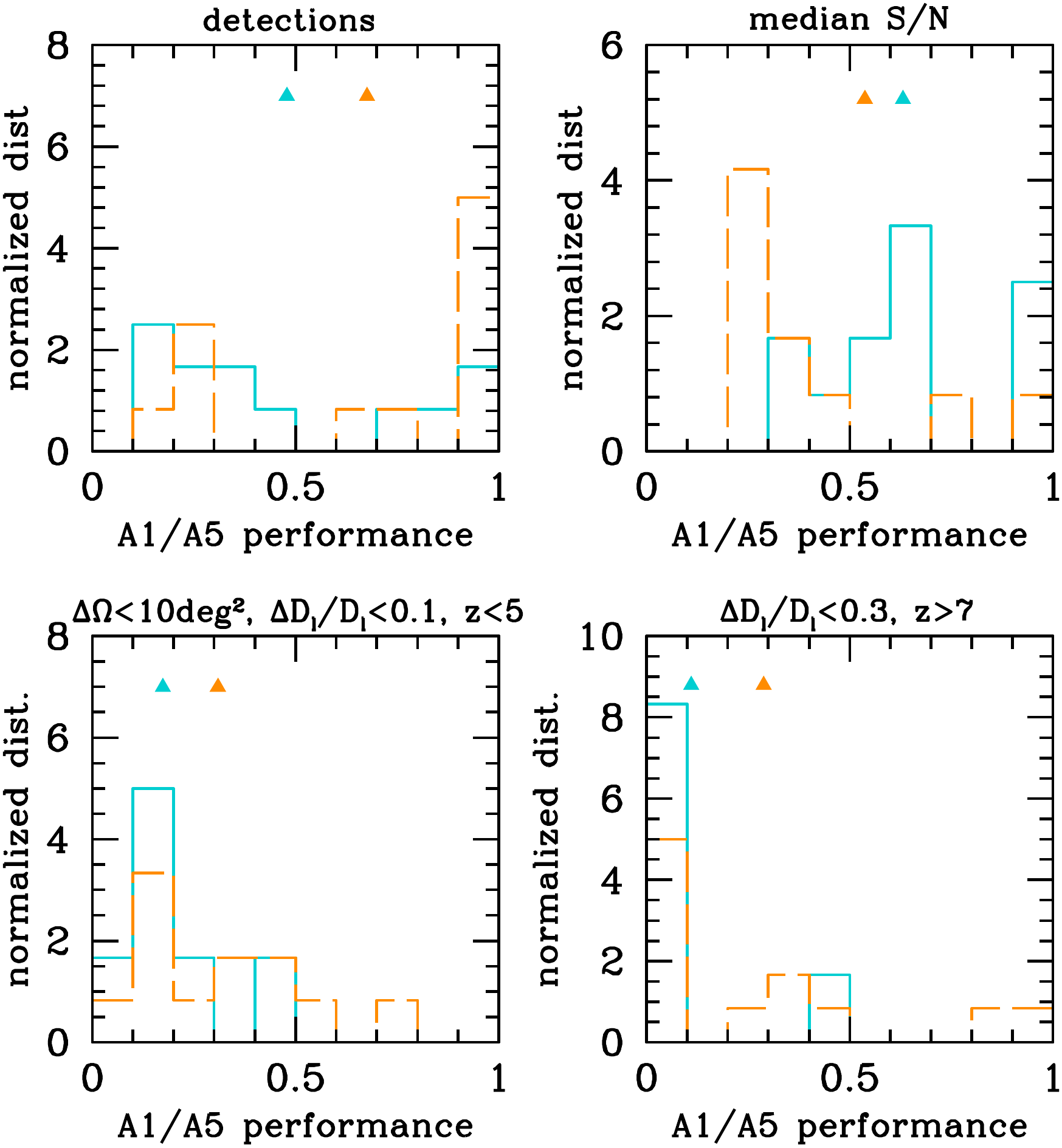}\\
\end{tabular}
\caption{Science loss as a function of specific design choices. In each plot, we 
compare the results of simulations
that only differ by a specific design element (see description in the main 
text). Those are: number of links L4$/$L6 (18 simulations, upper left); low 
frequency noise N1$/$N2 (18 simulations, upper right); arm length A2$/$A5 (12 
simulations, lower left); arm length A1$/$A5 (12 simulations, lower right). 
Each 
of the four plots visualizes the science loss for specific design choices 
according to four different indicators: total number of detections (upper left 
quadrant), median SNR of detected sources (upper right quadrant), number of low 
$z$ sources with good enough sky localization for counterpart searches and good
distance determination (lower left quadrant), and number of high $z$ sources with good distance measurement (lower right quadrant). In each panel we plot the histograms of the indicators over all the simulations, and the mean
value (triangles). Blue is for inspiral waveforms only, orange is for IMR-rescaled waveforms.} 
\label{loss}
\end{figure*}

Our ``success metrics'' to assess the science capabilities of various
mission designs are the expected number of observed binaries that meet one or more of
the following conditions:

\noindent(i) Both redshifted masses ($m_{1z}$ and $m_{2z}$) are
measured with a relative statistical error of 1\% or better:
$\Delta m_{1z}/m_{1z}<0.01$ and $\Delta m_{2z}/m_{2z}<0.01$. This
metric is useful to gauge the mission's capability to probe MBH growth
across cosmic history.

\noindent(ii) Spin magnitudes and directions are measured accurately.
For the spin magnitude, we require that either the spin parameter
$\chi_1$ of the more massive black hole (the ``primary'') be measured
with absolute statistical error of 0.01 or better, or that the spin
parameter $\chi_2$ of the less massive black hole (the ``secondary'')
be measured with absolute statistical error of 0.1 or better. Note
that we use different thresholds because the secondary's spin is
typically harder to measure. As for the spin direction, we require
that the angles of both spins with respect to the orbital angular
momentum of the system ($\theta_{\chi_{1}}$, $\theta_{\chi_{2}}$) be
determined to within an error of 10 degrees or less. Spin magnitudes
and directions are related to the global accretion history and to the
local dynamics of the accretion flow
\cite{Berti:2008af,mymodel,spin_model}, as well as to the interaction
between the MBHs and the gas in the nuclear region via the
Bardeen-Petterson
effect~\cite{1975ApJ...195L..65B,2007ApJ...661L.147B,Gerosa:2015xya}.
Therefore this metric is useful to gauge the mission's capability to
probe the nature of MBH feeding by discriminating the role of various
growth mechanisms (coherent vs chaotic gas accretion, mergers,
etc.).

\noindent(iii) The remnant spin parameter $\chi_{r}$ is measured with
(statistical) absolute precision of 0.1 or better. This is
particularly useful for tests of GR and tests of the black hole
no-hair theorem.

\noindent(iv) The statistical sky position error is
$\Delta\Omega<10$~deg$^2$, the statistical relative error on the
luminosity distance is of 10\% or better ($\Delta{D_l}/D_l<0.1$) and
$z<5$. These systems have sky localization error comparable to (or
smaller than) the SKA and LSST fields of view, and they are close
enough that a possible transient electromagnetic counterpart might be
identified, thus allowing us to measure their redshift and potentially
test the cosmological $D_l(z)$ relation (and therefore the Hubble
parameter, the composition of the Universe, and the equation of state
of dark energy).

\noindent(v) The statistical error on the luminosity distance is
$\Delta{D_l}/D_l<0.3$ and $z>7$. These are the systems that will
provide the most information on the formation of MBH seeds at high
$z$. Clearly, in order to test competing scenarios for MBH seeds we
should be able to detect MBHBs at high redshift, but we should
also be able to ascertain that they \textit{are} indeed at high
redshift (hence the requirement on $\Delta{D_l}/D_l$, which can be
translated into a requirement on the redshift error by assuming a
standard $\Lambda$CDM cosmology).

The results for these figures of merit are reported for all
configurations in Tables~\ref{tabmbhspin2} and \ref{tabmbhsky2} (for a
two-year mission) and in Tables~\ref{tabmbhspin5} and \ref{tabmbhsky5}
(for a five-year mission). The same results are also represented graphically in
Figs.~\ref{fig:omega_10degsq}--\ref{fig:omega_bis}, and are briefly
described below.

\noindent(i) All configurations allow the precise measurement of MBHB
masses for at least a few systems (see Figure~\ref{fig:m12}). Note,
however, that these numbers get dangerously close to one for the worst
performing mission designs. The number of links does not have a strong
impact on this metric. However it is essential to accumulate cycles in
band, either with a long arm length (A5) or by preserving the target
low-frequency noise (N2).

\noindent(ii) Similar considerations apply to spin measurements. In
order to determine the spin magnitude (Figure~\ref{fig:s12}), it is
essential to achieve the target low-frequency noise (N2), since we
typically predict less than a single good spin measurement (especially
of the secondary $\chi_2$) for some population models and baselines
with sub-optimal low-frequency noise (N1). Spin directions
(Figure~\ref{fig:angle}) are also much better determined using
N2A5M$k$L6 baselines, for which we have more than 10 satisfactory
measurements for all MBHB population models considered in this study.
The errors are generally a factor of $\approx2$ worse for 4 links
(L4), and quickly deteriorate by shortening the arm length, if
sub-optimal low-frequency noise (N1) is assumed.

\noindent(iii) Measurements of the remnant spin $\chi_r$ are quite
insensitive to the details of the detector, as shown in
Figure~\ref{fig:chir}.  This is because the individual spins before
merger are measured from their imprint on the adiabatic inspiral, and those measurements are 
sensitive to the detector's arm length and low-frequency
noise. Conversely, the post-coalescence MBH mass
and spin is measured from the quasinormal modes of the remnant,
which lie at higher frequencies, where differences between the various
configurations are smaller (cf. Fig.~\ref{fig1}).

\noindent(iv) The identification of systems suitable for
electromagnetic follow-up is heavily dependent on the number of
links, as clearly shown in the left panel of
Figure~\ref{fig:omega_bis}. Notice that any six-link configuration
(even with pessimistic low-frequency noise N1) performs better than
NGO. We also notice that adding merger and ringdown (SUA IMR)
mitigates the gap across designs {\it in the six-link case only}. In the four-link case, 
the availability of a more complete waveform model does not significantly improve the performance
of the modestly performing baselines with respect to this particular metric.

\noindent(v) Similar arguments apply to the identification of
high-redshift systems (right panel of Figure~\ref{fig:omega_bis}). As
before, adding merger and ringdown (SUA IMR) partially improves the capability of the
worst performing six-link designs, but it
does not have as much of an impact on the performance of the four-link designs.

From these results we can draw several conclusions about the
benefits/drawbacks of specific design choices with respect to concrete
science goals. Configurations N1A1M2L4 and N1A1M2L6 will only detect a
few systems with mass estimates that are precise at the 1\% level or
better. Their performance is even worse for the MBH spins. The number
of systems with precise measurements of the individual MBH redshifted
masses ($m_{1z}, m_{2z}$), spin magnitudes ($\chi_1, \chi_2$) and spin
orientation angles at the innermost stable circular orbit
($\theta_{\chi_{1}}$, $\theta_{\chi_{1}}$) varies by a factor
$\sim 30$--$100$ across different configurations. However, this is not
true for the remnant spin $\chi_r$, for the reasons explained above.
This means that our ability to probe fundamental physics (e.g. black
hole horizons, no hair theorem, etc.) by measuring ringdown modes of
merging MBHBs is more intimately related to an adequate knowledge of
the waveform and to the intrinsic rate of MBHB mergers in the
Universe. The specific detector baseline (at least in
the range considered in this study) will affect the results by less than a factor of four. Only some of the six-link
configurations ensure localization of a sufficient number of sources
to allow for electromagnetic follow-ups and hence either multimessenger
astronomy or (potentially) studies of the dark-energy equation of
state, for which $\gtrsim 10$ sources would be required. Finally, six-link
configurations with good low-frequency noise (N2) are necessary to
detect high-redshift systems with a relatively small distance error,
thus probing the early epoch of MBH formation. In general,
configurations with worse low-frequency noise (N1) and short arm
length are expected to provide very few detections at high redshifts.
The ability to detect high-redshift MBHs and to localize systems for
systematic electromagnetic follow-up are of crucial
importance for the ``traditional'' astronomy community. Our
  results suggest that six links are a firm requirement
  to achieve these goals. On the other hand, precise mass and spin
measurements are not so sensitive to the number of links, and part of
the scientific potential of the mission is preserved even in a four-link
scenario.

A different way of comparing instrument performance for different
design choices is presented in Fig.~\ref{loss}. Here we have
quantified the impact of each single baseline element on selected
figures of merit. We generated this figure as follows. For each of the three MBHB
population models we considered all pairs of eLISA configurations that
differ only by a specific element (number of links, arm length, low
frequency noise) and compared the results of the respective analyses,
focusing for simplicity on five-year missions (M5). For example, in
assessing the impact of four vs six links (upper-right panel in
Fig.~\ref{loss}), we compared the results of our Fisher matrix analysis for
each of the three MBHB models (popIII, Q2-nod Q2d) using the six
detector pairs N$i$A$j$M5L4 and N$i$A$j$M5L6 (i.e., N1A1M5L4 vs
N1A1M5L6, N1A2M5L4 vs N1A2M5L6, etc.), for a total of 18
comparisons. This procedure yielded 18 comparisons of L4 vs L6 and N1
vs N2, and 12 comparisons of A2 vs A5 and A1 vs A5. For each metric
we plotted a histogram of the ratio of the number of sources satisfying
that particular metric across the comparisons. Note that we always
take the ratio of the worse over the better configuration (L4/L6,
N1/N2, A2/A5, A1/A5), i.e., we quantify the ``science loss'' related
to a specific descoping option.
The two top panels show that either dropping the third arm (L6
$\rightarrow$ L4, top-left panel) or not meeting the target low
frequency sensitivity (N2 $\rightarrow$ N1, top-right panel) can
seriously jeopardize the mission potential. Indeed, the average number
(blue triangle) of high-redshift detections and potential
electromagnetic follow-up targets might drop by a factor 10, harming
the astrophysical impact of the mission. On the other hand,
shortening the arm length to 2~Gm (A5 $\rightarrow$ A2, bottom-left
panel) seems to preserve most of the performance metrics within a
factor of two. Further shortening the arm length to 1~Gm (A2
$\rightarrow$ A1, bottom-right panel), however, is potentially
damaging. Note that these average figures of merit are somewhat
waveform-dependent. As mentioned above, adding merger and ringdown
tends to reduce the gap between different designs, mitigating the
average science loss for several of the figures of merit

\section{Conclusions}
\label{conclusions}

We have investigated the relative performance of different eLISA
designs for the study of the formation and growth of MBHs throughout
the Universe. We have explored the complex aspects of this problem to
the best of our current capabilities, which are necessarily limited by
our incomplete knowledge of MBH astrophysics and waveform models. For
example, work by Littenberg et al.~\cite{Littenberg:2012uj} using
effective-one-body models to improve on the parameter estimation
findings of~\cite{Babak:2008bu,McWilliams:2011zs} found that
systematic errors on NR waveforms were too large to draw any
conclusions. It is therefore urgent and important to revisit our study 
using state-of-the-art waveforms and models for detector design when these are available. In
particular, our assessment of the impact of merger and ringdown should
be seen as a preliminary estimate.

For the reader's convenience, we summarize here the main science
questions that we have addressed (and the figures and tables that
summarize our findings):

\noindent
(i) Can we measure MBH masses and probe the growth of MBHs across
cosmic history? (Figure~\ref{fig:m12}, Tables~\ref{tabmbhspin2} and
\ref{tabmbhspin5}.)

\noindent
(ii) Can we measure the pre-merger spins and probe the nature of MBH
feeding? (Figures~\ref{fig:s12} and \ref{fig:angle},
Tables~\ref{tabmbhspin2} and \ref{tabmbhspin5}.)

\noindent 
(iii) Can we measure the remnant spin (which allows further tests on
MBH feeding, as well as potential tests of the black hole no-hair
theorem)? (Figure~\ref{fig:chir}, Tables~\ref{tabmbhspin2} and
\ref{tabmbhspin5}.)

\noindent 
(iv) Is the sky localization error small enough for close sources that
possible transient electromagnetic counterparts might be identified,
thus allowing us to measure the source redshift and potentially test
the cosmological $D_l(z)$ relation? (Left panel of
Figure~\ref{fig:omega_bis} , Tables~\ref{tabmbhsky2} and
\ref{tabmbhsky5}.)

\noindent 
(v) Can we test seed formation scenarios by detecting MBHBs at high
redshift {\em and} ascertain that they are indeed at high redshift?
(Right panel of Figure~\ref{fig:omega_bis}, Tables~\ref{tabmbhsky2}
and \ref{tabmbhsky5}.)

Our study is somewhat similar in spirit (but different in many
details, ranging from instrumental design to MBH modeling and
waveforms) to a similar investigation that was carried out in 2012 in
the US by the NASA Physics of the Cosmos (PCOS) Gravitational-Wave
Mission Concept Study, whose final report is available
online~\cite{PCOS}.
The PCOS report considered several mission concepts, including ``SGO
High'' (essentially the LISA concept modified to include all known
cost savings, but with the same science performance); ``SGO Mid,''
where the scalable parameters -- the arm length, distance from the
Earth, telescope diameter, laser power, and duration of science
operations -- were all reduced for near maximum cost savings; and
``SGO Low,'' which eliminates one of the measurement arms, giving a
similar performance to ESA's NGO concept.

Here we have adopted the NGO concept as a baseline (N2A1M2L4) and
investigated how different ``science metrics'' vary as we tune
different variables in the mission design. We broadly agree with some
of the main conclusions of the PCOS study, namely that: (i)
scientifically compelling mission concepts exist that have worse sensitivity than the ``classic LISA''
design; (ii) scaling down the three-arm LISA architecture by shortening the
measurement baseline and the mission lifetime (``SGO Mid'' in the PCOS
report terminology) preserves compelling science -- provided the
low-frequency target sensitivity can be achieved (N2 in our
notation) -- and does not increase risk; (iii) eliminating a
measurement arm reduces the science return, as well as increasing mission risk.

In Figure~\ref{loss} we have also quantified the impact of each single
baseline element on selected figures of merit. Our studies suggest that
the cost-saving intervention that preserves the most science is
shortening the arm length from 5~Gm to 2~Gm (A5 $\to$ A2). In this case,
the detector performance in each specific figure of merit is degraded
by at most a factor of two. On the other hand, either dropping the
third arm (L6 $\to$ L4) or not meeting the target low frequency
sensitivity (N2 $\to$ N1) can seriously jeopardize the mission
potential. High-redshift detections and potential electromagnetic
follow-up targets might drop by a factor $\sim 5$--$10$, correspondingly 
reducing the likelihood of coincident observations with
``traditional'' astronomical instruments. Further shortening the arm
length to 1Gm (A2 $\to$ A1) is also potentially damaging. Our results
indicate that compromising on arm length might be the best way to save
on mission costs while preserving most of the original LISA MBHB
science.

Our study also suggests that in order to achieve the mission's science
goals while cutting cost, a significant effort must be put into modeling
the merger and ringdown of the MBHB waveforms: a complete knowledge of
the MBHB IMR waveforms can compensate, at least partially, for cost
reductions in detector design, but only for six-link configurations.
Our study is incomplete in this
respect, since we have included the merger/ringdown by simply rescaling the
angular resolution and distance determination errors by appropriate
powers of the SNR, so further investigations in this direction will be 
particularly
valuable.

\acknowledgments

E.~Barausse acknowledges support from the European
Union's Seventh Framework Programme (FP7/PEOPLE-2011-CIG) through the
Marie Curie Career Integration Grant GALFORMBHS PCIG11-GA-2012-321608, and
from the H2020-MSCA-RISE-2015 Grant No. StronGrHEP-690904.
A.~Klein and E.~Berti are supported by NSF CAREER Grant
No.~PHY-1055103.
E.~Berti is supported by FCT contract IF/00797/2014/CP1214/CT0012
under the IF2014 Programme.
F. Ohme is supported by the European Research Council Consolidator
Grant 647839,
A.~Sesana is supported by a University Research Fellowship of the Royal
Society.
B.~Wardell is supported by the Irish Research Council,
which is funded under the National Development Plan for Ireland.
Computations were
performed on the GPC supercomputer at the SciNet HPC Consortium,
on the Datura cluster at the AEI,
on the XSEDE network (allocation TG-MCA02N014),
and on the ``Projet Horizon Cluster'' at the Institut d'Astrophysique de Paris.
This work was supported in part by the DFG under Grant SFB/Transregio~7
``Gravitational-Wave Astronomy''.

\appendix

\section{Precessing black hole binary hybrid waveforms}
\label{app:Hybrid}

Our study makes use of a set of complete inspiral-merger-ringdown waveforms, each
constituting a ``hybrid'' of an analytical PN prediction and the result of a
fully relativistic NR simulation. In this appendix we shall briefly outline their
construction.

Our waveforms are computed from numerical solutions of the full
Einstein equations in which binary black hole initial data is evolved
using a 3+1 approach through inspiral, merger and ringdown.
The initial data is of Bowen-York
type \cite{Bowen:1980yu,Brandt:1997tf}, and the initial parameters are
chosen (i) for minimal eccentricity ($e \lesssim 10^{-4}$), using a
variation of the method presented in \cite{Purrer:2012wy}, and (ii)
for maximal precession, with the spin directions chosen to maximize
the angle between the orbital and total angular momentum, using an
iterative effective-one-body method.  See Table
\ref{tab:configurations} for a summary of the physical parameters of
the configurations studied.
The initial data is evolved using the BSSN formulation of the Einstein
equations with 8th order finite-differencing using components of the
\texttt{Einstein Toolkit}
\cite{Goodale:2002a,Cactuscode:web,Schnetter:2003rb,Schnetter:2006pg,CarpetCode:web,Ansorg:2004ds,Loffler:2011ay,EinsteinToolkit:web}
in combination with the \texttt{Llama} multipatch
code~\cite{Pollney:2009yz}.  In comparison to purely Cartesian
numerical grids, the use of constant angular resolution grids leads to
high accuracy in the wave zone for not only the dominant $\ell=2, m=\pm 2$
modes, but also the higher modes which are important for precessing
systems.

The waveforms, consisting of 10--16 GW cycles, are
constructed in terms of the standard complex Newman-Penrose scalar
\begin{equation}
 \Psi_4(t) = \frac{\partial^2}{\partial t^2} \left[ h_+(t) - i h_\times (t) \right] = \vert \Psi_4 (t) \vert e^{i  \phi(t)}~,
\end{equation}
where $h_+$ and $h_\times$ are the GW polarizations in the source
frame, which is determined by our numerical simulations.  The
waveforms are measured on coordinate spheres at finite radius, and
extrapolated to future null infinity using standard methods.  The
simulations were all performed at several numerical resolutions to
measure the effect of numerical truncation error.

The ``stitching'' of a numerically obtained signal $\Psi_4^{\rm NR}$
to inspiral data of a particular PN approximant is a well studied procedure commonly applied
to nonprecessing signals
\cite{Ajith:2007kx,Boyle:2008ge,Boyle:2009dg,Sant10,Hannam:2010ky,MacDonald:2011ne,Boyle:2011dy,Ohme:2011zm,Ajith:2012tt}.
However, combining multiple harmonic modes of precessing signals presents
additional challenges, and rapid progress in using precessing hybrids as
well as comparing analytical and numerical waveforms has been reported
recently
\cite{Gualtieri:2008ux,Campanelli:2008nk,Ajith:2009bn,Schmidt:2012rh,
Boyle:2014ioa,Hannam:2013oca,Pan:2013rra,Ossokine:2015vda}.

\begingroup
\squeezetable
\begin{table}
 \begin{tabular}{c|c|cc|cc}
\hline
\hline
~ Config & $D/M$
  & $m_{1}^{\text{h}}$ & $[S_1^x, S_1^y, S_1^z]$
  & $m_{2}^{\text{h}}$ & $[S_2^x, S_2^y, S_2^z]$\\
\hline
~ Q1 & $9$
  & $0.5$ & $[-0.02,-0.01,0.15]$
  & $0.5$ & $[-0.01,-0.01,0.14]$\\
~ Q2 & $9$
  & $0.67$ & $[-0.02,-0.01,0.27]$
  & $0.33$ & $[-0.00,0.01,0.06]$\\
~ Q4 & $9$
  & $0.8$ & $[-0.02,0.02,0.38]$
  & $0.2$ & $[0.00,0.00,0.02]$\\
~ Q2a & $9$
  & $0.67$ & $[0,0,0.27]$
  & $0.33$ & $[0,0,0.06]$\\
\hline
~ Q2HP & $9$
  & $0.67$ & $[0.07,-0.18,0.11]$
  & $0.33$ & $[-0.01,0.02,-0.05]$\\
~ Q4HP & $9$
  & $0.8$ & $[-0.24,-0.30,-0.01]$
  & $0.2$ & $[0.00,-0.00,-0.00]$\\
\hline
\end{tabular}
\caption{Summary of the configurations studied.  All quantities are measured at
  the point where the NR waveform begins and are given in units where they
  have been a-dimensionalized by $M$, the sum of the initial irreducible
  masses of the black holes. $D/M$ is the separation,
  $m_i^\text{h}$ is the irreducible mass,
  and $[S_i^x, S_i^y, S_i^z]$ are the spin vectors.}
\label{tab:configurations}
\end{table}
\endgroup

Here, we follow an approach that is close in spirit to the treatment of 
nonprecessing binaries \cite{Ajith:2012tt}, with the additional complication that 
in the presence of precession we have to track more than just the evolution of 
the binary's orbital frequency. The black hole spins and the orbital plane 
constantly change direction, which leads to an extended set of PN equations that 
have to be integrated. We choose to employ the adiabatic TaylorT4 
approximant (see Section~\ref{sec:SUA} for details).

Apart from the number of equations to integrate, there is the further
difficulty of finding appropriate PN initial data. The spins and the orbital
plane constantly change their orientation, and it is not clear a
priori which PN initial conditions evolve to the same setup as
assumed by the respective NR simulation. One could of course
approach the problem the other way around, namely start with some PN
initial data, evolve the system up to a smaller separation and let
the NR code ``take over'' by feeding in the appropriate quantities
from the end of the PN evolution. This idea was explored already by
Campanelli \textit{et al.}\ \cite{Campanelli:2008nk}, who found that
although the results from PN and NR agree reasonably in the early
inspiral, they quickly differ considerably with progressing
simulation time. The cause of this disagreement is manifold. Apart
from the fact that a truncated PN series will always deteriorate
close to the merger, the disagreement potentially stems from the
different frameworks used in PN and NR to define physical quantities, and in particular from the
transition from Bowen-York initial data
\cite{Bowen:1980yu,Brandt:1997tf} to the actually modeled system in
the NR simulation. 

We overcome this issue here by reading the ``initial'' values $\bm{S}_i(t_{\rm 
ini})$, $\bm{L}(t_{\rm ini})$ and $\omega_{\rm orb}(t_{\rm ini})$ off the NR 
simulation, at a time $t_{\rm ini}$ when the junk radiation has left the system 
and we observe a reasonably clean evolution of the numerical solution. Together with the 
time-independent mass ratio, these quantities complete the set of parameters we 
need to specify in order to integrate the PN equations both forward and backward 
in time. Note that both mass and spin measures in NR typically employ
the formalism of quasilocal horizons \cite{lrr-2004-10}, and
determining the spin direction is a coordinate-dependent process
\cite{Jasiulek:2009zf,Campanelli:2006fy}. When combining PN and NR
descriptions for precessing binaries, however, we are anyway forced
to relate different coordinate systems with each other.
If the black holes are still far enough separated in the simulation, we
can hope to sensibly identify the NR measurements with PN
parameters, in due consideration of the appropriate spin supplementary
condition in PN that ensures constant spin magnitudes
\cite{Faye:2006gx,Kidder:1995zr}. The orbital frequency as well as the direction of the
Newtonian orbital angular momentum are estimated simply through the
coordinate motion of the punctures and the Euclidean vector product of the separation and
the relative velocity. Again, this is a coordinate-dependent measure,
but we merely extract from the NR simulation that the modeled system
is (approximately) characterized at some instant by the specified values.
We also tested the idea of optimizing the initial orbital frequency by a 
least-squares fit, and found slightly better agreement between the PN and NR 
evolutions (of all relevant quantities) when we set $\omega_{\rm orb} (t_{\rm 
ini}) = {\hat \omega}_{\rm orb}$, with ${\hat \omega}_{\rm orb}$ determined in 
turn by minimizing the PN-NR difference of, e.g., the spin of the heavier black 
hole over a few hundred $M$ of evolution time.

Having calculated the PN evolution of $\omega_{\rm orb}(t)$, $\bm{L}(t)$ and 
$\bm{S}_i(t)$, we obtain the GW strain $h = h_+ - i h_\times$ by applying the 
explicit expressions provided in the appendix of \cite{Arun:2008kb}. The 
Newman-Penrose scalar $\Psi_4^{\rm PN}$ follows through two numerical time 
derivatives. However, even assuming that we have modeled the same system
analytically and numerically, we cannot immediately combine the
two waveform parts due to additional subtleties. Firstly, there is
another initial parameter, the initial phase,
which does not enter the waveform simply as $e^{i \phi_0}$ (this is
just
the lowest order effect); there are higher-order amplitude
corrections that
depend on $\phi_0$ \cite{Arun:2008kb}. Knowing them analytically,
however, we can still fit for an optimal $\phi_0$ between the PN and
NR parts of the waveform. Secondly, although finding the initial
parameters also
relates the time between the PN and NR evolution, there is the
problem that the physical quantities affect the PN waveform
\emph{immediately}, whereas if we consider a waveform extracted at
some finite radius in NR, there is a time lag between the spin
evolution and the observed GW signal. As already discussed in
\cite{Campanelli:2008nk}, this time lag approximately corresponds to
the travel time between source and observer, but gauge effects will
spoil this relation, and we shall determine both $\phi_0$ and $t_0$ by
an additional least-squares fit of the GW phase, just as in the
nonprecessing case.

In short, we (1) make sure to simulate the same physical system
numerically and analytically by reading the PN parameters off the NR
simulation, and then (2) combine the NR and PN parts of the
waveform by minimizing the phase difference in $\Psi_4$ over a certain
length of evolution time. The choice of this interval would ideally be
based on considerations concerning the hybridization accuracy (see
e.g.~\cite{Sant10,MacDonald:2011ne}). Our goal, however, is not to
produce highly accurate template waveforms to be eventually used in GW
searches; we merely want to complete the PN description in a
reasonably well motivated and robust way. Hence, we simply overlay the
PN and NR waveforms in a region of approximately $250M$ length, as
early as the NR simulation permits. Within this interval, the GW
frequency $\omega(t) = d \phi(t)/dt$ evolves from
$M\omega \approx 0.065$ to $M \omega \approx 0.08$, which is more than
the minimal frequency evolution suggested in \cite{MacDonald:2011ne}
to ensure an unambiguous matching.

Once we have matched the dominant spherical harmonic $\ell =2, m= 2$ mode, all other
modes are aligned as well. There is no additional freedom left to apply any time or
phase shift to individual modes. We can only check that the agreement is similar to
the dominant mode, and indeed, we find that the phase difference between PN and NR in
the matching region is comparable ($<0.1\, {\rm rad}$) for all spherical harmonic
modes. The accuracy of the PN amplitude, however, degrades towards higher modes
(higher spherical harmonic modes enter at different PN orders and are thus determined
to lower relative expansion order \cite{Blanchet:2008je,Arun:2008kb}), which
effectively limits our matching procedure to modes with $\ell \leq 4$ and $m = \pm
\ell$.

The final hybrid waveform is now constructed mode by mode as a smooth connection of the PN and NR parts of the signal,
\begin{equation}
 \left\vert \Psi_4^{\rm hyb}(t) \right\vert = \left \vert \Psi_4^{\rm PN}(t) \right \vert \big[1 - \mathcal T(t) \big] +
 \left \vert \Psi_4^{\rm NR}(t) \right \vert \mathcal T(t)\,.
\end{equation}
An equivalent transition is also used separately for the phase $\phi(t)$, and $\mathcal T$ is a blending function. We employ a form of the Planck taper function,
\begin{equation}
 \mathcal T (t) = \left \{
\begin{array}{cl}
 0, & t \leq t_1 \\
\left[\exp \left( \frac{t_2 - t_1}{t-t_1} + \frac{t_2 -
t_1}{t-t_2}\right) + 1 \right]^{-1} , & t_1 < t < t_2 \\
1 , & t> t_2
\end{array}
\right.
\end{equation}
as suggested by \cite{McKechan:2010kp}.  The parameters $t_1$ and $t_2$ used for constructing the phase are defined by the matching interval that determined the optimal time and phase shift between the PN and NR parts of the waveform. However, we find that a slightly larger value of $t_2$ results in a smoother amplitude transition, which in turn avoids artifacts in the transformation from $\Psi_4^{\rm hyb}$ to $h^{\rm hyb}$. Finally, we obtain $h^{\rm hyb}$ by two time integrations in the Fourier domain, as suggested by \cite{Sant10,Reisswig:2010di}. A comparison with the original PN waveform $h^{\rm PN}$ ensures that the transformation is accurate, and we generally find
 \begin{equation}
 \left \vert\frac{ h^{\rm PN}(t) - h^{\rm hyb}(t)}{ h^{\rm
hyb}(t) } \right
\vert  < 1\% 
\end{equation}
for the dominant mode. This is not merely a statement about the
amplitude accuracy, it also confirms that the phases agree very
accurately over thousands of $M$ in evolution time. We further
confirmed the robustness of our hybrids by producing NR data at three
different resolutions for each configuration. The hybrids we obtain
are consistent among all NR resolutions, and we used the highest
resolution for the results presented here.

\section{Estimate of the error on the remnant spin}
\label{app:ringdown}

To estimate the errors on the remnant mass and spin we
follow~\cite{Berti:2005ys}. The post-merger waveform is dominated by
quasinormal ringing. The dominant oscillation modes have large quality
factor, so one can use an approximation where each mode is replaced by
a $\delta$-function at the appropriate oscillation frequency and
compute the SNR as in Eq.~(3.16) of~\cite{Berti:2005ys}, where we
include redshift factors and substitute the Euclidean distance $r$ by
the luminosity distance $D_l$ as appropriate:
\beq
\label{rhoanalytic}
\rho_{\rm FH} &=& \left (\frac{2}{5} \right )^{1/2} 
	\left (\frac{1}{\pi {\cal F}_{lmn} D_l} \right )
	\left (\frac{\epsilon_{\rm rd}}{S_h^{\rm NSA}(f_{lmn})} \right
        )^{1/2} \nonumber \\
        &\times& \left[M(1+z)\right]^{3/2}
	\frac{2Q_{lmn}}{\sqrt{1+4Q_{lmn}^2}} \,.
\eeq
Here $M$ is the remnant mass in the source frame,
${\cal F}_{lmn}=M\omega_{lmn}$ is the dimensionless oscillation
frequency, and $Q_{lmn}$ is the quality factor of a quasinormal mode
with angular indices $(l,m)$ and overtone number $n$. Note that this
expression involves the {\it non-sky-averaged} noise curve:
see~\cite{Berti:2005ys} for details. For the fundamental mode with
$l=m=2$, the frequency and damping time of the oscillations are well
fitted by (cf.~Table VIII of~\cite{Berti:2005ys})
\beq
{\cal F}_{lmn}&=&
f_1+f_2(1-\chi_r)^{f_3}\,,\\
Q_{lmn}&=&q_1+q_2(1-\chi_r)^{q_3}\,,
\eeq
where $f_1=1.5251$, $f_2=-1.1568$, $f_3=0.1292$, $q_1=0.7000$,
$q_2=1.4187$, $q_3=-0.4990$, and $\chi_r$ is the dimensionless spin of the
final black hole.
Then the errors on mass and spin, in the Flanagan-Hughes
convention~\cite{Flanagan:1997sx}, can be estimated as in Eqs.~(4.12a)
and (4.12b) of~\cite{Berti:2005ys}:
\begin{subequations}
\label{rii}
\beq
\Delta \chi_r &=&
\frac{1}{\rho_{\rm FH}}
\left|2\frac{Q_{lmn}}{Q_{lmn}'}
\left(1+\frac{1+4\beta}{16\Qlm^2}\right)\right|\,, \label{erra}\\
\frac{\Delta M}{M} &=&
\frac{1}{\rho_{\rm FH}}
\left|2\frac{Q_{lmn}
f_{lmn}'}{f_{lmn}Q_{lmn}'}
\left(1+\frac{1+4\beta}{16 \Qlm^2}\right)\right|\,, \label{errm}
\eeq
\end{subequations}
where a prime denotes a derivative with respect to $\chi_r$. In general we would
have
\be
\beta = \sin^2 \psi \cos 2\ph^\times - \cos^2 \psi \cos 2\ph^+\,
\ee
\vspace{0.025in}

\noindent with 
%
$\cos \psi \equiv (1+N_\times^2)^{-1/2}$, 
%
$\sin \psi \equiv N_\times (1+N_\times^2)^{-1/2}$. 
The parameter $N_\times$ is the ratio between plus and cross
polarization amplitudes. Following Flanagan and
Hughes~\cite{Flanagan:1997sx} we set $N_\times=1$,
$\phi^+=\phi^\times=0$, and therefore $\beta=0$, which simplifies the
expressions even further.

As an estimate of the ringdown efficiency $\epsilon_{\rm rd}$
we use the ``matched-filtering based'' estimate of Eq.~(4.17)
in~\cite{Berti:2007fi}:
\be
\epsilon_{\rm rd}=0.44\frac{q^2}{(1+q)^4}\,,
\ee
where $q=m_1/m_2>1$ is the mass ratio of the binary (see also \cite{Nagar:2006xv,Bernuzzi:2010ty,Barausse:2012qz} for similar scalings).  This estimate is
conservative, in the sense that it is appropriate for nonspinning
binary mergers.
Spin corrections should modify the efficiency by an amount which is
roughly proportional to the sum of the components of the binary spins
along the orbital angular momentum~\cite{Kamaretsos:2012bs}. These
spin-dependent corrections will change $\epsilon_{\rm rd}$ by at most
a factor $\approx 2$, which is within the scope of our
order-of-magnitude calculation. We plan to improve the accuracy of
these estimates in future work.

\bibliography{SMBHB_paper,PRD_Ao_Ba_Hi_Oh_Pe_Se_Wa}

\end{document}